\documentclass[a4paper,11pt]{article}
\usepackage{jheppub} 

\pdfoutput=1

\usepackage{mathrsfs,amsmath,amsthm,amssymb,xcolor,epstopdf,verbatim,hyperref,enumerate,graphicx,hyperref,subfigure,bm,bbm,amsfonts,subdepth,cleveref,setspace,booktabs,tabularx,placeins,multirow,mathtools,units}
\usepackage{cleveref}
\usepackage[english]{babel}
\usepackage[sort&compress]{natbib}
\usepackage[normalem]{ulem}
\usepackage{dcolumn}
\graphicspath{{Figs/}}
\usepackage{float}

\usepackage{tabularx}

\newcommand{\drawsquare}[2]{\hbox{%
		\rule{#2pt}{#1pt}\hskip-#2pt
		\rule{#1pt}{#2pt}\hskip-#1pt
		\rule[#1pt]{#1pt}{#2pt}}\rule[#1pt]{#2pt}{#2pt}\hskip-#2pt
	\rule{#2pt}{#1pt}}

\newcommand{\fund}{\raisebox{-.5pt}{\drawsquare{6.5}{0.4}}}
\newcommand{\Ysymm}{\raisebox{-.5pt}{\drawsquare{6.5}{0.4}}\hskip-0.4pt%
	\raisebox{-.5pt}{\drawsquare{6.5}{0.4}}}
\newcommand{\Yasymm}{\raisebox{-3.5pt}{\drawsquare{6.5}{0.4}}\hskip-6.9pt%
	\raisebox{3pt}{\drawsquare{6.5}{0.4}}}
\newcommand{\antifund}{\overline{\fund}}

 \renewcommand{\thetable}{\arabic{table}} 
 \def\ov{\overline}
\usepackage{jheppub} 
\begin{document}

\title{String-scale Gauge Coupling Relations in the Supersymmetric Pati-Salam Models from Intersecting D6-branes}

\author[a,b,c]{Tianjun Li,}
\author[d]{Rui Sun}
\author[,a]{and Lina Wu\footnote{Corresponding author}}

\affiliation[a]{School of Sciences, Xi'an Technological University, Xi'an 710021,  P. R. China}
\affiliation[b]{CAS Key Laboratory of Theoretical Physics, Institute of Theoretical Physics,\\
		Chinese Academy of Sciences, Beijing 100190, P. R. China}
\affiliation[c]{School of Physical Sciences, University of Chinese Academy of Sciences,\\
		No.19A Yuquan Road, Beijing 100049, P. R. China}
\affiliation[d]{Korea Institute for Advanced Study,\\
		85 Hoegiro, Dongdaemun-Gu, Seoul 02455, Korea}

\emailAdd{tli@itp.ac.cn}
\emailAdd{sunrui@kias.re.kr}
\emailAdd{wulina@xatu.edu.cn}

\abstract{
We have constructed all the three-family ${\cal N} = 1$ supersymmetric Pati-Salam models from intersecting D6-branes, and obtained 33 independent models in total. But how to realize the string-scale gauge coupling relations in these models is a big challenge. We first discuss how to decouple the exotic particles in these models. In addition, we consider the adjoint chiral mulitplets for $SU(4)_C$ and $SU(2)_L$ gauge symmetries, the Standard Model (SM) vector-like particles from D6-brane intersections, as well as the vector-like particles from the ${\cal N}=2$ subsector. We show that  the gauge coupling relations at string scale can be achieved via two-loop renormalization group equation running for all these supersymmetric Pati-Salam models. Therefore, we propose a concrete way to obtain the string-scale gauge coupling relations for the generic intersecting D-brane models.
}

\maketitle

\section{Introduction}

The three-family ${\cal N} = 1$ supersymmetric Pati-Salam models from intersecting D6-branes have been studied extensively since we can generate the Yukawa couplings for the Standard Model (SM) fermions, and break the Pati-Salam gauge symmetry down to the SM gauge symmetry  via the D-brane splitting and supersymmetry preserving Higgs mechanism~\cite{Lykken:1998ec,Cvetic:1999hb,Cvetic:2000st,Aldazabal:2000sa,Berkooz:1996km,Cvetic:2004ui, Chen:2007ms,Aldazabal:2000cn,Cvetic:2001tj,Blumenhagen:2002gw,Chen:2005aba,Blumenhagen:2005mu, Aldazabal:2000dg,Ibanez:2001nd, Sabir:2022hko}. 
Realizing the SM in D-brane vacua from Orbifold configuration and Gepner configurations were also stuided in~\cite{Honecker:2012jd, Honecker:2017air, Anastasopoulos:2005ba, Anastasopoulos:2006da,Anastasopoulos:2009mr,Anastasopoulos:2010ca, Anastasopoulos:2010hu,Anastasopoulos:2011zz, Ecker:2015vea, Marchesano:2022qbx, Antoniadis:2004dt, Antoniadis:2010zze, Kiritsis:2007zz, Kiritsis:2003mc}. 
However, deriving all the possible Pati-Salam models is a challenging topic due to the large amount of possible wrapping numbers~\cite{Chen:2007px,Chen:2007zu, Cvetic:2003xs,Cvetic:2002pj,Cvetic:2002qa,Cvetic:2002wh,Cvetic:2003yd,Cvetic:2003ch,Honecker:2003vq,Chen:2005mj}. 
The full landscape of intersecting D-brane models have been under investigation via different perspectives~\cite{Douglas:2006xy,Halverson:2019tkf, Loges:2021hvn, Loges:2022mao, Cvetic:2004nk, Cvetic:2001nr, Li:2019nvi, Li:2021pxo, Li:2019miw,Li:2022fzt}. 
In the Type IIA  string theory on $T^6/(\mathbb{Z}_2\times \mathbb{Z}_2)$ orientifold with intersecting D6-branes, we propose a systematic method to construct all the possible 202752 ${\cal N}=1$ supersymmetric Pati-Salam models~\cite{He:2021kbj, He:2021gug}.  After modding out the equivalent classes, we obtain 33 physical independent models with 33 types of gauge coupling relations. 
The main idea is to solve the common solutions of three generation conditions, and RR tadpole cancellation conditions iteratively.  Thus, we complete the landscape for one particular type of intersecting D-brane models for the first time.  However, how to realize the string-scale gauge coupling relations in these models is still a big challenge. 

It is well-known that gauge coupling unification can be achieved in the Minimal Supersymmetric SM (MSSM)~\cite{Ellis:1990wk, Langacker:1991an, Amaldi:1991cn}, which strongly suggested the Grand Unified Theory (GUT). The unification scale or say GUT scale is around $2\times 10^{16}$ GeV. On the other hand, the string scale is about one order larger. 
For example, in the weakly coupled heterotic string theory, the string scale $M_{\rm string}$ is given by~\cite{Dienes:1996du}
\begin{eqnarray} 
M_{\rm string} = g_{\rm string} \times 5.27\times 10^{17}~{\rm GeV}~,~\,
\end{eqnarray} 
where $g_{\rm string}$ is string coupling constant. Because $g_{\rm string} \sim {\cal O} (1)$ is an order one coupling constant, we obtain
\begin{eqnarray} 
M_{\rm string} =  5\times 10^{17}~{\rm GeV}~.~\,
\end{eqnarray} 
And then we have a factor about 25 between the GUT scale and string scale. Thus, how to realize the string-scale gauge
coupling unification is an important question in string phenomenology. In principle, the string scale can be low due to the large volume of the extra-dimension \cite{Antoniadis:2002qm,Antoniadis:2011bi}. However, recall that the Yang-Mills gauge coupling on the D6-brane stack $x$ at string scale is given by~\cite{Dienes:1996du}  
\begin{equation}
    (g_{YM}^x)^2=\dfrac{\sqrt{8\pi}M_{\rm string}}{M_{\rm Pl}}\dfrac{1}{\prod_{i=1}^3\sqrt{(n^i_x)^2\chi_i^{-1}+(2^{-\beta_i l^i_x})^2\chi_i}},
\end{equation}
where $M_{\rm Pl}$ is the 4-dimensional Planck scale. 
To have the proper Yang-Mills coupling $g_{YM}\sim {\cal O}(1)$ at order one, the string scale is expected to be closer to the Planck scale $M_{\rm Pl}$ to have proper gauge couplings. Therefore, we set the string scale at $M_{\rm string} =  5\times 10^{17}~{\rm GeV}.$

Gauge coupling unification has been studied extensively previously from various physics point of view~\cite{Bachas:1995yt, Lopez:1996gd, Blumenhagen:2003jy, Barger:2005qy, Jiang:2006hf, Barger:2007qb, Jiang:2008xrg, Jiang:2009za,Kokorelis:2016ckp, Chen:2017rpn, Chen:2018ucf}. In general, to achieve the string-scale gauge coupling unification, we need to introduce the new particles at the intermediate scale, such as the SM vector-like particles, and the $SU(3)_C$/$SU(2)_L$ adjoint particles. In particular, we need to obtain these new particles from string model building as well. Moreover, there are two kinds of the string-scale gauge coupling unification: one-step unification~\cite{Bachas:1995yt}, and two-step unification for flipped $SU(5)\times U(1)_X$ models~\cite{Lopez:1996gd, Jiang:2006hf, Jiang:2008xrg, Jiang:2009za}, where the $SU(3)_C\times SU(2)_L$ gauge symmetries are unified around the traditional GUT scale, and then $SU(5)\times U(1)_X$ gauge symmetries are unified at the string scale.

In the three-family ${\cal N} = 1$ supersymmetric Pati-Salam models from the Type IIA string theory on $T^6/(\mathbb{Z}_2\times \mathbb{Z}_2)$ orientifold with intersecting D6-branes~\cite{He:2021kbj, He:2021gug}, there is one and only one model with gauge coupling unification at the string scale. For the rest 32 models, 
the gauge couplings for $SU(4)_C$, $SU(2)_L$, and $SU(2)_R$ are not unified at the string scale, and then
we have the gauge coupling relations at the string scale as follows
\begin{equation}
	k_3 g_a^2=k_2g_b^2=k_Y g_Y^2 = k_y\left(\frac{5}{3} g_Y^2\right)=g_{\text{U}}^2 \sim g_{\text{string}}^2~,~\,
\end{equation}
where $g_a$, $g_b$, and $g_Y$ are respectively the gauge couplings for $SU(3)_C$, $SU(2)_L$, and $U(1)_Y$, and $k_3$, $k_2$, $k_Y$, and $k_y$ are rational numbers. The canonical normalization in $SU(5)$ and $SO(10)$ models give $k_3=1$, $k_2=1$ and $k_Y=5/3$. For simplicity, we shall choose $k_3=1$.

In the string model building, we generically have exotic particles, and thus need to study how to decouple the exotic particles first. We explain that the exotic particles in most of our models can be decoupled except  Model 4, 23, and 32, which have chiral multiplets under $SU(4)_C$ symmetric representation. To realize the string-scale gauge coupling relations, we assume that the exotic particles in these three models can be decoupled as well.
Moreover, we can have two scenarios to obtain the string-scale gauge coupling relations: traditional one step, and two steps where the SM gauge symmetry becomes Pati-Salam gauge symmetry at the intermediate scale, and then their gauge couplings satisfy the gauge coupling relations at string scale. In this paper, we only consider the first scenario, and will study the second scenario in the future. Furthermore, we consider the adjoint chiral mulitplets for $SU(4)_C$ and $SU(2)_L$ gauge symmetries, the SM vector-like particles from D6-brane intersections, as well as the vector-like particles from the ${\cal N}=2$ subsector. We show that  the gauge coupling relations at string scale can be achieved via two-loop renormalization group equation (RGE) running for all these supersymmetric Pati-Salam models. Therefore, we propose a concrete way to obtain the string-scale gauge coupling relations for the generic intersecting D-brane models.

This paper is organized as follows. In Section II we first review the construction of supersymmetric Pati-Salam models and present the relevant knowledge.  In Section III, we discuss how to decouple the exotic particles. 
In Section IV, we perform the two-loop RGEs of gauge couplings as well as one-loop RGEs of Yukawa couplings.  We show that the string-scale gauge coupling relations are achieved after taking into account the contributions from extra vector-like particles. In Section V we finally conclude the results and make an outlook.

\section{The Supersymmetric Pati-Salam Models from  Type IIA $T^6/(\mathbb{Z}_2\times \mathbb{Z}_2)$  Orientifolds with Intersecting D6-branes}
	
The supersymmetric Pati-Salam models have been constructed on Type IIA $T^6/(\mathbb{Z}_2\times \mathbb{Z}_2)$  orientifolds with D6-branes intersecting at generic angles.
The orbifold group $\mathbb{Z}_{2} \times \mathbb{Z}_{2}$ results in generators $\theta$ and $\omega$ and associated with  twist vectors $(1/2,-1/2,0)$ and $(0,1/2,-1/2)$ respectively. 
They act on the complex coordinates $z_i$ as~\cite{Cvetic:2001nr, Cvetic:2002pj}
\begin{eqnarray} 
& \theta: & (z_1,z_2,z_3) \mapsto (-z_1,-z_2,z_3)~,~ \\ \nonumber 
& \omega: & (z_1,z_2,z_3) \mapsto (z_1,-z_2,-z_3)~.~\,
\label{orbifold} 
\end{eqnarray} 
The orientifold projection acts by gauging the $\Omega R$ symmetry, where $\Omega$ is world-sheet parity and $R$ acts on the complex coordinates as
\begin{equation}
 R: (z_1,z_2,z_3) \mapsto ({\ov z}_1,{\ov z}_2,{\ov	z}_3)~.~\, 
\end{equation}
Overall there are four kinds of orientifold 6-planes (O6-planes) contributes to the actions of $\Omega R$, $\Omega R\theta$, $\Omega R\omega$, and $\Omega R\theta\omega$ respectively~\cite{Blumenhagen:2000ea,Chen:2007zu,Cvetic:2001nr, Cvetic:2002pj}. 
Three stacks of $N_a$ D6-branes wraps on the factorized three-cycles cancelling the RR charges of these O6-planes.

The homology classes of these three cycles are wrapped by the D6-brane stack, taking the form for a rectangular torus as $n_a^i[a_i]+m_a^i[b_i]$  and for a tilted torus as $n_a^i[a'_i]+m_a^i[b_i]$, with $[a_i']=[a_i]+\frac{1}{2}[b_i]$.
We label the generic the generic one cycle by $(n_a^i,l_a^i)$ in terms of the so-called wrapping numbers, with $l_{a}^{i}\equiv m_{a}^{i}$ for a rectangular and $l_{a}^{i}\equiv 2\tilde{m}_{a}^{i}=2m_{a}^{i}+n_{a}^{i}$ for  tilted two-torus respectively. Here $l_a^i-n_a^i$ is even for tilted two-tori, yet odd for rectangular torus.
Therefore, the wrapping number for stack $a$ of D6-branes along the cycle are denoted by $(n_a^i,l_a^i)$, while the $\Omega R$ images ${a'}$ stack of $N_a$ D6-branes can be labeled by wrapping numbers $(n_a^i,-l_a^i)$. 
The homology three-cycles for $a$ stack of D6-branes and  its orientifold image  $a'$  results in 
\begin{eqnarray*} \label{homo}
&[\Pi_a]=\prod_{i=1}^{3}\left(n_{a}^{i}[a_i]+2^{-\beta_i}l_{a}^{i}[b_i]\right),\\
&[\Pi_{a'}]=\prod_{i=1}^{3} \left(n_{a}^{i}[a_i]-2^{-\beta_i}l_{a}^{i}[b_i]\right),
\end{eqnarray*}
in which $\beta_i=0$ when the $i$-th two-torus is rectangular while $\beta_i=1$ for tilted two-torus.
Furthermore, the homology three-cycles wrapped by the four O6-planes leads to 
\begin{eqnarray}  \label{orienticycles}
		&\Omega R: [\Pi_{\Omega R}]= 2^3 [a_1]\times[a_2]\times[a_3]~,~\, \nonumber\\
		& \Omega R\omega: [\Pi_{\Omega	R\omega}]=-2^{3-\beta_2-\beta_3}[a_1]\times[b_2]\times[b_3]~,~\nonumber\\
		&\Omega R\theta\omega: [\Pi_{\Omega	R\theta\omega}]=-2^{3-\beta_1-\beta_3}[b_1]\times[a_2]\times[b_3]~,~\nonumber\\
		&\Omega R\theta:  [\Pi_{\Omega		R}]=-2^{3-\beta_1-\beta_2}[b_1]\times[b_2]\times[a_3]~.~
\end{eqnarray}
Based on these, the RR tadpole cancellation condition provides the restriction rule~\cite{Cvetic:2001nr, Aldazabal:2000dg,Ibanez:2001nd}
\begin{eqnarray}  \label{tap2}
		-2^k N^{(1)}+\sum_a N_a A_a=-2^k N^{(2)}+\sum_a N_a
		B_a= \nonumber\\ -2^k N^{(3)}+\sum_a N_a C_a=-2^k N^{(4)}+\sum_a
		N_a D_a=-16~,
\end{eqnarray}
where $2 N^{(i)}$ is the number of
D6-branes wrapping along the $i$-th O6-plane\,(filler branes) and the last term represents the O6-planes  with $-4$ RR charges in D6-brane charge unit. Note that for simplification, we denote 
	$ A_a \equiv -n_a^1n_a^2n_a^3,\,B_a \equiv n_a^1l_a^2l_a^3,\,
	C_a \equiv l_a^1n_a^2l_a^3,\, D_a \equiv l_a^1l_a^2n_a^3, \,
	\tilde{A}_a \equiv -l_a^1l_a^2l_a^3,\,  \tilde{B}_a \equiv
	l_a^1n_a^2n_a^3,  \,\tilde{C}_a \equiv n_a^1l_a^2n_a^3,\, 
	\tilde{D}_a \equiv n_a^1n_a^2l_a^3$. These filler branes represent the $USp$ group carrying the wrapping numbers with one of the O6-planes as shown in table~\ref{orientifold}.
	
\renewcommand{\arraystretch}{1}
\begin{table}[t]  
	\begin{center}
		\begin{tabular}{c|c|c}
			\hline\hline
			{\bf	Orientifold action} & 	{\bf O6-Plane }& $(n^1,l^1)\times (n^2,l^2)\times	(n^3,l^3)$\\
			\hline	
			$\Omega R$& 1 & $(2^{\beta_1},0)\times (2^{\beta_2},0)\times
			(2^{\beta_3},0)$ \\
			\hline
			$\Omega R\omega$& 2& $(2^{\beta_1},0)\times (0,-2^{\beta_2})\times
			(0,2^{\beta_3})$ \\
			\hline
			$\Omega R\theta\omega$& 3 & $(0,-2^{\beta_1})\times
			(2^{\beta_2},0)\times
			(0,2^{\beta_3})$ \\
			\hline
			$\Omega R\theta$& 4 & $(0,-2^{\beta_1})\times (0,2^{\beta_2})\times
			(2^{\beta_3},0)$ \\
			\hline \hline
		\end{tabular}
	\end{center}
	\caption{The wrapping numbers for four O6-planes.}\label{orientifold}
\end{table}

To have three family of chiral fermions under $SU(4)_C\times SU(2)_L\times SU(2)_R$ gauge symmetries, further  constraints were imposed on the intersection numbers expressed in terms of the wrapping numbers
\begin{eqnarray}  \small \label{intersection}
	&I_{ab}=[\Pi_a][\Pi_b]=2^{-k}\prod_{i=1}^3(n_a^il_b^i-n_b^il_a^i)~,~\nonumber\\
	&I_{ab'}=[\Pi_a][\Pi_{b'}]=-2^{-k}\prod_{i=1}^3(n_{a}^il_b^i+n_b^il_a^i)~,~\nonumber\\
	&I_{aa'}=[\Pi_a][\Pi_{a'}]=-2^{3-k}\prod_{i=1}^3(n_a^il_a^i)~,~\nonumber\\
	&I_{aO6}=[\Pi_a][\Pi_{O6}]=	
	2^{3-k}(-l_a^1l_a^2l_a^3	+l_a^1n_a^2n_a^3+n_a^1l_a^2n_a^3+n_a^1n_a^2l_a^3)
\end{eqnarray}
where $k=\beta_1+\beta_2+\beta_3$ is the total number of tilted two-tori, and $[\Pi_{O6}]=[\Pi_{\Omega R}]+[\Pi_{\Omega R\omega}]+[\Pi_{\Omega R\theta\omega}]+[\Pi_{\Omega R\theta}]$ is the sum of  four O6-plane homology three-cycles.
And the intersection numbers shall follow
\begin{eqnarray}\small
	\label{3gen}
	I_{ab} + I_{ab'}=3~,~
	I_{ac} =-3~,~ I_{ac'}=0~,~\,
\end{eqnarray} 
with $c$ and $c'$ exchange as well.

The massless particle spectrum for the supersymmetric Pati-Salam is listed in table \ref{spectrum}, where the gauge symmetry results from $\mathbb Z_2\times \mathbb Z_2$ orbifold projection~\cite{Blumenhagen:2003jy}.
The representations refer to $U(N_a/2)$ when the intersecting D6-branes are of number $N_a=8, N_b=4, N_c=4$.
Note that the chiral supermultiplets not only represents the scalars but also the fermions in this supersymmetric constructions. Moreover, positive intersection numbers refer to the left-handed chiral supermultiplets. In our later discussion, some of the introduced particles for gauge coupling unification can be read off from the spectrum table as well.
{\small \begin{table}[h]
	\renewcommand{\arraystretch}{1.25}
	\begin{center}
		\begin{tabular}{c|c}
			\hline\hline {\bf Sector} & 
			{\bf
				Representation}
			\\
			\hline
			$aa$   & $U(N_a/2)$ vector multiplet  \\
			& 3 adjoint chiral multiplets  \\
			\hline
			$ab+ba$   & $I_{ab}$ $(\fund_a,\antifund_b)$ fermions   \\
			\hline
			$ab'+b'a$ & $I_{ab'}$ $(\fund_a,\fund_b)$ fermions \\
			\hline $aa'+a'a$ &$\frac 12 (I_{aa'} - \frac 12 I_{a,O6})\;\;
			\Ysymm\;\;$ fermions \\
			& $\frac 12 (I_{aa'} + \frac 12 I_{a,O6}) \;\;
			\Yasymm\;\;$ fermions \\
			\hline\hline
		\end{tabular}
	\end{center}
\caption{General massless particle spectrum for intersecting D6-branes at generic angles.}\label{spectrum}
\end{table}
}

In addition to the three generation conditions and tadpole cancellation condition, we further need to require $N=1$ supersymmetry preservation in four dimension, with the equality and inequality conditions~\cite{Berkooz:1996km,Cvetic:2002pj}
 \begin{eqnarray}
		\label{eq:susy}
		x_A\tilde{A}_a+x_B\tilde{B}_a+x_C\tilde{C}_a+x_D\tilde{D}_a=0,
		\nonumber\\ A_a/x_A+B_a/x_B+C_a/x_C+D_a/x_D<0,
		\label{susyconditions}
\end{eqnarray} where $x_A=\lambda,\;
x_B=\lambda 2^{\beta_2+\beta3}/\chi_2\chi_3,\; x_C=\lambda
2^{\beta_1+\beta3}/\chi_1\chi_3,\; x_D=\lambda
2^{\beta_1+\beta2}/\chi_1\chi_2$.
Here $\chi_i=R^2_i/R^1_i$ represent the complex structure moduli of the $i$-th two-torus, and $\lambda$  is introduced as a positive parameter  to put all the variables $A,\,B,\,C,\,D$ at equal footing. 

The supersymmetric Pati-Salam gauge symmetry $SU(4)_C\times SU(2)_L\times SU(2)_R$ can then be broken down to the SM gauge symmetry via the D-brane splitting  and Higgs mechanism preserving supersymmetry. 
Concrete supersymmetric Pati-Salam models have been constructed~\cite{Li:2019nvi, Li:2021pxo}, 
and such D-brane models have also been investigated with powerful reinforcement machine learning methods~\cite{Halverson:2019tkf, Loges:2021hvn}. 
In~\cite{He:2021gug, He:2021kbj}, we for the first time propose a systematic method to construct all the three-family ${\cal N}=1$ supersymmetric Pati-Salam models and conclude that there are in total 202752 models with 33 different gauge coupling relations.  Among which, there is one class of models with gauge coupling unification at GUT scale, while the others do not. 
The absence of gauge coupling unification at the string scale appear be a generic problem. In~\cite{He:2021kbj}, we propose that by introducing vector-like particles from ${\cal N}=2$ subsector the gauge unification problem may be solved. In particular, the number of these exotic particles are fully determined by the brane intersection number. 
These additional particles can be decoupled as discussed in~\cite{Chen:2007px}, and its gauge coupling relation  can be realized at string scale via two-loop RGE running. 
This leads to our exploration that whether this solved for all the supersymmetric Pati-Salam models in general. We will discuss in details for all the 33 representative models.

\section{Decoupling of the Exotic Particles}

In the string model building, there exist exotic particles in general. So we need to discuss how to decouple the exotic particles first. In our Pati-Salam models which are given in Appendix A, we can decouple most of the exotic particles via Higgs mechanism and instanton effects, etc, except the chiral multiplets under $SU(4)_C$ symmetric representation. The key point is the gauge anomaly cancellation. First, the chiral multiplets under $SU(4)_C$ anti-symmetric representation do not contribute to the gauge anomaly. Their mass terms are forbidden by the anomalous $U(1)$ gauge symmetries, and  can be generated  via the  instanton effects~\cite{Blumenhagen:2006xt, Haack:2006cy, Florea:2006si}. Thus, they can be decoupled. Second, for the models without the chiral multiplets under $SU(4)_C$ symmetric representation, all the exotic particles can be decoupled via Higgs mechanism and instanton effects. For a concrete example, please see~\cite{Chen:2007px,Chen:2007zu}. Third, for  the models with the chiral multiplets under $SU(4)_C$ symmetric representation, it seems to us that we cannot decouple the exotic particles. Therefore, we cannot decouple the exotic particles only in Model 4, 23, and 32 in Appendix~\ref{apdx-data}. To study the gauge coupling unification, we assume the extoic particles in these models can be decoupled as well.

\section{String-scale Gauge Coupling Relations}

As presented with details in~\cite{He:2021gug,He:2021kbj}, the full list of supersymmetric Pati-Salam models have $33$ different gauge coupling relations, see Appendix~\ref{apdx-data}. In these models, $a$ stack of D6-branes gives the $U(4)$ gauge symmetry, $b$ stack of D6-branes gives  the $U(2)_L$ gauge symmetry, and $c$ stack of D6-branes gives the $U(2)_R$ gauge symmetry.   
In particular, there is one class of models with gauge coupling unification at GUT scale. Namely,
the strong, weak and hypercharge gauge couplings $g^2_a,  g^2_b$ and $\frac{5}{3}g^2_Y$ satisfy $g^2_a=g^2_b=g^2_c=(\frac{5}{3}g^2_Y)=4\sqrt{\frac{2}{3}} \pi  e^{\phi ^4}$, where $\phi^4$ represents the dilaton field.
However, not all the supersymmetry Pati-Salam models are with gauge coupling unification at GUT scale.
To have string-scale gauge coupling relation, we utilize the
Renormalization Group Equations\,(RGEs) evolution~\cite{Chen:2017rpn,Chen:2018ucf,He:2021kbj}.

The RGEs for the gauge couplings at the two-loop level are given by \cite{Barger:2004sf,Barger:2007qb, Barger:2005qy,Gogoladze:2010in} 
{\small \begin{equation}
	\frac{d}{d\ln \mu} g_i=\frac{b_i}{(4\pi)^2}g_i^3 +\frac{g_i^3}{(4\pi)^4}
	\left[ \sum_{j=1}^3 B_{ij}g_j^2-\sum_{\alpha=u,d,e} d_i^\alpha
	{\rm Tr}\left( h^{\alpha \dagger}h^{\alpha}\right) \right],\label{eq:rge}
\end{equation}}
where $g_i(i=1,2,3)$ are the SM gauge couplings and $h^{\alpha}(\alpha=u,d,e)$ are the Yukawa couplings. 
The coefficients of beta functions in SM \cite{Machacek:1983tz,Machacek:1983fi,Machacek:1984zw,Cvetic:1998uw} and supersymmetric models \cite{Barger:1992ac,Barger:1993gh,Martin:1993zk} are represented by
{ \begin{eqnarray}
		&&b_{\rm SM}=\left(\frac{41}{6} \frac{1}{k_Y},-\frac{19}{6}\frac{1}{k_2},-7\right) ,~
		B_{\rm SM}=\begin{pmatrix}
			\frac{199}{18} \frac{1}{k_Y^2} &
			\frac{27}{6} \frac{1}{k_Y k_2} &\frac{44}{3} \frac{1}{k_Y} \cr 
			\frac{3}{2} \frac{1}{k_Y k_2} & \frac{35}{6}\frac{1}{k_2^2}&12\frac{1}{k_2} \cr
			\frac{11}{6} \frac{1}{k_Y} &\frac{9}{2}\frac{1}{ k_2}&-26 
		\end{pmatrix},~\\
		&&d^u_{\rm SM}=\left(\frac{17}{6} \frac{1}{k_Y} ,\frac{3}{2}\frac{1}{k_2},2\right),~
		d^d_{\rm SM}=0,~
		d^e_{\rm SM}=0, \,\\
		&&b_{\rm SUSY}=\left(11 \frac{1}{k_Y},\frac{1}{k_2},-3\right) ,~ 
		B_{\rm SUSY}=
		\begin{pmatrix}
			\frac{199}{9}
			\frac{1}{k_Y^2}&  9\frac{1}{k_Yk_2}&\frac{88}{3} \frac{1}{k_Y} \cr
			3\frac{1}{k_Yk_2} & 25\frac{1}{k_2^2}&24\frac{1}{k_2} \cr
			\frac{11}{3}\frac{1}{k_Y} & 9\frac{1}{k_2} & 14
		\end{pmatrix},~ \\
		&&d^u_{\rm SUSY}=\left(\frac{26}{3} \frac{1}{k_Y},6\frac{1}{k_2},4\right) ,~
		d^d_{\rm SUSY}=0,~
		d^e_{\rm SUSY}=0,
\end{eqnarray}}
where $k_Y$ and $k_2$ are general normalization factors.
By solving the two-loop RGEs for SM gauge couplings, we perform numerically calculations including the one-loop RGEs for Yukawa couplings and taking into account the new physics contributions and threshold. The general one-loop RGEs for Yukawa couplings can be found in~\cite{Gogoladze:2010in}.  Starting from the electroweak theory, we run the couplings up from Z boson mass scale $M_Z$ to high energies with the boundary conditions for these equations at $M_Z$ as
\begin{equation}
	g_1(M_Z)=\sqrt{k_Y}\frac{g_{em}}{\cos\theta_W}~,~g_2(M_Z)=\sqrt{k_2}\frac{g_{em}}{\sin\theta_W}~,~g_3(M_Z)=\sqrt{4\pi \alpha_s}~.
\end{equation}
From $M_Z$ up to a supersymmetry breaking scale $M_{\rm S}$, we consider only the non-supersymmetric SM spectrum including a top quark pole mass at $m_t=173.34$ GeV. From $M_{\rm S}$ scale, we perform the supersymmetric RGEs with all the states including the introduced exotic vector-like particles at $M_V$.
Moreover, the Z boson mass, the Higgs vacuum expectation value, strong coupling constant, fine structure constant, and weak mixing angle at $M_Z$ are choosen to be \cite{ParticleDataGroup:2018ovx,ParticleDataGroup:2020ssz}
\begin{equation}
	\begin{split}
		&M_Z=91.1876 {\rm~GeV},~
		m_t=173.34\pm 0.27({\rm{stat}})\pm 0.71 ({\rm{syst}}){\rm{~GeV}},~
		v=174.10 {\rm{~GeV}},\\
		&\alpha_s(M_Z)=0.1181\pm 0.0011,~
		\alpha_{em}^{-1}(M_Z)=128.91\pm 0.02,~
		\sin^2\theta_W(M_Z)=0.23122.
	\end{split}
\end{equation}
Based on the experimental lower limits of supersymmetry and gauge hierarchy preservation, we have the supersymmetry breaking scale $M_{\rm S}$ at TeV scale, such as $2.5$ TeV or $3.0$ TeV. The difference  has effect less than $5\%$ on the scale of unification $M_{\rm U}$ while the larger value for $M_{\rm S}$ reduces the unification scale.

Utilizing the current precision electroweak data and setting  the supersymmetry breaking scale to be $M_{\text{S}}\simeq3.0$ TeV, we study the gauge coupling unification around string scale   
by solving two-loop  RGEs, with
\begin{equation}
	M_\text{U}\sim M_{\text{string}}\simeq5\times10^{17} {~\text{GeV}}.
\end{equation}
Recall that the generic gauge coupling relations at string scale for supersymmetric Pati-Salam models read
\begin{equation}
	g_a^2=k_2g_b^2=k_Yg_Y^2=g_{\text{U}}^2 \sim g_{\text{string}}^2,
\end{equation}
where $k_Y$ and $k_2$ are constants for each model and will determine the value of the mixing angel $\sin \theta_W$ at the string scale.
The string-scale gauge coupling relation in the evolution  is realized via 
\begin{equation}
\alpha_{\text{U}}^{-1}\equiv \alpha_1^{-1}=(\alpha_2^{-1}+\alpha_3^{-1})/2
\end{equation}
with 
$\alpha_1 \equiv k_Yg_Y^2/4\pi$, $\alpha_2 \equiv k_2g_b^2/4\pi$, $\alpha_3 \equiv g_a^2/4\pi$, and the accuracy $\Delta=|\alpha_1^{-1}-\alpha_2^{-1}|/\alpha_1^{-1}$ is limited to be less than $1.0\%$. Here, $\alpha_1$ and $\alpha_2$ are traditional gauge couplings if and only if $k_Y=5/3$ and $k_2=1$, respectively.

As discussed in~\cite{He:2021kbj}, Model 1 in table \ref{tb:model1} has a canonical hypercharge normalization $k_Y=5/3$, and then the gauge coupling unification is naturally achieved at most $M_{\rm U}\sim 2\times 10^{16}$ GeV, similar to the predicted unification scale in the MSSM. However, the unification scale is about one order of magnitude smaller than string scale, where the discrepancy can be diminished by the introducing additional vector-like particles \cite{Blumenhagen:2003jy}.

Interestingly, we also find that the precise string-scale gauge coupling relation can be achieved at two-loop level by introducing the extra vector-like particles from ${\cal N}=2$ subsector or four-dimensional chiral sectors in other models. The quantum numbers for the vector-like particles are the same as those of the SM fermions and their Hermitian conjugates. While, the number $n_V$ and the quantum numbers of these particles are highly model dependent and can be determined by brane construction. 
Particularly, in our calculations, multi-pair of extra particles appears naturally, and we only need to fine tune their masses  to unify the gauge coupling near the string scale $M_{\text{string}}$.  From the numerical results, we find that the mass of these extra particles approach to the string scale as the number of these particles increase. 

The quantum numbers of vector-like particles under $SU(3)_C\times SU(2)_L\times U(1)_Y$   and their contributions to one-loop level beta functions\cite{Jiang:2006hf,Barger:2007qb} are 
\begin{eqnarray}
	&& XQ + {\overline{XQ}} = {\mathbf{(3, 2, {1\over 6}) + ({\bar 3}, 2,
			-{1\over 6})}}\,, \quad \Delta b =({1\over 5}, 3, 2)\,;\label{eq:XQ}\\ 
	&& XU + {\overline{XU}} = {\mathbf{ ({3},
			1, {2\over 3}) + ({\bar 3},  1, -{2\over 3})}}\,, \quad \Delta b =
	({8\over 5}, 0, 1)\,;\\
	&& XD + {\overline{XD}} = {\mathbf{ ({3},
			1, -{1\over 3}) + ({\bar 3},  1, {1\over 3})}}\,, \quad \Delta b =
	({2\over 5}, 0, 1)\,;
\end{eqnarray}
\begin{eqnarray}
	&& XL + {\overline{XL}} = {\mathbf{(1,  2, {1\over 2}) + ({1},  2,
			-{1\over 2})}}\,, \quad \Delta b = ({3\over 5}, 1, 0)\,;\\
	&& XE + {\overline{XE}} = {\mathbf{({1},  1, {1}) + ({1},  1,
			-{1})}}\,, \quad \Delta b = ({6\over 5}, 0, 0)\,; \\
	&& XG = {\mathbf{({8}, 1, 0)}}\,, \quad \Delta b = (0, 0, 3)\,;\\ 
	&& XW = {\mathbf{({1}, 3, 0)}}\,, \quad \Delta b = (0, 2, 0)\,. \label{eq:XW} 
\end{eqnarray}

In our numerical calculations, the two-loop level beta functions from these particles are used in supersymmetric models from Eqs.\,(B8)-(B11) in~\cite{Barger:2007qb}. We integrate the renormalization group equations in Eq. \eqref{eq:rge} from $M_Z$ up to $M_{\rm string}$ with three sections, from $M_Z$ to $M_{\rm S}$, from $M_{\rm S}$ to $M_V$, and from $M_V$ to $M_{\rm string}$. 
The precise energy scale where we realize the string-scale gauge coupling relations depends on the number of the introduced particle, the coefficients $b_i$, $B_{ij}$ and $d_i^{\alpha}$ of the beta-function. 
The values of $M_{XG}$, $M_{XW}$ are determined by the intersection energy scale of section 2(from $M_{\rm S}$ to $M_V$), and section 3(from $M_V$ to $M_{\rm string}$) in the plot, where these exotic particles are introduced. Thus, the mass of the exotic particles are determined by both the number of the particles, the beta functions and the precise energy scale where we realize the string-scale gauge coupling relations.

Once these particles are included with mass $M_{V}$, the evolution of $\alpha_i^{-1}(t)$ is depressed if $\Delta b_i\neq 0$ and the turning point is at $\mu=M_{V}$. 
Taking Model 1 as example, we find that there are two bending points for the running of gauge couplings in figure \ref{fig:model1}; the first one is corresponding to the supersymmetry breaking scale $M_{\rm S}\simeq 3.0$ TeV, while the second one corresponds to the introduced particle $M_{V}$ with chiral multiplets $XG$ and $XW$ obtain the same mass, \textit{i.e.}, $M_{XG}=M_{XW}=M_{V}$. Recall that $\Delta b_2(XW)=2$, the plotted lines for $\alpha_{2}^{-1}$ bend at $M_{XW}$ where the particle $XW$ is introduced. In the same way, the lines for $\alpha_{3}^{-1}$  bend at the point $M_{XG}$ due to $\Delta b_3(XG)=3$. 
Moreover, the degree of depression will increase as $\Delta b_i$ increases. That's to say, the inverse hypercharge coupling $\alpha_1^{-1}$ decreases more rapidly in models with the particles $XU+\overline{XU}$  than that in models with $XD+\overline{XD}$. And the depression of the inverse strong coupling $\alpha_3^{-1}$ from both pairs of vector-like particles are the same. Note that when both particles $XU+\overline{XU}$ and $XD+\overline{XD}$ are added, there will be two bending points at $M_{\rm S}$ and $M_V$ as these particles have the same mass $M_V=M_{XU}=M_{XD}$, while there will be three bending points at $M_{\rm S}$, $M_{XU}$ and $M_{XD}$ as the mass of these particles are different.

\subsection{Chiral Multiplets from Adjoint Representations of $SU(4)_C$ and $SU(2)_L$}

\begin{figure}[]
	\centering
	\subfigure[]{\includegraphics[width=0.45\linewidth]{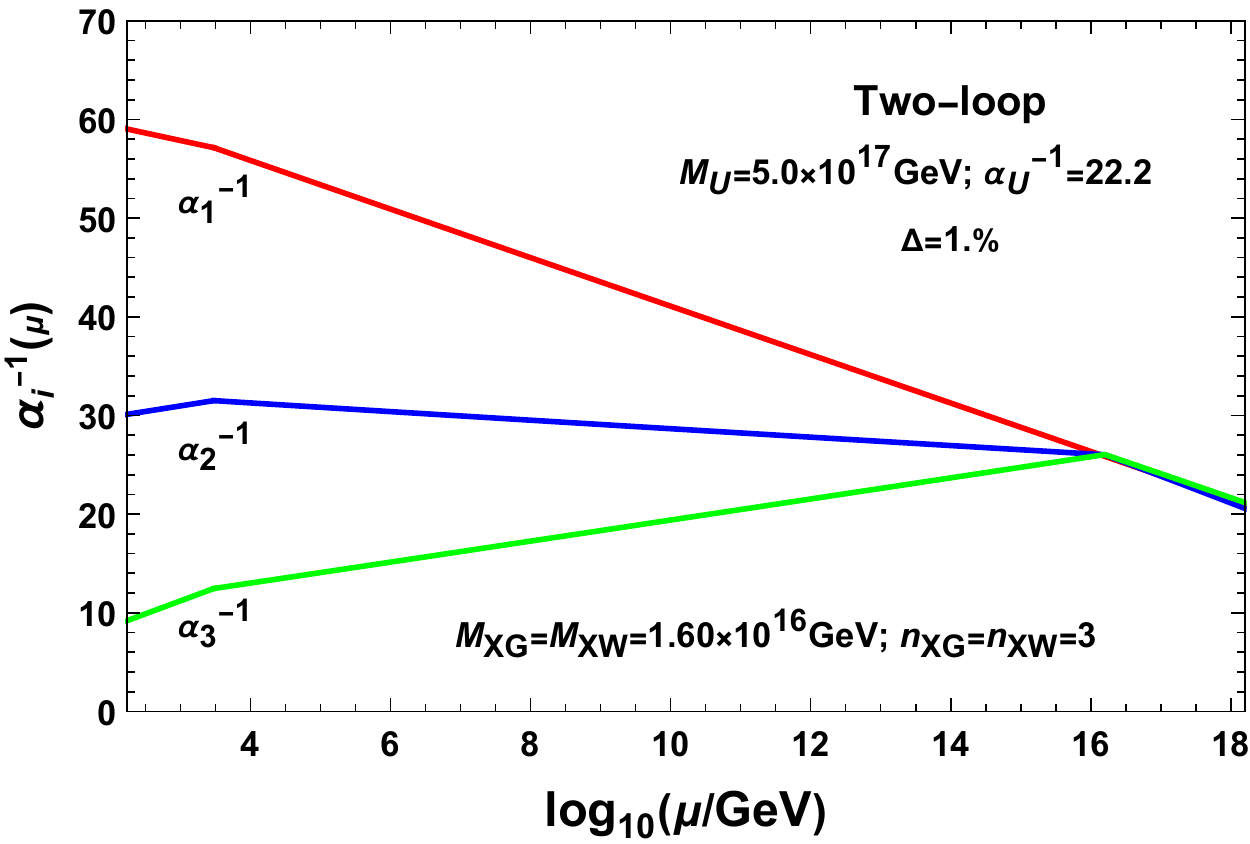}}\\
	\subfigure[]{\includegraphics[width=0.45\linewidth]{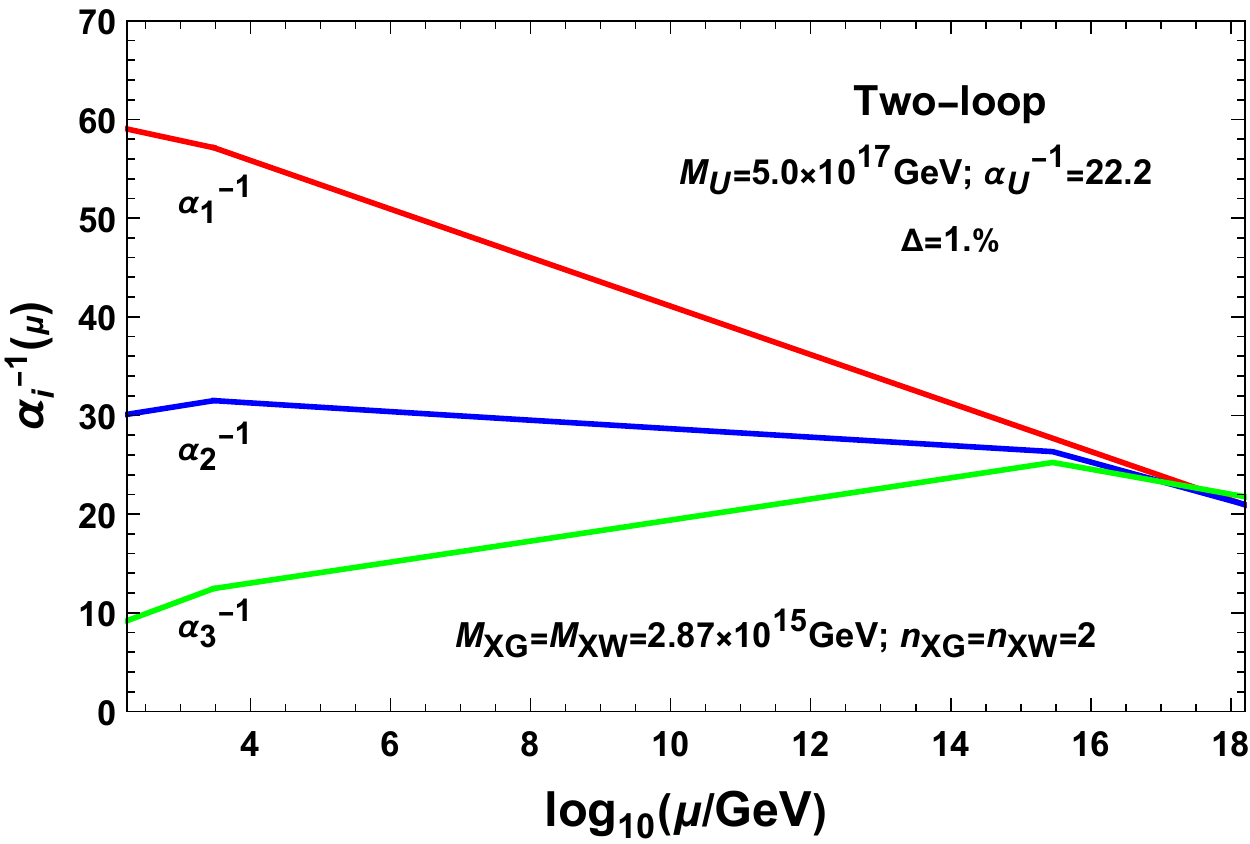}}\qquad
	\subfigure[]{\includegraphics[width=0.45\linewidth]{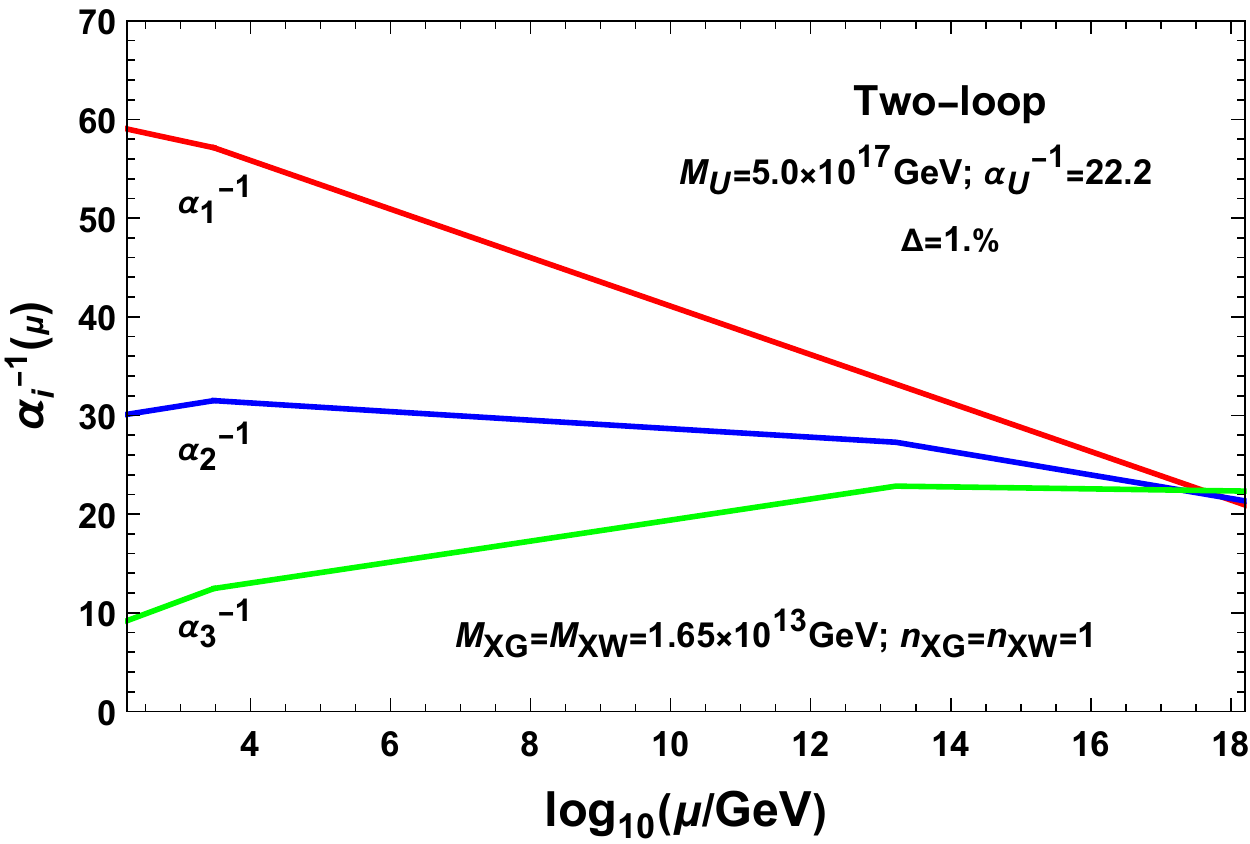}}
	\caption{Two-loop evolution of gauge couplings for the \textbf{Model 1}  with vector-like particles. In the model,  $k_Y= 1\times\frac{5}{3}$ and $k_2=1$. The string-scale gauge coupling relations can be achieved by adding $3(XG+XW)$ (a), $2(XG+XW)$ (b) and $(XG+XW)$ (c).}
	\label{fig:model1}
\end{figure}

\begin{figure}[]\centering
	\subfigure[]{\includegraphics[width=0.45\linewidth]{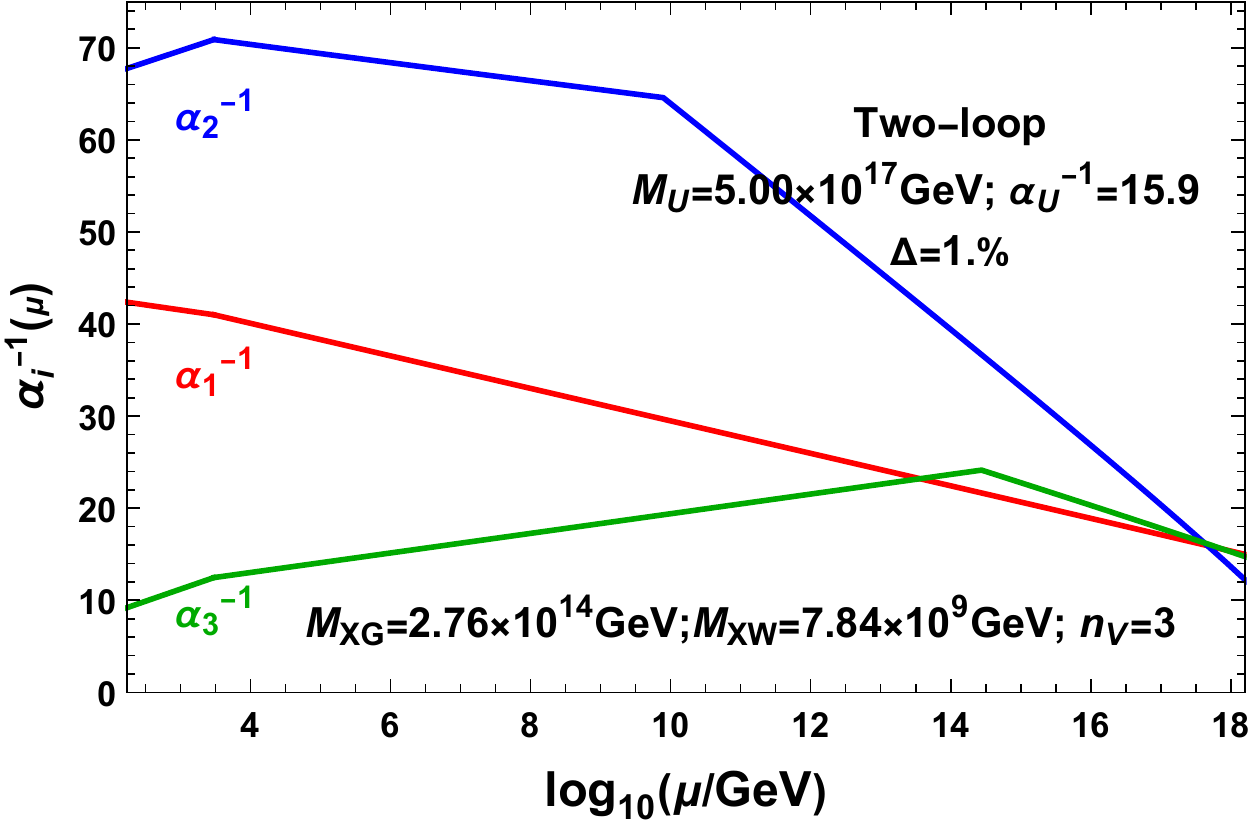}}\qquad
	\subfigure[]{\includegraphics[width=0.45\linewidth]{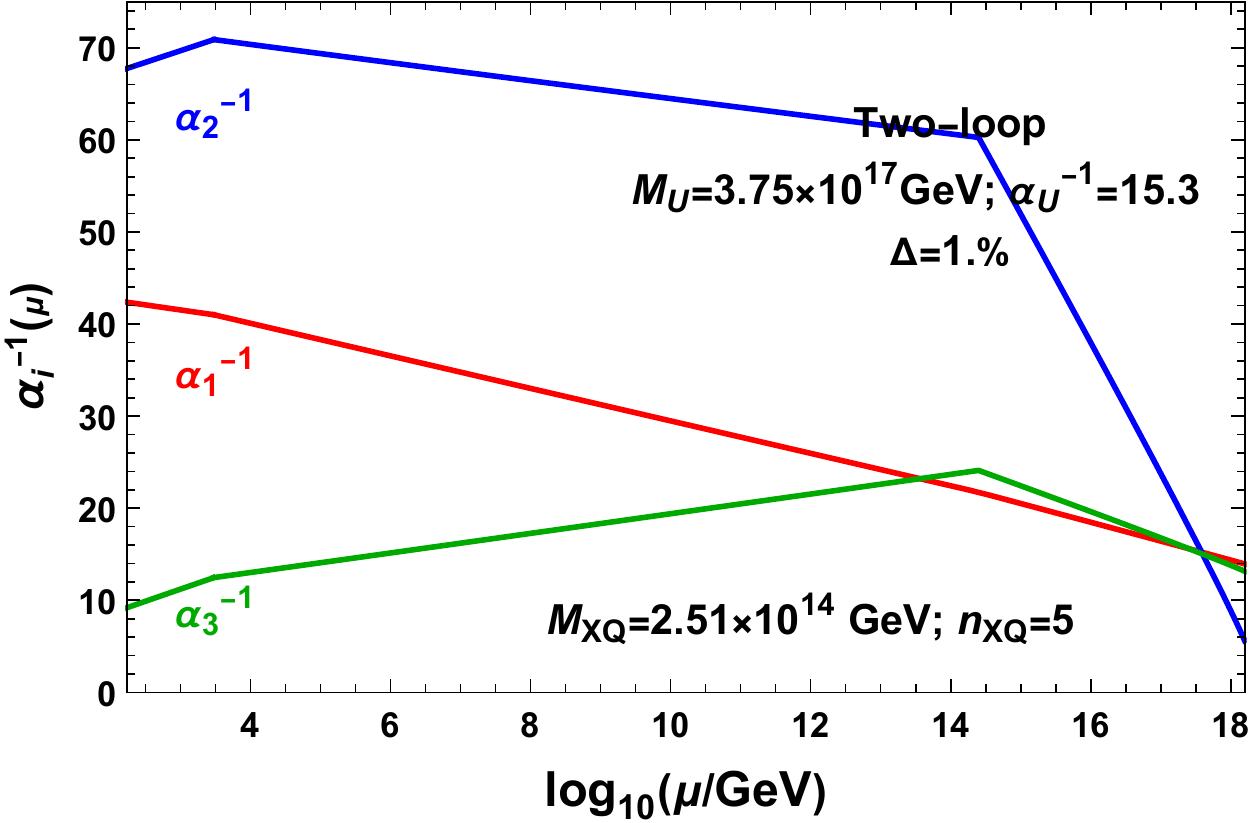}}
	\caption{Two-loop evolution of gauge couplings for the \textbf{Model 2} with vector-like particles. In the model,  $k_Y=\frac{85}{61}\times\frac{5}{3}$ and $k_2=\frac{4}{9}$. The string-scale gauge coupling relations can be achieved by adding $3(XW+XG)$ (a) and $5(XQ+\overline{XQ})$ (b).}
	\label{fig:model28}
\end{figure}

For the Model 1 with $k_y=k_2=1$, the gauge couplings naturally unifies at the traditional GUT scale, one order of magnitude smaller than the string scale. 
Instead of introducing the vector-like particle from a ${\cal N} =2$ subsector, we consider up to three chiral multiplets in the adjoint representations of $SU(4)_C$ and $SU(2)_L$ with masses  $M_{V}(V=XG ~\text{and}~ XW)$. These two particles $XG$ and $XW$ arise from  the $aa$ and $bb$ sectors in the spectrum table~\ref{spectrum}, and thus the maximum number is 3.
The gauge couplings can be unified at close to the string scale, $M_{\rm U}=5.0\times10^{17}$ GeV. The two-loop RGE running for the guage couplings are shown in figure~\ref{fig:model1}.  
As the number of particles increases, the mass of the particles increases and approaches the string scale. Moreover, the splitting between the $XW$ and $XG$ masses will be reduced. Thus in the following calculations, we choose to add 3 $(XG+XW)$. The energy scale, number of the particles, and the mass of the particles for the Model 1-12 are shown in table \ref{tab:XGXW}.
Above $M_{V}$, the running of $SU(3)_C$ coupling and $SU(2)_L$ coupling are reducing due to the non-zero beta functions $\Delta b_3(XG)$ and $\Delta b_2(XW)$.  
\begin{table*}[h]\centering
	\footnotesize
		\begin{tabular}{p{45pt}p{35pt}p{35pt}p{35pt}p{60pt}p{60pt}p{60pt}}\hline\hline
			Model No.& $k_y$ & $k_2$ & $n_{V}$ &$M_{XG}$(GeV)&$M_{XW}$(GeV) & $M_{\rm U}$(GeV)\\
			\hline
			\multirow{3}{*}{1}& \multirow{3}{*}{1} & \multirow{3}{*}{1} & 3 & $1.60\times10^{16}$  &  $1.60\times10^{16}$ & $5.00\times10^{17}$  \\
			& & & 2 & $2.87\times10^{15}$ & $2.87\times10^{15}$ & $5.00\times10^{17}$ \\
			& & & 1 & $1.65\times10^{13}$ & $1.65\times10^{13}$ & $5.00\times10^{17}$ \\\hline
			2& 85/61 &4/9  &3 &$2.76\times 10^{14}$&$7.84\times10^{9}$&$5.00\times10^{17}$\\ 
			3& 65/44 &1/2  &3 &$1.28\times10^{14}$&$1.34\times10^{10}$&$5.00\times10^{17}$\\ 
			4& 35/32 &5/6  & 3 &$3.50\times10^{15}$&$1.23\times10^{14}$&$5.00\times10^{17}$\\	
			5& 10/7 &2/3  &3 &$2.87\times10^{14}$&$2.00\times10^{11}$&$5.85\times10^{17}$\\ 
			6& 11/8 &5/6  &3 &$3.13\times10^{14}$&$3.06\times10^{12}$&$5.00\times10^{17}$\\ 
			7& 25/19 &1   &3 &$3.98\times10^{14}$&$1.19\times10^{14}$&$5.00\times10^{17}$\\ 
			8& 10/7  &1   &3 &$1.73\times10^{14}$&$3.17\times10^{13}$&$5.00\times10^{17}$\\ 
			9& 11/8  &1/6  &3 &$2.63\times10^{14}$&$2.06\times 10^{8}$ &$5.00\times10^{17}$\\ 
			10& 50/47 &4/9  &3 &$5.10\times10^{15}$&$6.90\times 10^{10}$ &$5.00\times10^{17}$\\ 
			11& 1     &1/3  &3 &$1.21\times10^{16}$&$1.18\times10^{12}$&$5.00\times10^{17}$\\ 
			12& 35/32 &1/6  &3 &$1.22\times10^{16}$&$4.49\times10^{8}$&$5.00\times10^{17}$\\ 
            13 & 11/8 & 5/14 & 3 
            & $3.18\times10^{14}$ &$2.45\times10^9$ &$5.00\times10^{17}$ \\ 
            14& 35/32& 35/66  &3 &$4.62\times10^{14}$&$2.33\times10^{11}$&$5.00\times10^{17}$\\ 
	    \hline	\hline
      \end{tabular}
	\caption{String-scale gauge coupling relations achieved with $XG+XW$, from adjoint representation of $SU(4)$ and $SU(2)$.}\label{tab:XGXW}
\end{table*}
For the models with $k_y>1$ and $k_2<1$, the string-scale gauge coupling relation can also be achieved by adding $XW+XG$ from adjoint representation of $SU(4)$ and $SU(2)$. The two-loop RGE running for the gauge couplings of the Model 2 is plotted in figure~\ref{fig:model28}.  Here we include the contributions from 3$(XW+XG)$ to reduce the mass splitting of these added particles. And the two-loop RGE running of the gauge couplings for Model 3-5 are shown in figures~\ref{fig:model29}, \ref{fig:model3}, \ref{fig:model24}, for Model 6-12 in figures~\ref{fig:model17}, \ref{fig:model7}, \ref{fig:model14}, \ref{fig:model33}, \ref{fig:model12}, \ref{fig:model20}, \ref{fig:model31} in Appendix~\ref{apdx-RGE}, for Model 13-14 in figures~\ref{fig:model30} and \ref{fig:model11}. 

\subsection{The Vector-Like Particles from $\mathcal{N}=2$ Subsector with $k_y>1$ and $k_2<1$}

For the models like Model 2 in table \ref{tb:model28}, the evolution of $\alpha_1^{-1}(\mu)$ is depressed when $k_y>1$ and the evolution of $\alpha_2^{-1}(t)$ is raised when $k_2<1$. They induces that the intersection point of $\alpha_1^{-1}(t)$ and $\alpha_3^{-1}(t)$ lines is below the line of $\alpha_2^{-1}(t)$, as illustrated in  figure \ref{fig:model28}(a).  In this case, to get an string-scale gauge coupling relation, we need  introduce extra particles $XQ+\overline{XQ}$ or $XW$ as well as $XG$ into the models, which will mainly modify the evolution of the electroweak  and strong couplings without substantially affecting the U(1) coupling. This is owing to the large contributions to $\Delta b_2$ and $\Delta b_3$, rather relatively small contributions to $\Delta b_1$.  So as the energy rises from mass scale $M_{XQ}$, $\alpha_2^{-1}$ and $\alpha_3^{-1}$ reduce rapidly. Therefore, string-scale gauge coupling relations are achieved near $M_{\rm U}=3.75\times 10^{17}$ GeV by adding 5 sets of $XQ+\overline{XQ}$ at $M_{XQ}=2.51\times 10^{14}$ GeV. The evolution of gauge couplings for Model 2 with $k_y=85/61$ and $k_2=4/9$ is shown in figure \ref{fig:model28}(b) and the energy scale, the number and mass of the added vector-like particles $XQ+\overline{XQ}$ are list in table \ref{tab:XQ-n2}.

\begin{table*}[!t]\centering
	\footnotesize
		\begin{tabular}{p{45pt}p{30pt}p{30pt}p{30pt}p{55pt}p{30pt}p{55pt}p{55pt}} \hline\hline
			Model No.& $k_y$ & $k_2$ & $n_{XQ}$ &$M_{XQ}$(GeV)&  & & $M_{\rm U}$(GeV)\\
			\hline
			2& 85/61 &4/9  &5 &$2.51\times 10^{14}$&&&$3.75\times10^{17}$\\ 
			3& 65/44 &1/2  &3 &$1.26\times10^{12}$&&&$1.81\times10^{17}$\\ 
			4& 25/32 &5/6  &7 &$1.07\times10^{16}$&&&$1.09\times10^{17}$\\	
			\hline
			Model No.& $k_y$ & $k_2$ & $n_{XQ}$ &$M_{XQ}$(GeV)&$n_{XG}$&$M_{XG}$(GeV) & $M_{\rm U}$(GeV)\\\hline
			5& 10/7 &2/3  &5 &$1.03\times10^{15}$&3&$1.18\times10^{17}$&$5.00\times10^{17}$\\ 
			6& 11/8 &5/6  &7 &$1.51\times10^{16}$&3& $3.69\times10^{16}$&$5.00\times10^{17}$\\ 
			7& 25/19 &1   &3 &$1.43\times10^{14}$&3&$1.41\times10^{16}$&$5.00\times10^{17}$\\ 
			8& 10/7  &1   &3 &$4.73\times10^{14}$&3&$1.50\times10^{16}$&$5.00\times10^{17}$\\ 
			\hline
			Model No.& $k_y$ & $k_2$ & $n_{XQ}$ &$M_{XQ}$(GeV)&$n_{XW}$&$M_{XW}$(GeV) & $M_{\rm U}$(GeV)\\\hline 
			9& 11/8  &1/6  &5 &$3.84\times10^{14}$&3&$8.10\times 10^{15}$ &$5.00\times10^{17}$\\ 
			10& 50/47 &4/9  &5 &$5.13\times10^{15}$&3&$4.39\times 10^{16}$ &$5.00\times10^{17}$\\ 
			11& 1     &1/3  &5 &$1.60\times10^{16}$&3&$4.12\times10^{13}$&$5.00\times10^{17}$\\ 
			12& 35/32 &1/6  &5 &$1.66\times10^{16}$&3&$1.44\times10^{12}$&$5.00\times10^{17}$\\ 
		\hline\hline
        \end{tabular}
	\caption{String-scale gauge coupling relations achieved by adding vector-like particles $XQ+\overline{XQ}$ as well as $XG$ or $XW$, from $\mathcal{N}=2$ subsector. The number of vector-like particles are defined by the intersection number on \textit{a} and \textit{b} stacks, $n_v=I_{ab}$.}\label{tab:XQ-n2} 
\end{table*}

\begin{figure}[h]\centering
	\subfigure[]{\includegraphics[width=0.45\linewidth]{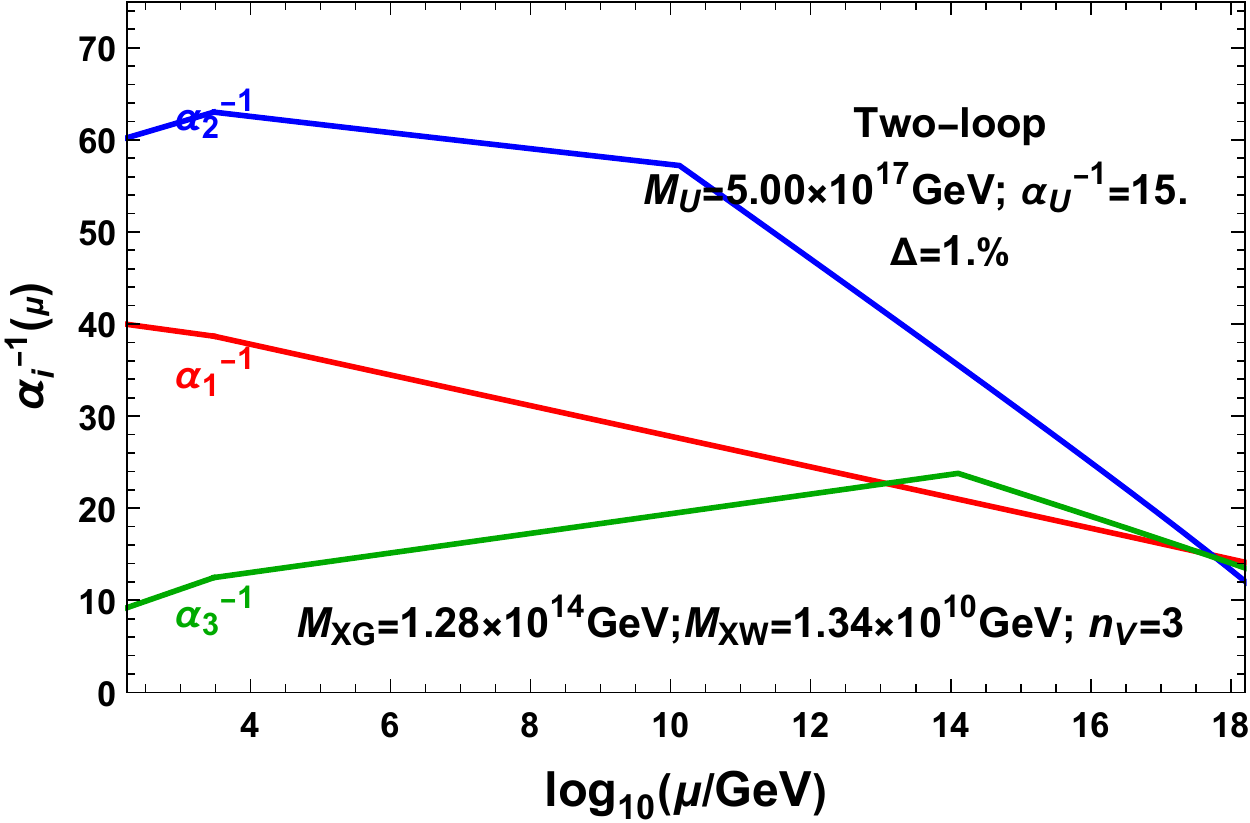}}\qquad
	\subfigure[]{\includegraphics[width=0.45\linewidth]{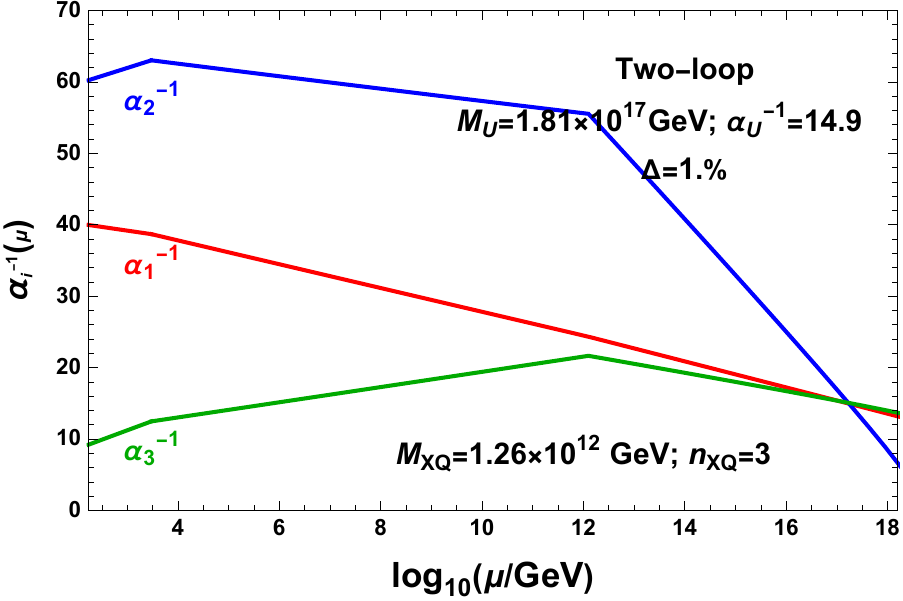}}
	\caption{Two-loop evolution of gauge couplings for the \textbf{Model 3} with vector-like particles. In the model, $k_Y=\frac{65}{44}\times\frac{5}{3}$ and $k_2=\frac{1}{2}$. The string-scale gauge coupling relations can be achieved by adding $3(XW+XG)$ (a) and $3(XQ+\overline{XQ})$ (b).}
	\label{fig:model29}
\end{figure}

\begin{figure}[h]\centering
	\subfigure[]{\includegraphics[width=0.45\linewidth]{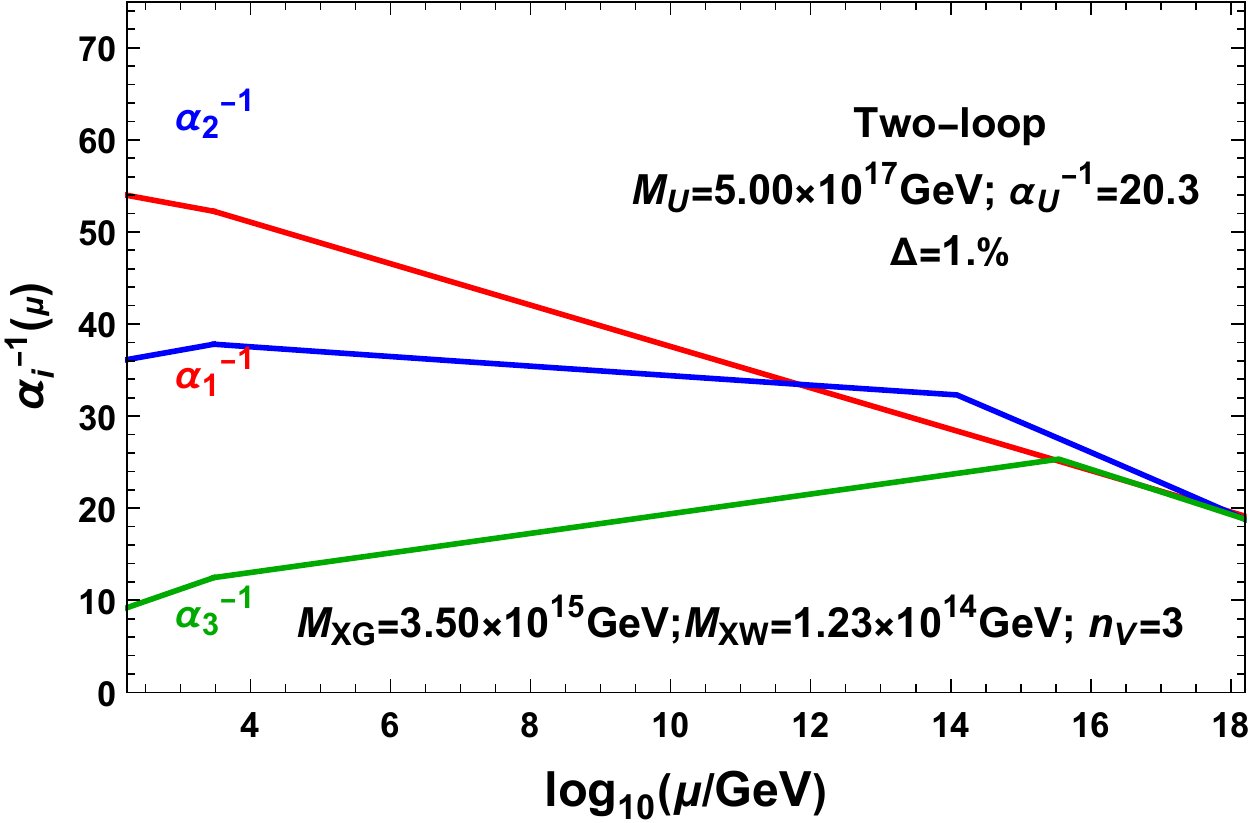}}\qquad
	\subfigure[]{\includegraphics[width=0.45\linewidth]{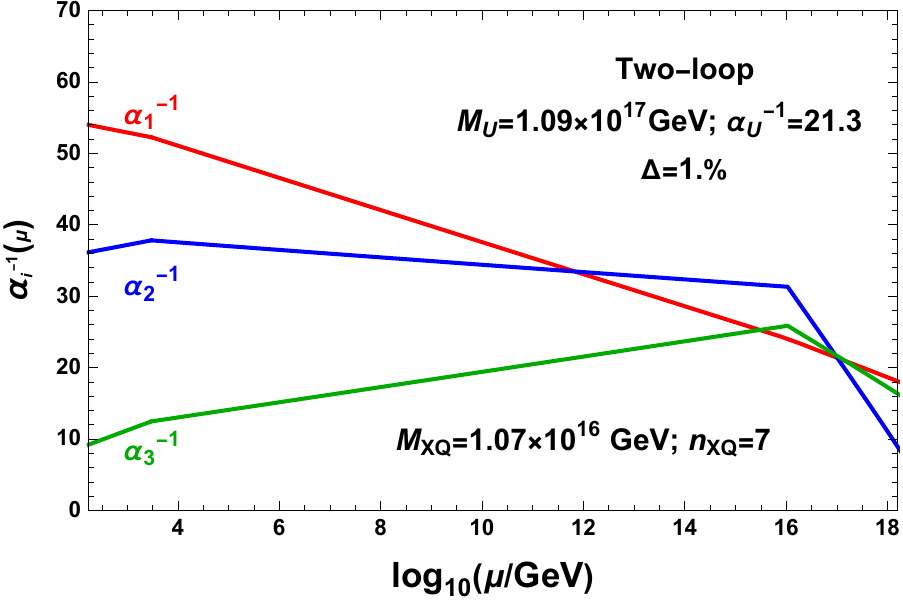}}
	\caption{Two-loop evolution of gauge couplings for the \textbf{Model 4} with vector-like particles. In the model, $k_Y=\frac{35}{32}\times\frac{5}{3}$ and $k_2=\frac{5}{6}$. The string-scale gauge coupling relations can be achieved by adding $3(XW+XG)$ (a) and $7(XQ+\overline{XQ})$ (b).}\label{fig:model3}
\end{figure}

Similarly, for Model 3 and 4, the non-canonical constants are $k_y=65/44,~k_2=1/2$  and $k_y=25/32,~k_2=5/6$. The string-scale gauge coupling relation are also achieved by bringing particles $XQ+\overline{XQ}$ into Model 3 at $1.26\times10^{12}$ GeV and Model 4 at $1.07\times 10^{16}$ GeV, respectively. The evolution of gauge couplings are shown in figures~\ref{fig:model29} and \ref{fig:model3}, in which the number of pairs of the new vector-like particles is 3 and 7, respectively. We note that the mass of the extra particles is related to the number of pairs of particles. As the number increases, the mass scale is pushed up to the high energy scale and  thus the vector-like particles decay at high energy scale.


Of course, the number of these extra vector-like particles is not random, yet from brane constructions. From  Eqs.~\eqref{eq:XQ} and \eqref{eq:XW}, the quantum numbers of these particles under $SU(3)_C\times SU(2)_L\times U(1)_Y$ are $XQ=(3,2,1/6)$, and $\overline{XQ}=(\bar{3},2,-1/6)$. 
In the supersymmetric Pati-Salam models, these vector-like particles $(XQ+\overline{XQ})$ arise from the intersections between $a$ and $b$ stacks of D6-branes or $a$ and $b'$ stacks of D6-brane. The particle $XW$ arise from $bb$ sector of the adjoint representation of $SU(2)_L$. 

Base on the brane construction, the number of vector-like particles $(XQ+\overline{XQ})$ can be determined by the intersection number of the $a$ and $b$ stacks of D6-brane, $I_{ab}$, or $a$ and $b'$ stacks of D6-brane, $I_{ab'}$.
For example, the intersection number is $I_{ab'}=-2^{0}(4+1)(1)=-5$ 
in Model 2 (table~\ref{tb:model28}) and the corresponding number of the additional particles $(XQ+\overline{XQ})$ is 5. 
If the wrapping numbers $(n_a^i,l_a^i)$ and $(n_b^i,l_b^i)$ have the same value, indicating that the D6-branes warpping on the $i-$ torus are parallel to each other. Therefore, there is no intersection on the $i-$torus, but only
on the other two torus. From table \ref{tb:model3}, we see that $(n_a^3,l_a^3)$ and $(n_b^3,l_b^3)$ for Model 3 are the same, thus the intersection number of the $a$ and $b$ stacks are only calculated on the first two torus, \textit{i.e.}, $I_{ab}=2^0(-2-5)(1)=7$. Namely, 7 pairs of $XQ+\overline{XQ}$ naturally arise from brane intersection.

\begin{figure}[]\centering
	\subfigure[]{\includegraphics[width=0.45\linewidth]{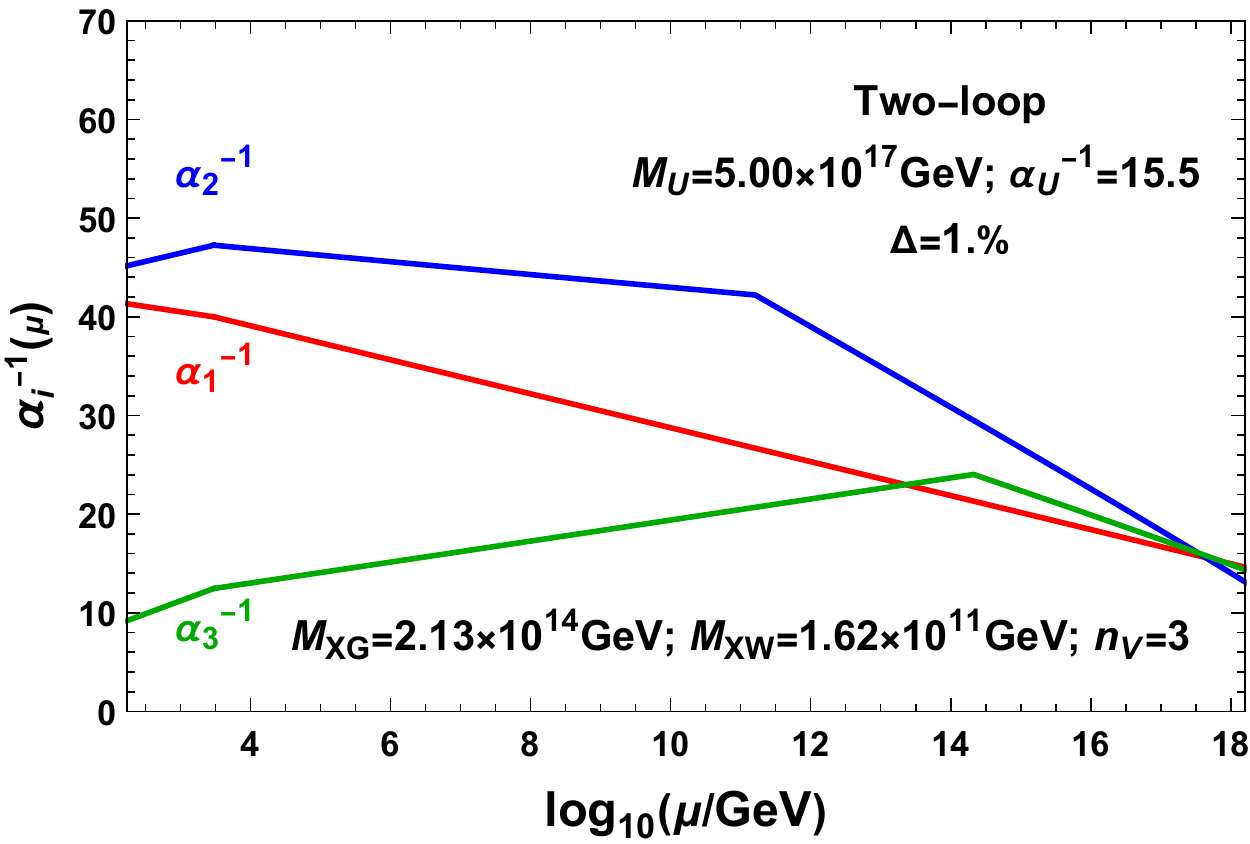}}\qquad
	\subfigure[]{\includegraphics[width=0.45\linewidth]{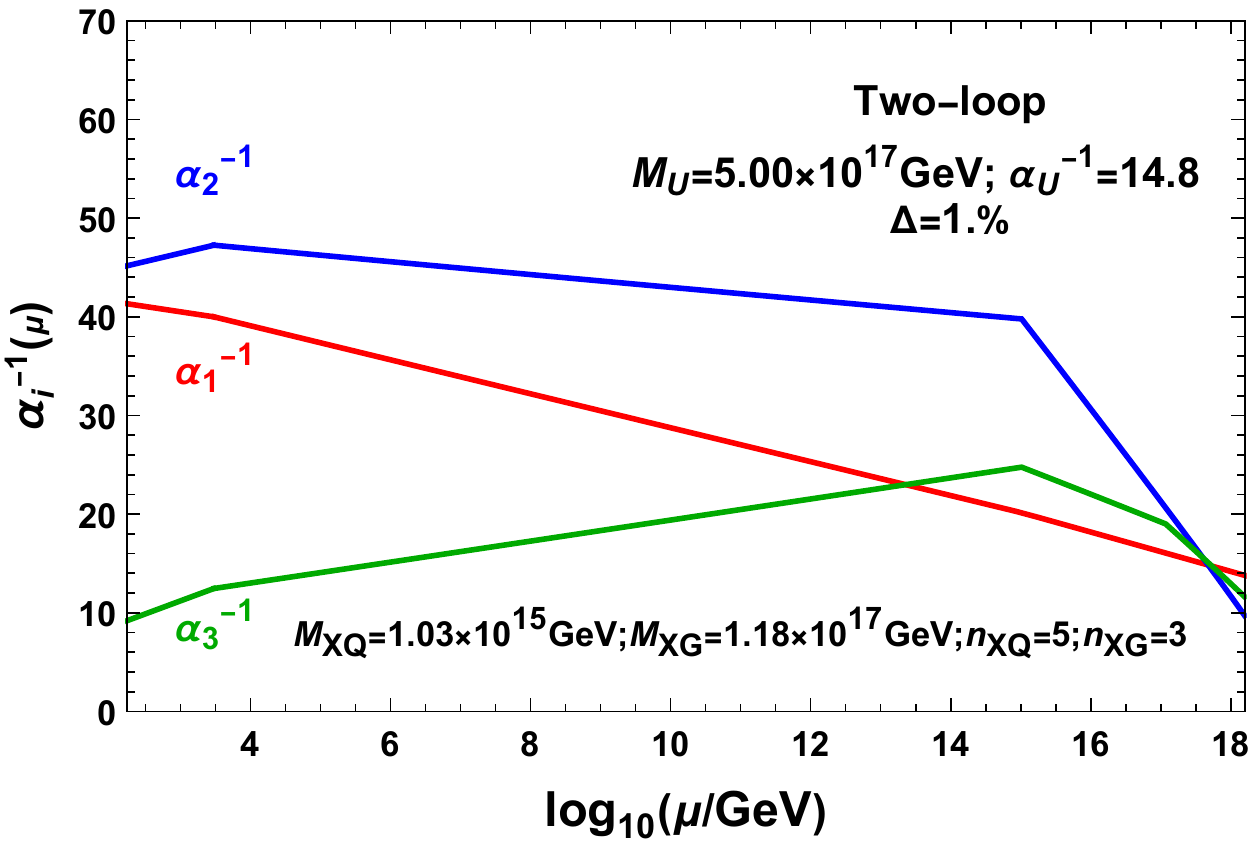}}
	\caption{Two-loop evolution of gauge couplings for the \textbf{Model 5} with vector-like particles. In the model,  $k_Y=\frac{10}{7}\times\frac{5}{3}$ and $k_2=\frac{2}{3}$. The string-scale gauge coupling relations can be achieved by adding $3(XW+XG)$ (a) and $5(XQ+\overline{XQ})+3XG$ (b). }
	\label{fig:model24}
\end{figure}

Based on the calculations, we know that the energy scale is pushed up to high energy scale as $k_y$ or $k_2$ deceases. Thus, for the models with $k_y>1$, to obtain a string-scale gauge coupling relation, the constant $k_2$ of the model should be smaller than $1/2$. Otherwise, the gauge couplings are unified at an intermediate energy scale $10^{14}-10^{16}$ GeV, little smaller than the string scale.  For Model 5 with $k_y>1$ and $k_2>1/2$, the gauge coupling relation can be achieved at the string scale $M_{\rm U}\simeq 5\times 10^{17}$ GeV by adding  5 pairs of $XQ+\overline{XQ}$ and 3 $XG$.  For Models 5-8, the parameter $k_y$ is almost equal. And the parameter $k_2$ in Models 6-8 is larger than that in Model 5. Thus, the energy scale is smaller than that in Model 5. To increase the energy scale, we also need to introduce adjoint particle $XG$ for Models 6-8.  The evolution of gauge couplings for Models 5-8 with vector-like particles and $XG$ are presented in figures~\ref{fig:model24}, \ref{fig:model17}, \ref{fig:model7} and \ref{fig:model14}, respectively. 
However, the parameter $k_2$ should not be too small either, as this would realize the gauge coupling relation beyond Planck level,  where we do not know how to quantize gravity. This thorny issue will arise when we deal with models 9-12. The evolution of gauge couplings for Models 9-12 is shown in figures~\ref{fig:model33}, \ref{fig:model12}, \ref{fig:model20}, \ref{fig:model31}, in Appendix~\ref{apdx-RGE}. Note that the U(1) and strong couplings are unified below the Planck scale. We find that the electroweak coupling can be unified with other two couplings below the Planck scale by introducing $XW$, which only affects the running of electroweak coupling due to $\Delta b_2\neq 0$. Take Model 11 as example, the U(1) and strong couplings are unified at $1\times10^{18}$ GeV, while the gauge couplings are unified at same energy scale after introducing 3 $XW$ at $1.4\times10^{12}$ GeV. To obtain  string-scale gauge coupling relation for these models, the additional particles are $XQ+\overline{XQ}$ as well as $XW$. The numbers and the masses of these particles are given in the gauge revolution figures.

\subsection{The Vector-Like Particles from Four-dimensional Chiral Sectors}

\begin{figure}[h]\centering
	\subfigure[]{\includegraphics[width=0.45\linewidth]{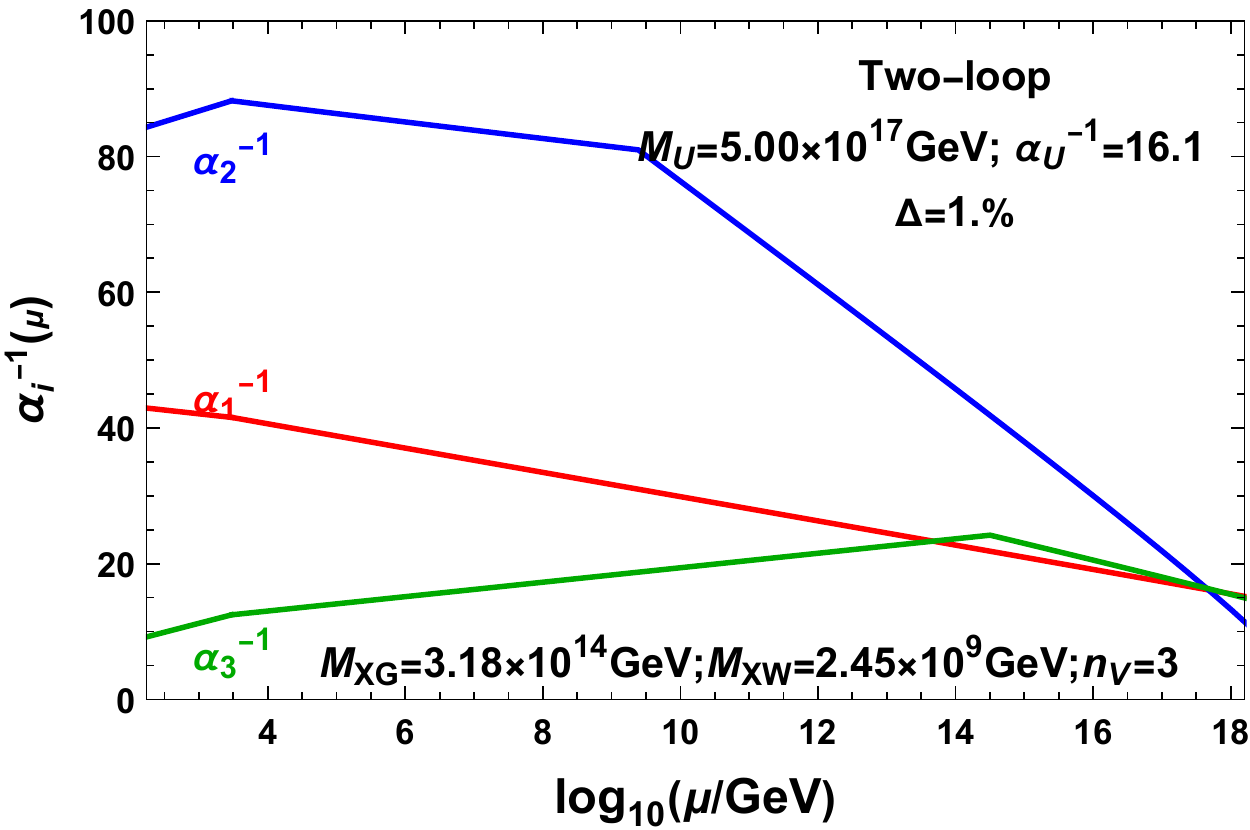}}\quad
	\subfigure[]{\includegraphics[width=0.45\linewidth]{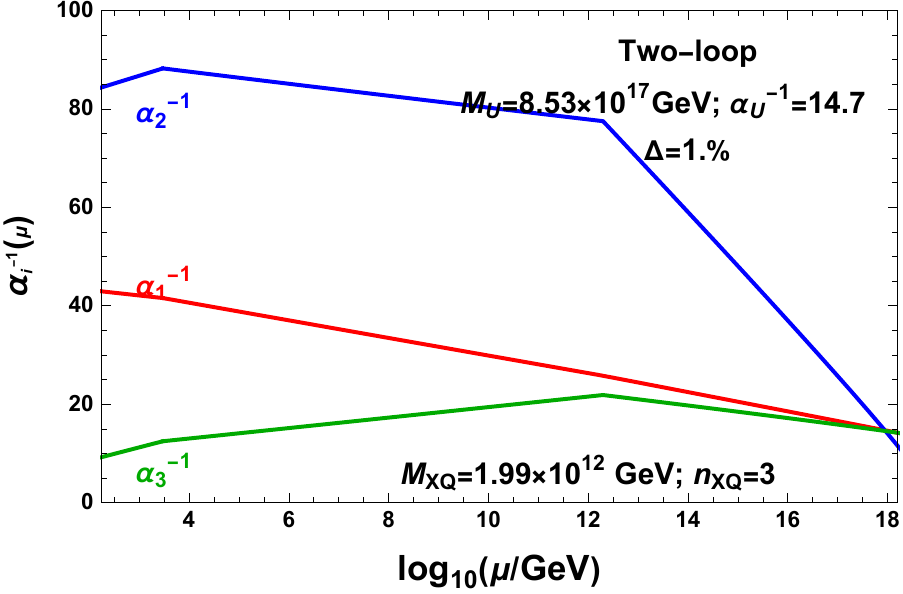}}
	\caption{Two-loop evolution of gauge couplings for the \textbf{Model 13} with vector-like particles. In the model, $k_Y=\frac{11}{8}\times\frac{5}{3}$ and $k_2=\frac{5}{14}$. The string-scale gauge coupling relations can be achieved by adding $3(XW+XG)$ (a) and $3(XQ+\overline{XQ})$ (b). }
	\label{fig:model30}
\end{figure}

\begin{figure}[]\centering
	\subfigure[]{\includegraphics[width=0.45\linewidth]{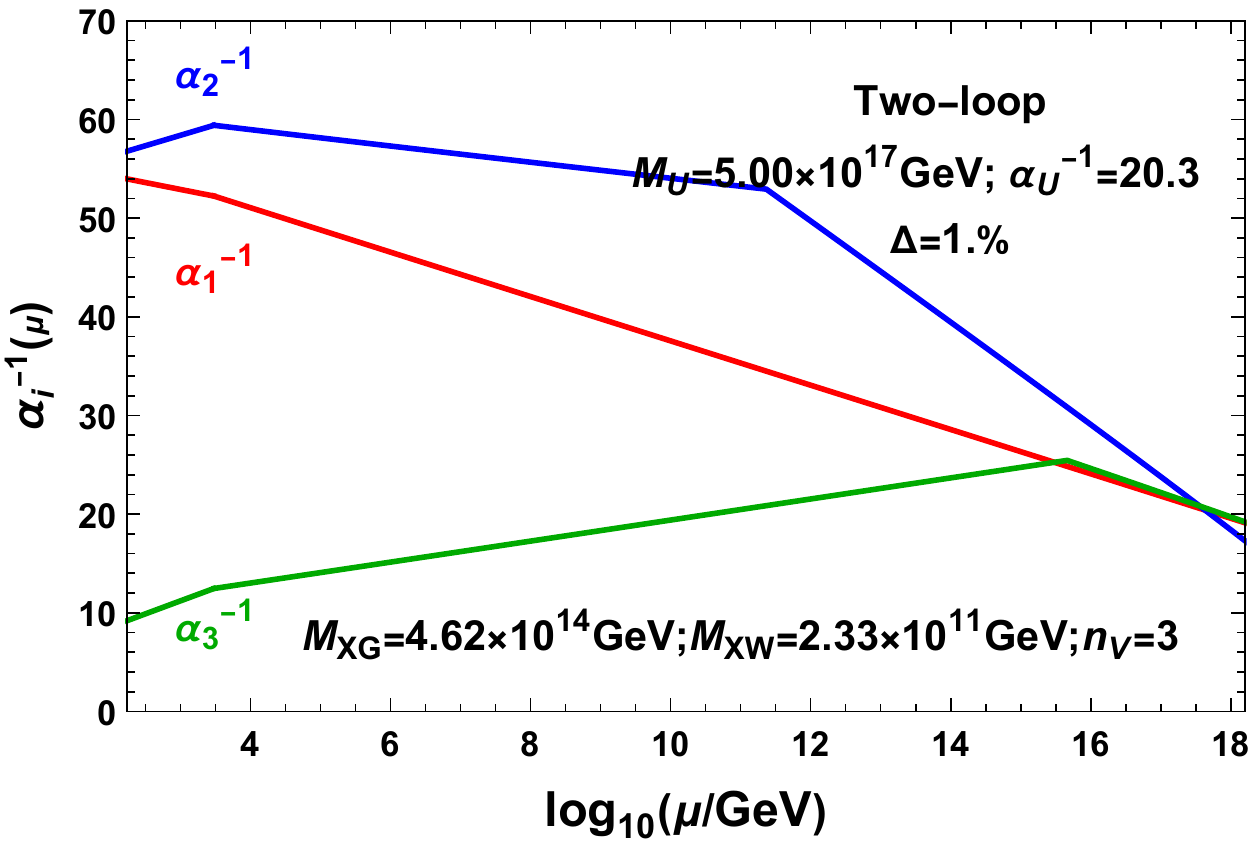}}\qquad
    \subfigure[]{\includegraphics[width=0.45\linewidth]{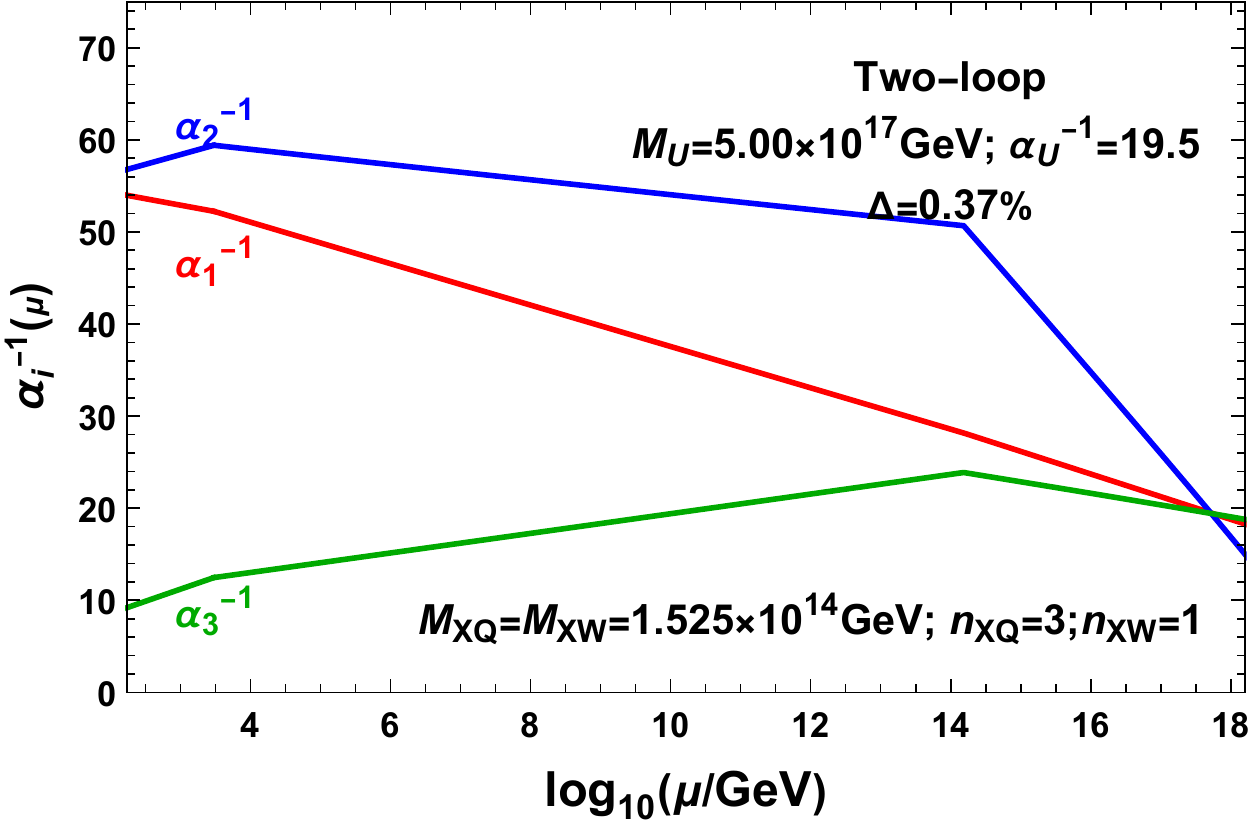}}
	\caption{Two-loop evolution of gauge couplings for the \textbf{Model 14} with vector-like particles. In the model, $k_Y=\frac{35}{32}\times\frac{5}{3}$ and $k_2=\frac{35}{66}$. The string-scale gauge coupling relation can be achieved by adding $3(XW+XG)$ (a) and $3(XQ+\overline{XQ})+XW$ (b).  } 
	\label{fig:model11}
\end{figure}

On the other hand, Model 13 in table \ref{tb:model30} with $k_y=11/8$ and $k_2=5/14$ also can achieve string-scale gauge coupling relation while the vector-like particles $XQ+\overline{XQ}$ are added. 
The number of these vector-like particles are defined by the fundamental minus anti-fundamental representation, with $6-3 \equiv 3$.
The evolution of gauge couplings for the model are shown in figure~\ref{fig:model30}. The energy scale is around $8\times10^{17}$ GeV, and the new vector-like particles decay to the corresponding SM fermions below $M_{XQ}\simeq2\times10^{12}$ GeV. The number and mass of new vector-like particles are also shown in the plot. Unlike Model 2-12 discussed above, the vector-like particles entered in this model come from the four-dimensional chiral sector of the brane building. Furthermore, Model 14 in table \ref{tb:model11} with $k_y=35/32$ and $k_2=35/66$, also have a energy scale at $1.8\times10^{18}$ GeV moderately larger than the string scale. As discussed before, the contributions of the particle $XW$ are also included in this model. The particle $XW$ comes from the \textit{bb} sector of the brane configuration. Comparison of the masses of vector-like particles with different $XW$ numbers shows that the mass splitting between $XQ$ and $XW$ increases with the number of $XW$. When $XW$(s) is\,(are) added, the $XW$ acquires the same mass as the $XQ$ from the naturalness point of view. Therefore, the number of $XW$ needs to be carefully chosen in order to reduce their mass splitting. The evolution of gauge couplings for the model are shown in figure~\ref{fig:model11}. The string-scale gauge coupling relations for Model 13-14 are listed in table.~\ref{tab:XQ-4d}.  
\begin{table*}[h]\centering
\footnotesize
		\begin{tabular}{p{45pt}p{30pt}p{30pt}p{30pt}p{55pt}p{30pt}p{55pt}p{55pt}}
        \hline\hline
			Model No. & $k_y$ & $k_2$ & $n_{XQ}$ &$M_{XQ}$ (GeV)&$n_{XW}$&$M_{XW}$& $M_{\rm U}$(GeV)\\
			\hline
			13& 11/8 & 5/14  &3 &$1.99\times10^{12} $&&&$8.53\times10^{17}$\\ 
						\hline
			\multirow{3}{*}{14}& \multirow{3}{*}{35/32}& \multirow{3}{*}{35/66}  &\multirow{3}{*}{3} &$1.525\times10^{14}$&1&$1.525\times10^{14}$&\multirow{3}{*}{$5.00\times10^{17}$}\\ 
			& & &  &$1.34\times 10^{14}$ &2 & $1.30\times10^{16}$ &\\
			& & &  &$1.70\times 10^{14}$ &3 & $2.75\times10^{16}$ &\\
         \hline\hline
		\end{tabular}
	\caption{String-scale gauge coupling relations achieved  by adding vector-like particles $XQ+\overline{XQ}$ from four-dimensional chiral sectors as well as $XW$ from \textit{bb} sector. } \label{tab:XQ-4d}
\end{table*}

\subsection{The Vector-Like Particles from $\mathcal{N}=2$ Subsector with $k_y<1$ and $k_2>1$} 

\begin{figure}[]\centering
	\centering
	\subfigure[]{\includegraphics[width=0.45\linewidth]{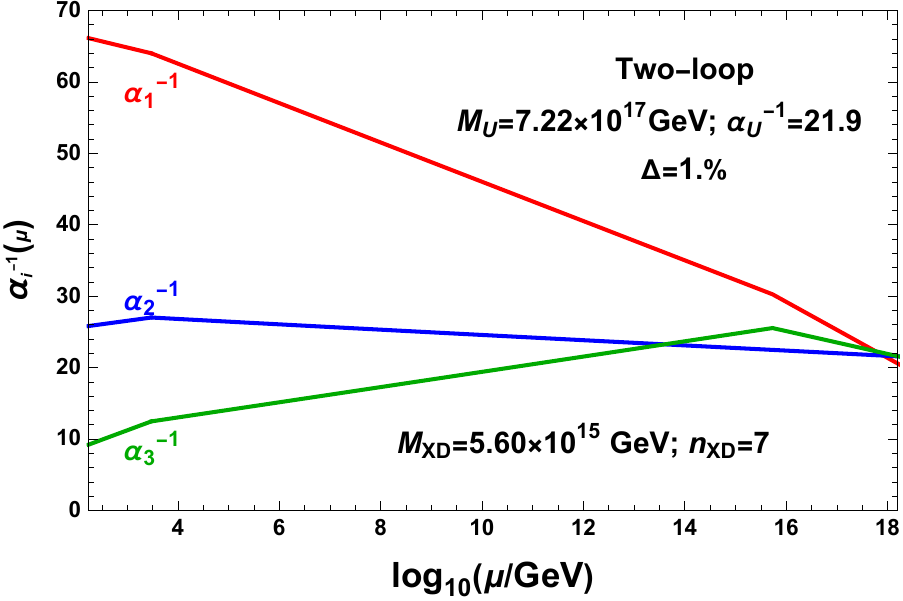}}
	\subfigure[]{\includegraphics[width=0.45\linewidth]{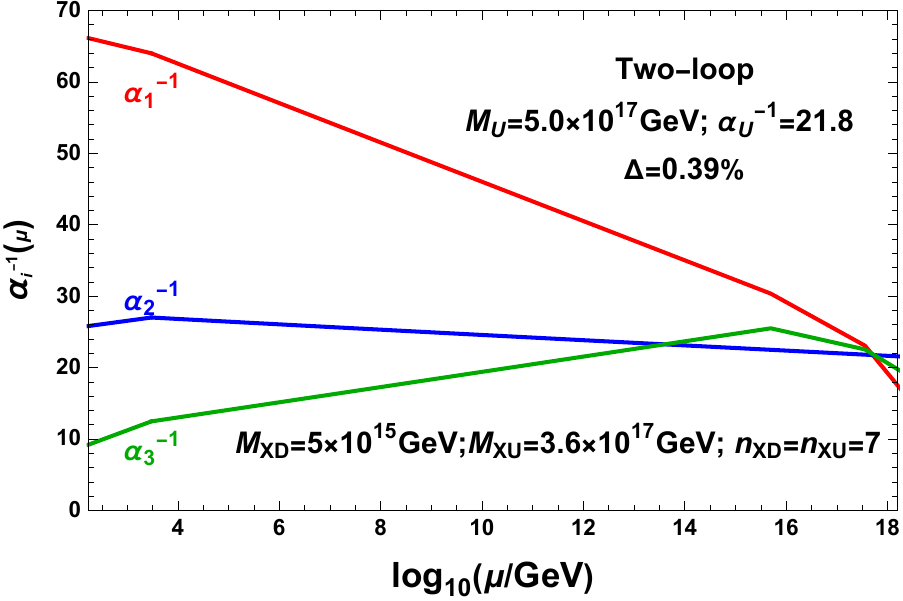}}
	\caption{Two-loop evolution of gauge couplings for the \textbf{Model 15}  with vector-like particles. In the model, $k_Y=\frac{25}{28}\times\frac{5}{3}$ and $k_2=\frac{7}{6}$. The string-scale gauge coupling relation can be achieved by adding $7(XD+\overline{XD})$ (a) as well as $7(XU+\overline{XE})$ (b).}
	\label{fig:model2}
\end{figure}

On the contrary, when $k_y<1$ or $k_2>1$, the intersection of $\alpha_1^{-1}$ and $\alpha_3^{-1}$ lines lies above the line of $\alpha_2^{-1}(t)$, which can be seen from Model 15 without vector-like particle in figure~\ref{fig:model2}(a). And thus, if we want to push or pull back the energy scale to string scale $M_{\rm string}$,  the extra particles arise from $\mathcal{N}=2$ subsector, like $(XD+\overline{XD})$ and $(XU+\overline{XU})$, which will modify the evolution of the $U(1)$ and strong couplings rather than electroweak coupling. This is due to the non-zero $\Delta b_1$ and $\Delta b_3$. As we mentioned earlier, since $\Delta b_1(XD)<\Delta b_1(XU)$, the suppression of $\alpha_1^{-1}$ is stronger in the models with $XD+\overline{XD}$ particles than that with $XU+\overline{XU}$ particles. And thus, the energy scale is higher in the models with $(XD+\overline{XD})$ than that with $(XU+\overline{XU})$. Of course, in some models with much smaller $k_y$ or much larger $k_2$, we need to add both particles to obtain the string-scale gauge coupling relation. Additional, there is mass splitting between these two particles. The energy scale, the number and mass of the added vector-like particles $(XD+\overline{XD})$ and $(XU+\overline{XU})$ are list in table \ref{tab:XDXU-n2}. Another vector-like particle $(XE+\overline{XE})$, from $\mathcal{N}=2$ subsector, will only modify the running of the $U(1)$ coupling due to $\Delta b_1(XE)\neq0$. When used with $XG$, from \textit{aa} sectors, the results are similar or even better than those of $(XD+\overline{XD})$ and $(XU+\overline{XU})$. The energy scale, the number and mass of the added vector-like particles $(XE+\overline{XE})$ and $XG$ are list in table \ref{tab:XEXG-n2}.
 
Base on the brane construction, the number of vector-like particles $(XD+\overline{XD})$, $(XU+\overline{XU})$ and $(XE+\overline{XE})$ can be determined by the intersection number of the $a$ and $c$ stacks $I_{ac}$.
For example, the intersection number is $I_{ac}=2^{0}(1)(-2-5)=-7$ in Model 15 (table \ref{tb:model2}) and the corresponding number of the vector-like particles $(XD+\overline{XD})$ is 7. In this model, the wrapping numbers $(n_a^3,l_a^3)$ and $(n_c^3,l_c^3)$ have the same value yet with opposite sign, indicating that the D6-branes warpping on the third torus are parallel to each other. Therefore, the intersection number of the $a$ and $c$ stacks are only calculated on the first two torus.

\begin{table*}\centering
	\footnotesize
		\begin{tabular}{p{45pt}p{35pt}p{35pt}p{35pt}p{60pt}p{60pt}p{60pt}}
  \hline\hline
			Model No. & $k_y$ & $k_2$ & $n_v$ &$M_{XD}$(GeV) &$M_{XU}$(GeV)& $M_{\rm U}$(GeV)\\
			\hline
			{15} & {25/28}&{7/6} &{7}& $5.60\times10^{16}$&&$7.22\times10^{17}$\\ 
			{16} & {10/7}&{2}  &{3} &$6.9\times10^4$& &$1.2\times10^{17}$\\ 
			\hline
			{17} & {1/4}&{11/6}  &{3} && $3.70\times10^{10}$&$1.43\times10^{17}$\\
			\hline
                {18} & {10/7}&{18/5}&{3}&$8.90\times10^3$& $1.60\times10^{14}$ & $3.25\times10^{17}$\\ 
			{19} & {1}&{5/3}  &{3} &$2.00\times10^8$&$1.40\times10^{17}$&$ 4.90\times10^{17}$\\
			{20} & {1}&{2}  &{3} &$3.71\times10^{7}$&$1.05\times10^{14}$&$5.00 \times10^{17}$\\
			{21} & {1}&{54/19}  &{3} &$2.13\times10^6$ &$2.43\times10^{13}$ & $5.00\times10^{17}$\\
			{22} & {1}&{9/5}  &{3} &$1.80\times10^8$&$6.20\times10^{15}$& $5.50\times10^{17}$\\
			{23} & {5/8}&{13/6}  &{3} &$8.20\times10^{10}$&$8.20\times10^{10}$&$3.60 \times10^{17}$\\
			{24} & {10/13}&{2}  &{5} &$9.00\times10^{12}$&$8.00\times10^{14}$&$5.58\times10^{17}$\\
			{25} & {4/7}&{17/9}  &{5} &$1.08\times10^{14}$&$1.08\times10^{14}$&$6.61 \times10^{17}$\\ 
			{26} & {25/28}&{11/6}  &{7} &$8.71\times10^{13}$&$2.75\times10^{15}$& $3.43 \times10^{17}$\\
   \hline\hline
		\end{tabular}
	\caption{String-scale gauge coupling relations achieved  by adding vector-like particles $XD+\overline{XD}$ or/and $XU+\overline{XU}$ from $\mathcal{N}=2$ subsector.  The number of vector-like particles are defined by the intersection number on \textit{a} and \textit{c} stacks, $n_v=I_{ac}$.}\label{tab:XDXU-n2}
\end{table*}	

For Model 15, the parameters $k_y=25/28$ and $k_2=7/6$ slightly deviate from 1, the string-scale gauge coupling relation can be achieved by introducing vector-like particles, \emph{e.g.}, a string-scale gauge coupling relation obtained at $M_{\rm U}=7.22\times 10^{17}$ GeV by adding 7 pair of $XD+\overline{XD}$. During the evolution of gauge couplings, the extra particles are introduced around $5.6\times10^{15}$ GeV. While, when the same number of vector-like particles $XU+\overline{XU}$ are added at $7\times10^{14}$ GeV, a GUT scale 
gauge coupling relation is obtained at $1.8\times10^{16}$ GeV. Furthermore, adding both particles and fine-tuning their masses, $M_{\rm U}=5.0\times10^{17}$ GeV can be achieved and the accuracy is as small as $0.39\%$. The evolution of gauge couplings for the model with vector-like particles are show in figure~\ref{fig:model2}. Models 16 is similar and the evolution of their gauge coupling is shown in figure \ref{fig:model15}.
\begin{figure}[]\centering
    \subfigure[]{\includegraphics[width=0.45\linewidth]{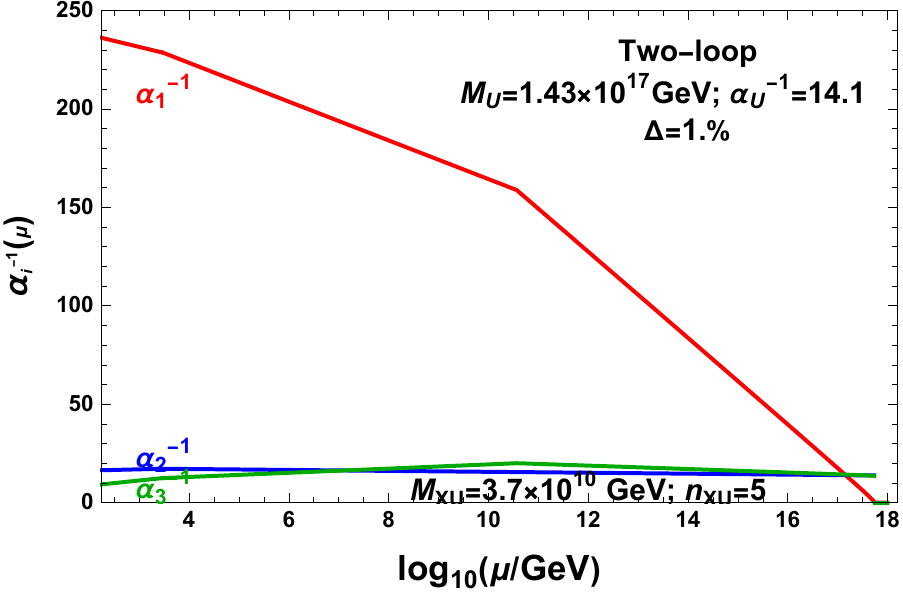}}
    \subfigure[]{\includegraphics[width=0.45\linewidth]{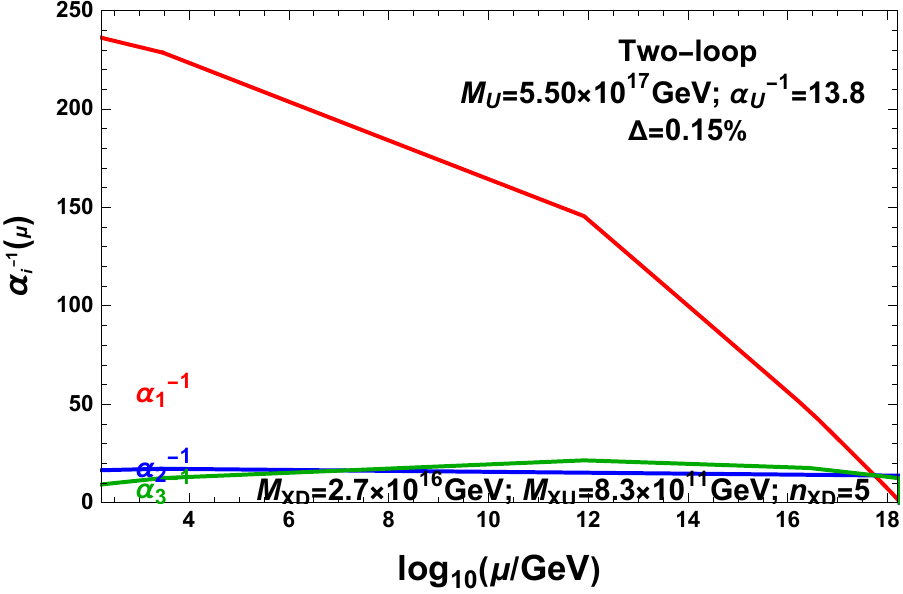}}
    \caption{Two-loop evolution of gauge couplings for the \textbf{Model 17} with vector-like particles. In the model, $k_Y=\frac{1}{4}\times\frac{5}{3}$ and $k_2=\frac{11}{6}$. The string-scale gauge coupling relations can be achieved by adding $5(XU+\overline{XU})$ (a) as well as $5(XD+\overline{XD})$ (b).} 
	\label{fig:model23}
\end{figure}

For Model 17, the parameters $k_y=1/4$ and $k_2=11/6$ deviate significantly from $1$, the evolution for U(1) coupling $\alpha_1^{-1}$ cannot intersect the other two couplings. In order to get a string-scale gauge coupling relation, we choose to introduce 5 pairs of $XU+\overline{XU}$ at  $3.7\times10^{10}$ GeV. The evolution of gauge couplings for this model without and with vector-like particles are shown in figure~\ref{fig:model23}. Note that the energy scale is pushed above the Planck scale  by adding $XD+\overline{XD}$. This is common for models with parameters $k_y\ll 1$ and $k_2\gg 1$. In such case, we introduce additional $XU+\overline{XU}$ particles to pull 
the energy scale  to intermediate scales $10^{14}-10^{15}$ GeV.  Therefore, a string-scale gauge coupling relation can be achieved by adding both $XD+\overline{XD}$ and $XU+\overline{XU}$ and fine-tuning their masses. The masses for these vector-like particles $M_V$ and the energy scales for Models 18-26 are listed in table~\ref{tab:XDXU-n2}, and the corresponding evolution of gauge couplings are illustrated in figures~\ref{fig:model4-6-16-5-21}, \ref{fig:model13-19} and \ref{fig:model8}, in Appendix~\ref{apdx-RGE}. 

\begin{figure}[H]\centering
\includegraphics[width=0.45\linewidth]{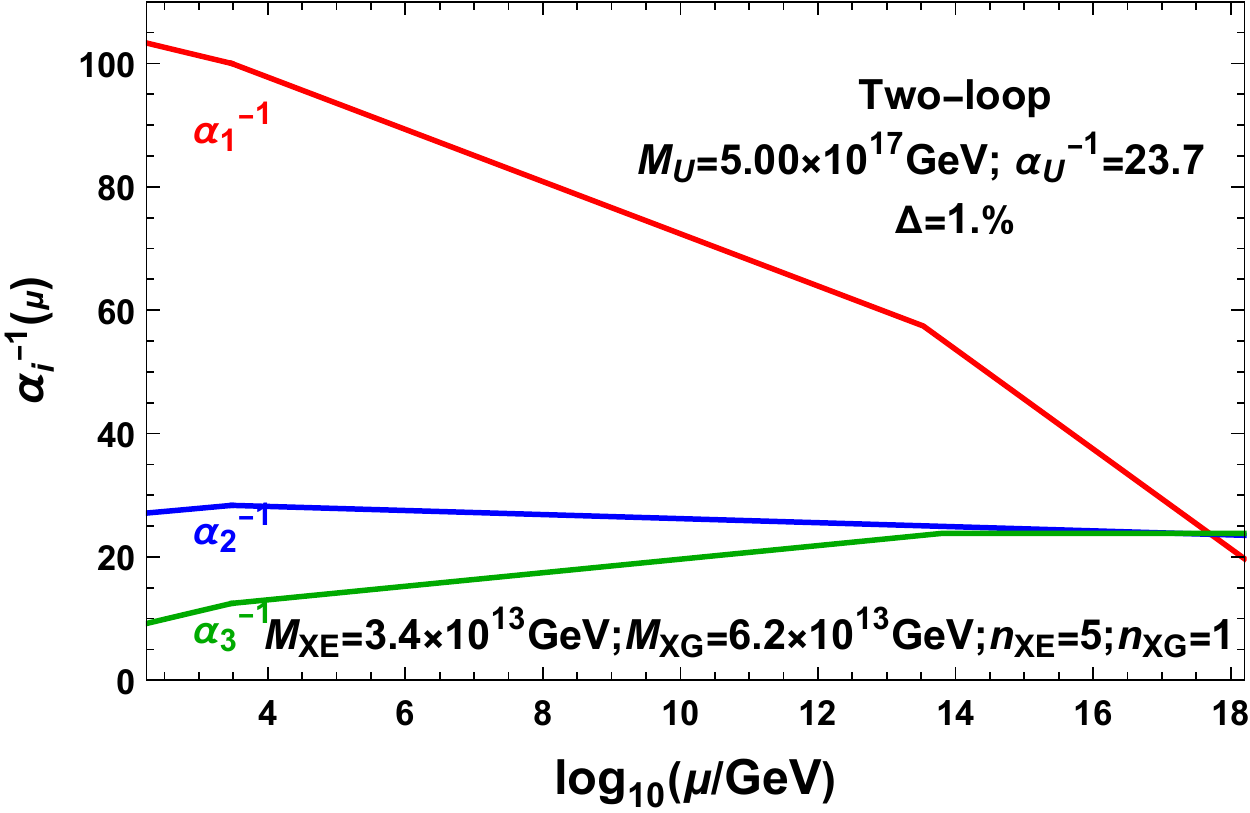}
	\caption{Two-loop evolution of gauge couplings for the \textbf{Model 27} with vector-like particles. In this model, $k_Y=\frac{4}{9}\times\frac{5}{3}$ and $k_2=\frac{10}{9}$. The string-scale gauge coupling relations can be achieved by adding $5(XE+\overline{XE})+XG$ around $10^{13}$ GeV.} 
	\label{fig:model9}
\end{figure}

However, if one and only one of parameters $k_y$ and $k_2$ deviates significantly from $1$, the situations are more complicated. When $k_2\sim 1$ and $k_y\ll 1$, by adding $(XD+\overline{XD})$ and $(XU+\overline{XU})$ the string-scale gauge coupling realtion is pushed too high and above the Planck scale.  Another kind of vector-like particles $(XE+\overline{XE})$ from $\mathcal{N}=2$ subsector as well as $XG$ from \textit{aa} sector are added in these models to obtain the string-scale gauge coupling relations. For Model 27 with $k_y=4/9$ and $k_2=10/9$, the gauge coupling relation can be realized at the string scale $M_{\rm U}=5.00\times10^{17}$ GeV by adding $5(XE+\overline{XE})+XG$ at $10^{13}$ GeV. 
The details for Model 27, and 28-29 are listed in table~\ref{tab:XEXG-n2} and the corresponding running of gauge couplings are plotted in figures~\ref{fig:model9}, and \ref{fig:model10-18}.  

\begin{table*}[!t]\centering
	\footnotesize
		\begin{tabular}{p{45pt}p{30pt}p{30pt}p{30pt}p{55pt}p{30pt}p{55pt}p{55pt}}
  \hline\hline
			Model No. & $k_y$ & $k_2$ & $n_{XE}$ &$M_{XE}$(GeV) & $n_{XG}$&$M_{XG}$(GeV)& $M_{\rm U}$(GeV)\\
			\hline		
			{27} & {4/9}&{10/9}  &{5} &$3.40\times10^{13}$&1&$6.20\times10^{13}$&$5.00\times10^{17}$\\ 
			{28} & {5/11}&{1}  &{5} &$5.50\times10^{12}$&1&$3.80\times10^{15}$&$5.00\times10^{17}$\\ 
			{29} & {1/4}&{7/6}  &{5} &$7.80\times10^{9}$&1&$1.70\times10^{12}$&$5.00\times10^{17}$\\ 
			{30} & {10/7}&{27/11}  &{3} &$1.51\times10^{12}$&3&$3.11\times10^{12}$&$ 5.00\times10^{17} $\\ 
			{31} & {5/3}&{13/5}  &{3} &$1.36\times10^{13}$&3&$4.98\times10^{12}$&$ 5.00\times10^{17} $\\ 
			{32} & {7/4}&{21/10}  &{3} &$5.72\times10^{16}$&3&$ 1.94\times10^{13} $&$5.00\times10^{17}$\\ 
			{33} & {2}&{26/5}  &{3} &$3.44\times10^{8}$&2&$4.92\times10^7$&$5.00\times10^{17}$\\ 
   \hline\hline
		\end{tabular}
    \caption{String-scale gauge coupling relations achieved  by adding vector-like particles $XE+\overline{XE}$ from $\mathcal{N}=2$ subsector and $XG$ from the adjiont sector.  The number of $XE+\overline{XE}$ is defined by the intersection number on \textit{a} and \textit{c} stacks, $n_{XE}=I_{ac}$.}\label{tab:XEXG-n2}
\end{table*}

For Model 30-33, even the parameter $k_y$ is greater than 1, the intersection of $\alpha_1^{-1}$ and $\alpha_3^{-1}$ is still above the line of $\alpha_2^{-1}$ because $k_2$ is too large. Thus, to get a string-scale gauge coupling relation, the vector-like particles added are $(XE+\overline{XE})$ from $\mathcal{N}=2$ subsector as well as $XG$ from \textit{aa} sectors. For Model 30, a string-scale gauge coupling relation is achieved at $5.00\times10^{17}$ GeV by fine-tuning the masses of $XE$ and $XG$ as well as the number of $XG$. The appropriate number of $XG$ are chosen to reduce the mass splitting.
For  Model 30-33, the energy scales, number and mass of these additional particles are listed in table~\ref{tab:XEXG-n2} and the corresponding evolution are figured in figure~\ref{fig:model22-25-27-32} in Appendix~\ref{apdx-RGE}.

\section{Discussion and Conclusions}

In~\cite{He:2021gug}, we have constructed all the three-family ${\cal N} = 1$ supersymmetric Pati-Salam models in the Type IIA  string theory on $T^6/(\mathbb{Z}_2\times \mathbb{Z}_2)$ orientifold with intersecting D6-branes, and obtained all the possible 33 independent models in total. However, how to realize the string-scale gauge coupling relations for these models is still a big challenge.  In this paper, we systematically studied the string-scale gauge coupling relations for all these models. First, we discussed how to decouple the exotic particles in these models. Second, utilizing the two-loop RGEs revolutions, we obtained string-scale gauge coupling relations by 
introducing additional particles from the adjoint representations of $SU(4)_C$ and $SU(2)_L$ gauge symmetries, SM vector-like particles from four-dimensional chiral sectors, as well as vector-like particles from $\mathcal{N}=2$ subsector. Although most of these supersymmetric Pati-Salam models do not directly have traditional gauge coupling unification at string scale, their gauge coupling relations can indeed be realized at string scale.
Therefore, we solved the string-scale gauge coupling relation problems for the generic intersecting D6-brane models.
It seems to us that this systematic method can be applied to the other intersecting D-brane model building as well.

\begin{acknowledgments}
TL is supported in part by the National Key Research and Development Program of China Grant No. 2020YFC2201504, by the Projects No. 11875062, No. 11947302, No. 12047503, and No. 12275333 supported by the National Natural Science Foundation of China, by the Key Research Program of the Chinese Academy of Sciences, Grant NO. XDPB15, by the Scientific Instrument Developing Project of the Chinese Academy of Sciences, Grant No. YJKYYQ20190049, and by the International Partnership Program of Chinese Academy of Sciences for Grand Challenges, Grant No. 112311KYSB20210012. RS is supported by KIAS Individual Grant PG080701 and PG080704.   

\end{acknowledgments}

\newpage
\appendix
\FloatBarrier

\section{Supersymmetric Pati-Salam Models}\label{apdx-data}
In this Appendix, we present the supersymmetric Pati-Salam models with $33$ types of allowed gauge coupling relations on the landscape of supersymmetric Pati-Salam model building.

\begin{table}[h]\scriptsize
    \begin{center}
	\begin{tabular}{|c|c|c|c|c|c|c|c|c|c|c|c|c|}
			\hline\rm{{Model 1}}  & \multicolumn{12}{c|}{$U(4)\times U(2)_L\times U(2)_R\times USp(2)^4 $}\\
			\hline \hline			\rm{stack} & $N$ & $(n^1,l^1)\times(n^2,l^2)\times(n^3,l^3)$ & $n_{\Ysymm}$& $n_{\Yasymm}$ & $b$ & $b'$ & $c$ & $c'$ & 1 & 2 & 3 & 4\\
			\hline
			$a$ & 8 & $(1,1)\times (1,0)\times (1,-1)$ & 0 & 0  & 3 & 0 & 0 & -3 & 0 & 1 & 0 & -1\\
			$b$ & 4 & $(-1,0)\times (-1,3)\times (1,1)$ & 2 & -2  & - & - & 0 & 0 & 0 & 0 & -3 & 1\\
			$c$ & 4 & $(0,1)\times (-1,3)\times (-1,1)$ & 2 & -2  & - & - & - & - & -3 & 1 & 0 & 0\\
			\hline
			1 & 2 & $(1, 0)\times (1, 0)\times (2, 0)$& \multicolumn{10}{c|}{$x_A = \frac{1}{3}x_B = x_C = \frac{1}{3}x_D$}\\
			2 & 2 & $(1, 0)\times (0, -1)\times (0, 2)$& \multicolumn{10}{c|}{$\beta^g_1=\beta^g_2=\beta^g_3=\beta^g_4=-3$}\\
			3 & 2 & $(0, -1)\times (1, 0)\times (0, 2)$& \multicolumn{10}{c|}{$\chi_1=1$, $\chi_2=\frac{1}{3}$, $\chi_3=2$}\\
			4 & 2 & $(0, -1)\times (0, 1)\times (2, 0)$& \multicolumn{10}{c|}{}\\
			\hline
		\end{tabular}
	\end{center}
 \caption{D6-brane configurations and intersection numbers of Model 1, and its gauge coupling relation is $g^2_a=g^2_b=g^2_c=(\frac{5}{3}g^2_Y)=4 \sqrt{\frac{2}{3}} \pi  e^{\phi ^4}$.}
\label{tb:model1}
\end{table}

\begin{table}[!h]\scriptsize
	\begin{center}
		\begin{tabular}{|c||c|c||c|c|c|c|c|c|c|c|c|}
			\hline\rm{{Model 2}} & \multicolumn{11}{c|}{$U(4)\times U(2)_L\times U(2)_R\times USp(2)^3 $}\\
			\hline \hline			\rm{stack} & $N$ & $(n^1,l^1)\times(n^2,l^2)\times(n^3,l^3)$ & $n_{\Ysymm}$& $n_{\Yasymm}$ & $b$ & $b'$ & $c$ & $c'$ & 2 & 3 & 4\\
			\hline
			$a$ & 8 & $(1,-1)\times (1,0)\times (1,1)$ & 0 & 0  & 3 & 0 & 0 & -3 & -1 & 0 & 1\\
			$b$ & 4 & $(-1,4)\times (0,1)\times (-1,1)$ & 3 & -3  & - & - & -7 & 0 & 0 & 1 & 0\\
			$c$ & 4 & $(-2,1)\times (-1,1)\times (1,1)$ & -2 & -6  & - & - & - & - & -1 & -2 & 2\\
			\hline
			2 & 2 & $(1, 0)\times (0, -1)\times (0, 2)$& \multicolumn{9}{c|}{$x_A = \frac{1}{9}x_B = \frac{1}{4}x_C = \frac{1}{9}x_D$}\\
			3 & 2 & $(0, -1)\times (1, 0)\times (0, 2)$& \multicolumn{9}{c|}{$\beta^g_2=-3$, $\beta^g_3=-3$, $\beta^g_4=-2$}\\
			4 & 2 & $(0, -1)\times (0, 1)\times (2, 0)$& \multicolumn{9}{c|}{$\chi_1=\frac{1}{2}$, $\chi_2=\frac{2}{9}$, $\chi_3=1$}\\
			\hline
		\end{tabular}
	\end{center}
 	\caption{D6-brane configurations and intersection numbers of Model 2, and its MSSM gauge coupling relation is $g^2_a=\frac{4}{9}g^2_b=\frac{17}{9}g^2_c=\frac{85}{61}(\frac{5}{3}g^2_Y)=\frac{32 \pi  e^{\phi ^4}}{15}$.}
	\label{tb:model28}
\end{table}

\begin{table}[!h]\scriptsize
	\begin{center}
		\begin{tabular}{|c|c|c|c|c|c|c|c|c|c|c|}
			\hline\rm{Model 3} & \multicolumn{10}{c|}{$U(4)\times U(2)_L\times U(2)_R\times USp(2)^2 $}\\
			\hline \hline			\rm{stack} & $N$ & $(n^1,l^1)\times(n^2,l^2)\times(n^3,l^3)$ & $n_{\Ysymm}$& $n_{\Yasymm}$ & $b$ & $b'$ & $c$ & $c'$ & 1 & 4\\
			\hline
			$a$ & 8 & $(-1,-1)\times (1,1)\times (1,1)$ & 0 & -4  & 0 & 3 & 0 & -3 & 1 & -1\\
			$b$ & 4 & $(-5,2)\times (-1,0)\times (1,1)$ & -3 & 3  & - & - & 0 & 1 & 0 & 5\\
			$c$ & 4 & $(-2,-1)\times (0,1)\times (1,1)$ & 1 & -1  & - & - & - & - & 1 & 0\\
			\hline
			1 & 2 & $(1, 0)\times (1, 0)\times (2, 0)$& \multicolumn{8}{c|}{$x_A = \frac{14}{5}x_B = 2x_C = 7x_D$}\\
			4 & 2 & $(0, -1)\times (0, 1)\times (2, 0)$& \multicolumn{8}{c|}{$\beta^g_1=-3$, $\beta^g_4=1$}\\
			& & & \multicolumn{8}{c|}{$\chi_1=\sqrt{5}$, $\chi_2=\frac{7}{\sqrt{5}}$, $\chi_3=\frac{4}{\sqrt{5}}$}\\
			\hline
		\end{tabular}
	\end{center}
 	\caption{D6-brane configurations and intersection numbers of Model 3, and its gauge coupling relation is $g^2_a=\frac{5}{6}g^2_b=\frac{7}{6}g^2_c=\frac{35}{32}(\frac{5}{3}g^2_Y)=\frac{8}{27} 5^{3/4} \sqrt{7} \pi  e^{\phi ^4}$.}
	\label{tb:model3}
\end{table}
\begin{table}[!h]\scriptsize
	\begin{center}
		\begin{tabular}{|c||c|c||c|c|c|c|c|c|c|}
			\hline\rm{{Model 4}}  & \multicolumn{9}{c|}{$U(4)\times U(2)_L\times U(2)_R\times USp(4) $}\\
			\hline \hline			\rm{stack} & $N$ & $(n^1,l^1)\times(n^2,l^2)\times(n^3,l^3)$ & $n_{\Ysymm}$& $n_{\Yasymm}$ & $b$ & $b'$ & $c$ & $c'$ & 4\\
			\hline
			$a$ & 8 & $(2,1)\times (1,0)\times (1,-1)$ & 1 & -1  & 3 & 0 & 0 & -3 & -2\\
			$b$ & 4 & $(1,0)\times (1,-3)\times (1,1)$ & 2 & -2  & - & - & -4 & 0 & 1\\
			$c$ & 4 & $(-1,2)\times (-1,1)\times (-1,1)$ & 2 & 6  & - & - & - & - & 1\\
			\hline
			4 & 4 & $(0, -1)\times (0, 1)\times (2, 0)$& \multicolumn{7}{c|}{$x_A = \frac{4}{3}x_B = 8x_C = \frac{8}{3}x_D$}\\
			& & & \multicolumn{7}{c|}{$\beta^g_4=0$}\\
			& & & \multicolumn{7}{c|}{$\chi_1=4$, $\chi_2=\frac{2}{3}$, $\chi_3=4$}\\
			\hline
		\end{tabular}
	\end{center}
 	\caption{D6-brane configurations and intersection numbers of Model 4, and its MSSM gauge coupling relation is $g^2_a=\frac{1}{2}g^2_b=\frac{13}{6}g^2_c=\frac{65}{44}(\frac{5}{3}g^2_Y)=\frac{16}{5} \sqrt{\frac{2}{3}} \pi  e^{\phi ^4}$.}
	\label{tb:model29}
\end{table}

\begin{table}[!h]\scriptsize
	\begin{center}
		\begin{tabular}{|c||c|c||c|c|c|c|c|c|c|}
			\hline\rm{Model 5}  & \multicolumn{9}{c|}{$U(4)\times U(2)_L\times U(2)_R\times USp(4) $}\\
			\hline \hline			\rm{stack} & $N$ & $(n^1,l^1)\times(n^2,l^2)\times(n^3,l^3)$ & $n_{\Ysymm}$& $n_{\Yasymm}$ & $b$ & $b'$ & $c$ & $c'$ & 3\\
			\hline
			$a$ & 8 & $(-1,1)\times (-1,0)\times (1,1)$ & 0 & 0  & 3 & 0 & 0 & -3 & 0\\
			$b$ & 4 & $(-1,4)\times (0,1)\times (-1,1)$ & 3 & -3  & - & - & -8 & 0 & 1\\
			$c$ & 4 & $(1,0)\times (2,-3)\times (1,1)$ & 1 & -1  & - & - & - & - & -3\\
			\hline
			3 & 4 & $(0, -1)\times (1, 0)\times (0, 2)$& \multicolumn{7}{c|}{$x_A = \frac{1}{6}x_B = \frac{1}{4}x_C = \frac{1}{6}x_D$}\\
			& & & \multicolumn{7}{c|}{$\beta^g_3=-2$}\\
			& & & \multicolumn{7}{c|}{$\chi_1=\frac{1}{2}$, $\chi_2=\frac{1}{3}$, $\chi_3=1$}\\
			\hline
		\end{tabular}
	\end{center}
 	\caption{D6-brane configurations and intersection numbers of Model 5, and its MSSM gauge coupling relation is $g^2_a=\frac{2}{3}g^2_b=2g^2_c=\frac{10}{7}(\frac{5}{3}g^2_Y)=\frac{16}{5} \sqrt{\frac{2}{3}} \pi  e^{\phi ^4}$.}
	\label{tb:model24}
\end{table}	
\begin{table}[!h]\scriptsize
	\begin{center}
		\begin{tabular}{|c||c|c||c|c|c|c|c|c|c|}
			\hline\rm{Model 6} & \multicolumn{9}{c|}{$U(4)\times U(2)_L\times U(2)_R\times USp(2) $}\\
			\hline \hline			\rm{stack} & $N$ & $(n^1,l^1)\times(n^2,l^2)\times(n^3,l^3)$ & $n_{\Ysymm}$& $n_{\Yasymm}$ & $b$ & $b'$ & $c$ & $c'$ & 3\\
			\hline
			$a$ & 8 & $(1,-1)\times (1,0)\times (1,1)$ & 0 & 0  & 3 & 0 & 0 & -3 & 0\\
			$b$ & 4 & $(-2,5)\times (0,1)\times (-1,1)$ & 3 & -3  & - & - & -8 & 0 & 2\\
			$c$ & 4 & $(-2,1)\times (-1,1)\times (1,1)$ & -2 & -6  & - & - & - & - & -2\\
			\hline
			3 & 2 & $(0, -1)\times (1, 0)\times (0, 2)$& \multicolumn{7}{c|}{$x_A = \frac{1}{6}x_B = \frac{2}{5}x_C = \frac{1}{6}x_D$}\\
			& & & \multicolumn{7}{c|}{$\beta^g_3=-2$}\\
			& & & \multicolumn{7}{c|}{$\chi_1=\sqrt{\frac{2}{5}}$, $\chi_2=\frac{\sqrt{\frac{5}{2}}}{6}$, $\chi_3=2 \sqrt{\frac{2}{5}}$}\\
			\hline
		\end{tabular}
	\end{center}
 	\caption{D6-brane configurations and intersection numbers of Model 6, and its MSSM gauge coupling relation is $g^2_a=\frac{5}{6}g^2_b=\frac{11}{6}g^2_c=\frac{11}{8}(\frac{5}{3}g^2_Y)=\frac{8 \sqrt[4]{2} 5^{3/4} \pi  e^{\phi ^4}}{7 \sqrt{3}}$.}
	\label{tb:model17}
\end{table}

\begin{table}[!h]\scriptsize
	\begin{center}
		\begin{tabular}{|c||c|c||c|c|c|c|c|c|c|c|}
			\hline\rm{Model 7} & \multicolumn{10}{c|}{$U(4)\times U(2)_L\times U(2)_R\times USp(4)^2 $}\\
			\hline \hline			\rm{stack} & $N$ & $(n^1,l^1)\times(n^2,l^2)\times(n^3,l^3)$ & $n_{\Ysymm}$& $n_{\Yasymm}$ & $b$ & $b'$ & $c$ & $c'$ & 2 & 4\\
			\hline
			$a$ & 8 & $(1,1)\times (1,0)\times (1,-1)$ & 0 & 0  & 3 & 0 & 0 & -3 & 1 & -1\\
			$b$ & 4 & $(1,0)\times (1,-3)\times (1,1)$ & 2 & -2  & - & - & -4 & 0 & 0 & 1\\
			$c$ & 4 & $(1,2)\times (-1,1)\times (-1,1)$ & -2 & -6  & - & - & - & - & 2 & -1\\
			\hline
			2 & 4 & $(1, 0)\times (0, -1)\times (0, 2)$& \multicolumn{8}{c|}{$x_A = \frac{1}{3}x_B = x_C = \frac{1}{3}x_D$}\\
			4 & 4 & $(0, -1)\times (0, 1)\times (2, 0)$& \multicolumn{8}{c|}{$\beta^g_2=-2$, $\beta^g_4=-2$}\\
			& & & \multicolumn{8}{c|}{$\chi_1=1$, $\chi_2=\frac{1}{3}$, $\chi_3=2$}\\
			\hline
		\end{tabular}
	\end{center}
 	\caption{D6-brane configurations and intersection numbers of Model 7, and its MSSM gauge coupling relation is $g^2_a=g^2_b=\frac{5}{3}g^2_c=\frac{25}{19}(\frac{5}{3}g^2_Y)=4 \sqrt{\frac{2}{3}} \pi  e^{\phi ^4}$.}
	\label{tb:model7}
\end{table}

\begin{table}[!h]\scriptsize
	\begin{center}
		\begin{tabular}{|c||c|c||c|c|c|c|c|c|c|c|c|}
			\hline\rm{Model 8} & \multicolumn{11}{c|}{$U(4)\times U(2)_L\times U(2)_R\times USp(2)^3 $}\\
			\hline \hline			\rm{stack} & $N$ & $(n^1,l^1)\times(n^2,l^2)\times(n^3,l^3)$ & $n_{\Ysymm}$& $n_{\Yasymm}$ & $b$ & $b'$ & $c$ & $c'$ & 1 & 2 & 4\\
			\hline
			$a$ & 8 & $(1,1)\times (1,0)\times (1,-1)$ & 0 & 0  & 3 & 0 & 0 & -3 & 0 & 1 & -1\\
			$b$ & 4 & $(-1,0)\times (-1,3)\times (1,1)$ & 2 & -2  & - & - & -3 & 0 & 0 & 0 & 1\\
			$c$ & 4 & $(0,1)\times (-2,3)\times (-1,1)$ & 1 & -1  & - & - & - & - & -3 & 2 & 0\\
			\hline
			1 & 2 & $(1, 0)\times (1, 0)\times (2, 0)$& \multicolumn{9}{c|}{$x_A = \frac{2}{3}x_B = 2x_C = \frac{2}{3}x_D$}\\
			2 & 2 & $(1, 0)\times (0, -1)\times (0, 2)$& \multicolumn{9}{c|}{$\beta^g_1=-3$, $\beta^g_2=-2$, $\beta^g_4=-3$}\\
			4 & 2 & $(0, -1)\times (0, 1)\times (2, 0)$& \multicolumn{9}{c|}{$\chi_1=\sqrt{2}$, $\chi_2=\frac{\sqrt{2}}{3}$, $\chi_3=2 \sqrt{2}$}\\
			\hline
		\end{tabular}
	\end{center}
 	\caption{D6-brane configurations and intersection numbers of Model 8, and its MSSM gauge coupling relation is $g^2_a=g^2_b=2g^2_c=\frac{10}{7}(\frac{5}{3}g^2_Y)=\frac{16 \sqrt[4]{2} \pi  e^{\phi ^4}}{3 \sqrt{3}}$.}
	\label{tb:model14}
\end{table}

\begin{table}[!h]\scriptsize
	\begin{center}
		\begin{tabular}{|c||c|c||c|c|c|c|c|c|c|c|}
			\hline\rm{Model 9} & \multicolumn{10}{c|}{$U(4)\times U(2)_L\times U(2)_R\times USp(2)^2 $}\\
			\hline \hline			\rm{stack} & $N$ & $(n^1,l^1)\times(n^2,l^2)\times(n^3,l^3)$ & $n_{\Ysymm}$& $n_{\Yasymm}$ & $b$ & $b'$ & $c$ & $c'$ & 1 & 4\\
			\hline
			$a$ & 8 & $(1,-1)\times (-1,1)\times (1,-1)$ & 0 & 4  & 3 & 0 & 0 & -3 & -1 & 1\\
			$b$ & 4 & $(1,-4)\times (1,0)\times (1,1)$ & 3 & -3  & - & - & -7 & 0 & 0 & 1\\
			$c$ & 4 & $(-2,1)\times (2,1)\times (-1,1)$ & -3 & -13  & - & - & - & - & -1 & -4\\
			\hline
			1 & 2 & $(1, 0)\times (1, 0)\times (2, 0)$& \multicolumn{8}{c|}{$x_A = 28x_B = \frac{28}{23}x_C = 7x_D$}\\
			4 & 2 & $(0, -1)\times (0, 1)\times (2, 0)$& \multicolumn{8}{c|}{$\beta^g_1=-3$, $\beta^g_4=1$}\\
			& & & \multicolumn{8}{c|}{$\chi_1=\sqrt{\frac{7}{23}}$, $\chi_2=\sqrt{161}$, $\chi_3=8 \sqrt{\frac{7}{23}}$}\\
			\hline
		\end{tabular}
	\end{center}
 	\caption{D6-brane configurations and intersection numbers of Model 9, and its MSSM gauge coupling relation is $g^2_a=\frac{1}{6}g^2_b=\frac{11}{6}g^2_c=\frac{11}{8}(\frac{5}{3}g^2_Y)=\frac{8}{405}  161^{3/4} \sqrt{2}\pi  e^{\phi ^4}$.}
	\label{tb:model33}
\end{table}
\begin{table}[!h]\scriptsize
	\begin{center}
		\begin{tabular}{|c||c|c||c|c|c|c|c|c|c|c|c|}
			\hline\rm{Model 10} & \multicolumn{11}{c|}{$U(4)\times U(2)_L\times U(2)_R\times USp(2)^2\times USp(4) $}\\
			\hline \hline			\rm{stack} & $N$ & $(n^1,l^1)\times(n^2,l^2)\times(n^3,l^3)$ & $n_{\Ysymm}$& $n_{\Yasymm}$ & $b$ & $b'$ & $c$ & $c'$ & 1 & 3 & 4\\
			\hline
			$a$ & 8 & $(-1,-1)\times (1,1)\times (1,1)$ & 0 & -4  & 0 & 3 & 0 & -3 & 1 & -1 & -1\\
			$b$ & 4 & $(-4,1)\times (-1,0)\times (1,1)$ & -3 & 3  & - & - & 0 & 2 & 0 & 0 & 4\\
			$c$ & 4 & $(-2,-1)\times (0,1)\times (1,1)$ & 1 & -1  & - & - & - & - & 1 & -2 & 0\\
			\hline
			1 & 4 & $(1, 0)\times (1, 0)\times (2, 0)$& \multicolumn{9}{c|}{$x_A = \frac{5}{2}x_B = 2x_C = 10x_D$}\\
			3 & 2 & $(0, -1)\times (1, 0)\times (0, 2)$& \multicolumn{9}{c|}{$\beta^g_1=-3$, $\beta^g_3=-2$, $\beta^g_4=0$}\\
			4 & 2 & $(0, -1)\times (0, 1)\times (2, 0)$& \multicolumn{9}{c|}{$\chi_1=2 \sqrt{2}$, $\chi_2=\frac{5}{\sqrt{2}}$, $\chi_3=\sqrt{2}$}\\
			\hline
		\end{tabular}
	\end{center}
 	\caption{D6-brane configurations and intersection numbers of Model 10, and its MSSM gauge coupling relation is $g^2_a=\frac{4}{9}g^2_b=\frac{10}{9}g^2_c=\frac{50}{47}(\frac{5}{3}g^2_Y)=\frac{16}{27} 2^{3/4} \sqrt{5} \pi  e^{\phi ^4}$.}
	\label{tb:model12}
\end{table}
\begin{table}[!h]\scriptsize
	\begin{center}
		\begin{tabular}{|c||c|c||c|c|c|c|c|c|c|c|}
			\hline\rm{Model 11} & \multicolumn{10}{c|}{$U(4)\times U(2)_L\times U(2)_R\times USp(2)\times USp(4) $}\\
			\hline \hline			\rm{stack} & $N$ & $(n^1,l^1)\times(n^2,l^2)\times(n^3,l^3)$ & $n_{\Ysymm}$& $n_{\Yasymm}$ & $b$ & $b'$ & $c$ & $c'$ & 1 & 3\\
			\hline
			$a$ & 8 & $(-1,1)\times (-1,0)\times (1,1)$ & 0 & 0  & 3 & 0 & 0 & -3 & 0 & 0\\
			$b$ & 4 & $(-1,4)\times (0,1)\times (-1,1)$ & 3 & -3  & - & - & -4 & 0 & -4 & 1\\
			$c$ & 4 & $(1,0)\times (1,-3)\times (1,1)$ & 2 & -2  & - & - & - & - & 0 & -3\\
			\hline
			1 & 2 & $(1, 0)\times (1, 0)\times (2, 0)$& \multicolumn{8}{c|}{$x_A = \frac{1}{12}x_B = \frac{1}{4}x_C = \frac{1}{12}x_D$}\\
			3 & 4 & $(0, -1)\times (1, 0)\times (0, 2)$& \multicolumn{8}{c|}{$\beta^g_1=-2$, $\beta^g_3=-2$}\\
			& & & \multicolumn{8}{c|}{$\chi_1=\frac{1}{2}$, $\chi_2=\frac{1}{6}$, $\chi_3=1$}\\
			\hline
		\end{tabular}
	\end{center}
 	\caption{D6-brane configurations and intersection numbers of Model 11, and its MSSM gauge coupling relation is $g^2_a=\frac{1}{3}g^2_b=g^2_c=(\frac{5}{3}g^2_Y)=\frac{16 \pi  e^{\phi ^4}}{5 \sqrt{3}}$.}
	\label{tb:model20}
\end{table}

\begin{table}[!h]\scriptsize
	\begin{center}
		\begin{tabular}{|c||c|c||c|c|c|c|c|c|c|c|}
			\hline\rm{Model 12} & \multicolumn{10}{c|}{$U(4)\times U(2)_L\times U(2)_R\times USp(2)\times USp(4) $}\\
			\hline \hline			\rm{stack} & $N$ & $(n^1,l^1)\times(n^2,l^2)\times(n^3,l^3)$ & $n_{\Ysymm}$& $n_{\Yasymm}$ & $b$ & $b'$ & $c$ & $c'$ & 1 & 4\\
			\hline
			$a$ & 8 & $(-1,-1)\times (2,1)\times (1,1)$ & 0 & -8  & 0 & 3 & 0 & -3 & 1 & -2\\
			$b$ & 4 & $(4,-1)\times (1,0)\times (1,1)$ & -3 & 3  & - & - & 0 & 2 & 0 & 4\\
			$c$ & 4 & $(-2,-1)\times (-1,1)\times (1,1)$ & 2 & 6  & - & - & - & - & 1 & 2\\
			\hline
			1 & 4 & $(1, 0)\times (1, 0)\times (2, 0)$& \multicolumn{8}{c|}{$x_A = \frac{13}{2}x_B = \frac{13}{8}x_C = 26x_D$}\\
			4 & 2 & $(0, -1)\times (0, 1)\times (2, 0)$& \multicolumn{8}{c|}{$\beta^g_1=-3$, $\beta^g_4=4$}\\
			& & & \multicolumn{8}{c|}{$\chi_1=\sqrt{\frac{13}{2}}$, $\chi_2=2 \sqrt{26}$, $\chi_3=\frac{\sqrt{\frac{13}{2}}}{2}$}\\
			\hline
		\end{tabular}
	\end{center}
 	\caption{D6-brane configurations and intersection numbers of Model 12, and its MSSM gauge coupling relation is $g^2_a=\frac{1}{6}g^2_b=\frac{7}{6}g^2_c=\frac{35}{32}(\frac{5}{3}g^2_Y)=\frac{16}{135} 26^{3/4} \pi  e^{\phi ^4}$.}
	\label{tb:model31}
\end{table}

\begin{table}[!h]\scriptsize
	\begin{center}
		\begin{tabular}{|c||c|c||c|c|c|c|c|c|c|c|}
			\hline\rm{Model 13} & \multicolumn{10}{c|}{$U(4)\times U(2)_L\times U(2)_R\times USp(2)^2 $}\\
			\hline \hline			\rm{stack} & $N$ & $(n^1,l^1)\times(n^2,l^2)\times(n^3,l^3)$ & $n_{\Ysymm}$& $n_{\Yasymm}$ & $b$ & $b'$ & $c$ & $c'$ & 1 & 4\\
			\hline
			$a$ & 8 & $(-1,-1)\times (1,1)\times (1,1)$ & 0 & -4  & 6 & -3 & 0 & -3 & 1 & -1\\
			$b$ & 4 & $(-1,2)\times (-1,0)\times (5,1)$ & 9 & -9  & - & - & -10 & -9 & 0 & 1\\
			$c$ & 4 & $(-2,-1)\times (0,1)\times (1,1)$ & 1 & -1  & - & - & - & - & 1 & 0\\
			\hline
			1 & 2 & $(1, 0)\times (1, 0)\times (2, 0)$& \multicolumn{8}{c|}{$x_A = 22x_B = 2x_C = \frac{11}{5}x_D$}\\
			4 & 2 & $(0, -1)\times (0, 1)\times (2, 0)$& \multicolumn{8}{c|}{$\beta^g_1=-3$, $\beta^g_4=-3$}\\
			& & & \multicolumn{8}{c|}{$\chi_1=\frac{1}{\sqrt{5}}$, $\chi_2=\frac{11}{\sqrt{5}}$, $\chi_3=4 \sqrt{5}$}\\
			\hline
		\end{tabular}
	\end{center}
 	\caption{D6-brane configurations and intersection numbers of Model 13, and its MSSM gauge coupling relation is $g^2_a=\frac{5}{14}g^2_b=\frac{11}{6}g^2_c=\frac{11}{8}(\frac{5}{3}g^2_Y)=\frac{8}{63} 5^{3/4} \sqrt{11} \pi  e^{\phi ^4}$.}
	\label{tb:model30}
\end{table}

\begin{table}[!h]\scriptsize
	\begin{center}
		\begin{tabular}{|c||c|c||c|c|c|c|c|c|c|}
			\hline\rm{Model 14} & \multicolumn{9}{c|}{$U(4)\times U(2)_L\times U(2)_R\times USp(2) $}\\
			\hline \hline			\rm{stack} & $N$ & $(n^1,l^1)\times(n^2,l^2)\times(n^3,l^3)$ & $n_{\Ysymm}$& $n_{\Yasymm}$ & $b$ & $b'$ & $c$ & $c'$ & 3\\
			\hline
			$a$ & 8 & $(1,-1)\times (1,0)\times (1,1)$ & 0 & 0  & -3 & 6 & 0 & -3 & 0\\
			$b$ & 4 & $(-2,1)\times (0,1)\times (-5,1)$ & -9 & 9  & - & - & 0 & 8 & 10\\
			$c$ & 4 & $(-2,1)\times (-1,1)\times (1,1)$ & -2 & -6  & - & - & - & - & -2\\
			\hline
			3 & 2 & $(0, -1)\times (1, 0)\times (0, 2)$& \multicolumn{7}{c|}{$x_A = \frac{5}{6}x_B = 10x_C = \frac{5}{6}x_D$}\\
			& & & \multicolumn{7}{c|}{$\beta^g_3=6$}\\
			& & & \multicolumn{7}{c|}{$\chi_1=\sqrt{10}$, $\chi_2=\frac{\sqrt{\frac{5}{2}}}{6}$, $\chi_3=2 \sqrt{10}$}\\
			\hline
		\end{tabular}
	\end{center}
 	\caption{D6-brane configurations and intersection numbers of Model 14, and its MSSM gauge coupling relation is $g^2_a=\frac{35}{66}g^2_b=\frac{7}{6}g^2_c=\frac{35}{32}(\frac{5}{3}g^2_Y)=\frac{8 \sqrt[4]{2} 5^{3/4} \pi  e^{\phi ^4}}{11 \sqrt{3}}$.}
	\label{tb:model11}
\end{table}

\begin{table}[h]\scriptsize
	\begin{center}
		\begin{tabular}{|c|c|c|c|c|c|c|c|c|c|c|}
			\hline\rm{Model 15} & \multicolumn{10}{c|}{$U(4)\times U(2)_L\times U(2)_R\times USp(2)^2 $}\\
			\hline \hline			\rm{stack} & $N$ & $(n^1,l^1)\times(n^2,l^2)\times(n^3,l^3)$ & $n_{\Ysymm}$& $n_{\Yasymm}$ & $b$ & $b'$ & $c$ & $c'$ & 1 & 4\\
			\hline
			$a$ & 8 & $(1,-1)\times (-1,1)\times (1,-1)$ & 0 & 4  & 0 & 3 & 0 & -3 & -1 & 1\\
			$b$ & 4 & $(0,1)\times (-2,1)\times (-1,1)$ & -1 & 1  & - & - & 0 & -1 & -1 & 0\\
			$c$ & 4 & $(-1,0)\times (5,2)\times (-1,1)$ & 3 & -3  & - & - & - & - & 0 & -5\\
			\hline
			1 & 2 & $(1, 0)\times (1, 0)\times (2, 0)$& \multicolumn{8}{c|}{$x_A = 2x_B = \frac{14}{5}x_C = 7x_D$}\\
			4 & 2 & $(0, -1)\times (0, 1)\times (2, 0)$& \multicolumn{8}{c|}{$\beta^g_1=-3$, $\beta^g_4=1$}\\
			& & & \multicolumn{8}{c|}{$\chi_1=\frac{7}{\sqrt{5}}$, $\chi_2=\sqrt{5}$, $\chi_3=\frac{4}{\sqrt{5}}$}\\
			\hline
		\end{tabular}
	\end{center}
 	\caption{D6-brane configurations and intersection numbers of Model 15, and its gauge coupling relation is $g^2_a=\frac{7}{6}g^2_b=\frac{5}{6}g^2_c=\frac{25}{28}(\frac{5}{3}g^2_Y)=\frac{8}{27} 5^{3/4} \sqrt{7} \pi  e^{\phi ^4}$.}
	\label{tb:model2}
\end{table}

\begin{table}[!h]\scriptsize
	\begin{center}
		\begin{tabular}{|c||c|c||c|c|c|c|c|c|c|c|}
			\hline\rm{Model 16} & \multicolumn{10}{c|}{$U(4)\times U(2)_L\times U(2)_R\times USp(2)^2 $}\\
			\hline \hline			\rm{stack} & $N$ & $(n^1,l^1)\times(n^2,l^2)\times(n^3,l^3)$ & $n_{\Ysymm}$& $n_{\Yasymm}$ & $b$ & $b'$ & $c$ & $c'$ & 2 & 4\\
			\hline
			$a$ & 8 & $(1,1)\times (1,0)\times (1,-1)$ & 0 & 0  & 2 & 1 & 0 & -3 & 1 & -1\\
			$b$ & 4 & $(-1,0)\times (-2,1)\times (1,3)$ & -5 & 5  & - & - & 8 & 8 & 0 & 6\\
			$c$ & 4 & $(0,1)\times (-2,3)\times (-1,1)$ & 1 & -1  & - & - & - & - & 2 & 0\\
			\hline
			2 & 2 & $(1, 0)\times (0, -1)\times (0, 2)$& \multicolumn{8}{c|}{$x_A = \frac{2}{3}x_B = \frac{1}{9}x_C = \frac{2}{3}x_D$}\\
			4 & 2 & $(0, -1)\times (0, 1)\times (2, 0)$& \multicolumn{8}{c|}{$\beta^g_2=-2$, $\beta^g_4=2$}\\
			& & & \multicolumn{8}{c|}{$\chi_1=\frac{1}{3}$, $\chi_2=2$, $\chi_3=\frac{2}{3}$}\\
			\hline
		\end{tabular}
	\end{center}
 	\caption{D6-brane configurations and intersection numbers of Model 16, and its MSSM gauge coupling relation is $g^2_a=\frac{18}{5}g^2_b=2g^2_c=\frac{10}{7}(\frac{5}{3}g^2_Y)=\frac{24 \pi  e^{\phi ^4}}{5}$.}
	\label{tb:model26}
\end{table}

\begin{table}[!h]\scriptsize
	\begin{center}
		\begin{tabular}{|c||c|c||c|c|c|c|c|c|c|c|}
			\hline\rm{Model 17} & \multicolumn{10}{c|}{$U(4)\times U(2)_L\times U(2)_R\times USp(2)^2 $}\\
			\hline \hline			\rm{stack} & $N$ & $(n^1,l^1)\times(n^2,l^2)\times(n^3,l^3)$ & $n_{\Ysymm}$& $n_{\Yasymm}$ & $b$ & $b'$ & $c$ & $c'$ & 2 & 4\\
			\hline
			$a$ & 8 & $(1,1)\times (1,0)\times (1,-1)$ & 0 & 0  & 3 & 0 & 0 & -3 & 1 & -1\\
			$b$ & 4 & $(-1,0)\times (-2,3)\times (1,1)$ & 1 & -1  & - & - & 0 & 0 & 0 & 2\\
			$c$ & 4 & $(0,1)\times (-2,3)\times (-1,1)$ & 1 & -1  & - & - & - & - & 2 & 0\\
			\hline
			2 & 2 & $(1, 0)\times (0, -1)\times (0, 2)$& \multicolumn{8}{c|}{$x_A = \frac{2}{3}x_B = x_C = \frac{2}{3}x_D$}\\
			4 & 2 & $(0, -1)\times (0, 1)\times (2, 0)$& \multicolumn{8}{c|}{$\beta^g_2=-2$, $\beta^g_4=-2$}\\
			& & & \multicolumn{8}{c|}{$\chi_1=1$, $\chi_2=\frac{2}{3}$, $\chi_3=2$}\\
			\hline
		\end{tabular}
	\end{center}
 	\caption{D6-brane configurations and intersection numbers of Model 17, and its MSSM gauge coupling relation is $g^2_a=2g^2_b=2g^2_c=\frac{10}{7}(\frac{5}{3}g^2_Y)=\frac{8 \pi  e^{\phi ^4}}{\sqrt{3}}$.}
	\label{tb:model15}
\end{table}

\begin{table}[!h]\scriptsize
	\begin{center}
		\begin{tabular}{|c||c|c||c|c|c|c|c|c|c|c|}
			\hline\rm{Model 18} & \multicolumn{10}{c|}{$U(4)\times U(2)_L\times U(2)_R\times USp(2)^2 $}\\
			\hline \hline			\rm{stack} & $N$ & $(n^1,l^1)\times(n^2,l^2)\times(n^3,l^3)$ & $n_{\Ysymm}$& $n_{\Yasymm}$ & $b$ & $b'$ & $c$ & $c'$ & 1 & 4\\
			\hline
			$a$ & 8 & $(-1,-1)\times (1,1)\times (1,1)$ & 0 & -4  & 3 & 0 & 0 & -3 & 1 & -1\\
			$b$ & 4 & $(-2,1)\times (2,1)\times (-1,1)$ & -3 & -13  & - & - & 7 & 0 & -1 & -4\\
			$c$ & 4 & $(1,-4)\times (1,0)\times (1,1)$ & 3 & -3  & - & - & - & - & 0 & 1\\
			\hline
			1 & 2 & $(1, 0)\times (1, 0)\times (2, 0)$& \multicolumn{8}{c|}{$x_A = 28x_B = \frac{28}{23}x_C = 7x_D$}\\
			4 & 2 & $(0, -1)\times (0, 1)\times (2, 0)$& \multicolumn{8}{c|}{$\beta^g_1=-3$, $\beta^g_4=1$}\\
			& & & \multicolumn{8}{c|}{$\chi_1=\sqrt{\frac{7}{23}}$, $\chi_2=\sqrt{161}$, $\chi_3=8 \sqrt{\frac{7}{23}}$}\\
			\hline
		\end{tabular}
	\end{center}
 	\caption{D6-brane configurations and intersection numbers of Model 18, and its MSSM gauge coupling relation is $g^2_a=\frac{11}{6}g^2_b=\frac{1}{6}g^2_c=\frac{1}{4}(\frac{5}{3}g^2_Y)=\frac{8}{405} \sqrt{2} 161^{3/4} \pi  e^{\phi ^4}$.}
	\label{tb:model23}
\end{table}

\begin{table}[!h]\scriptsize
	\begin{center}
		\begin{tabular}{|c||c|c||c|c|c|c|c|c|c|c|}
			\hline\rm{Model 19} & \multicolumn{10}{c|}{$U(4)\times U(2)_L\times U(2)_R\times USp(4)^2 $}\\
			\hline \hline			\rm{stack} & $N$ & $(n^1,l^1)\times(n^2,l^2)\times(n^3,l^3)$ & $n_{\Ysymm}$& $n_{\Yasymm}$ & $b$ & $b'$ & $c$ & $c'$ & 2 & 4\\
			\hline
			$a$ & 8 & $(1,1)\times (1,0)\times (1,-1)$ & 0 & 0  & 3 & 0 & 0 & -3 & 1 & -1\\
			$b$ & 4 & $(-2,1)\times (-1,1)\times (1,1)$ & -2 & -6  & - & - & 4 & 0 & -1 & 2\\
			$c$ & 4 & $(0,1)\times (-1,3)\times (-1,1)$ & 2 & -2  & - & - & - & - & 1 & 0\\
			\hline
			2 & 4 & $(1, 0)\times (0, -1)\times (0, 2)$& \multicolumn{8}{c|}{$x_A = \frac{1}{3}x_B = x_C = \frac{1}{3}x_D$}\\
			4 & 4 & $(0, -1)\times (0, 1)\times (2, 0)$& \multicolumn{8}{c|}{$\beta^g_2=-2$, $\beta^g_4=-2$}\\
			& & & \multicolumn{8}{c|}{$\chi_1=1$, $\chi_2=\frac{1}{3}$, $\chi_3=2$}\\
			\hline
		\end{tabular}
	\end{center}
 	\caption{D6-brane configurations and intersection numbers of Model 19, and its MSSM gauge coupling relation is $g^2_a=\frac{5}{3}g^2_b=g^2_c=(\frac{5}{3}g^2_Y)=4 \sqrt{\frac{2}{3}} \pi  e^{\phi ^4}$.}
	\label{tb:model4}
\end{table}
\begin{table}[!h]\scriptsize
	\begin{center}
		\begin{tabular}{|c||c|c||c|c|c|c|c|c|c|c|c|}
			\hline\rm{Model 20} & \multicolumn{11}{c|}{$U(4)\times U(2)_L\times U(2)_R\times USp(2)^3 $}\\
			\hline \hline			\rm{stack} & $N$ & $(n^1,l^1)\times(n^2,l^2)\times(n^3,l^3)$ & $n_{\Ysymm}$& $n_{\Yasymm}$ & $b$ & $b'$ & $c$ & $c'$ & 2 & 3 & 4\\
			\hline
			$a$ & 8 & $(1,1)\times (1,0)\times (1,-1)$ & 0 & 0  & 3 & 0 & 0 & -3 & 1 & 0 & -1\\
			$b$ & 4 & $(-1,0)\times (-2,3)\times (1,1)$ & 1 & -1  & - & - & 3 & 0 & 0 & -3 & 2\\
			$c$ & 4 & $(0,1)\times (-1,3)\times (-1,1)$ & 2 & -2  & - & - & - & - & 1 & 0 & 0\\
			\hline
			2 & 2 & $(1, 0)\times (0, -1)\times (0, 2)$& \multicolumn{9}{c|}{$x_A = \frac{1}{3}x_B = \frac{1}{2}x_C = \frac{1}{3}x_D$}\\
			3 & 2 & $(0, -1)\times (1, 0)\times (0, 2)$& \multicolumn{9}{c|}{$\beta^g_2=-3$, $\beta^g_3=-3$, $\beta^g_4=-2$}\\
			4 & 2 & $(0, -1)\times (0, 1)\times (2, 0)$& \multicolumn{9}{c|}{$\chi_1=\frac{1}{\sqrt{2}}$, $\chi_2=\frac{\sqrt{2}}{3}$, $\chi_3=\sqrt{2}$}\\
			\hline
		\end{tabular}
	\end{center}
 	\caption{D6-brane configurations and intersection numbers of Model 20, and its MSSM gauge coupling relation is $g^2_a=2g^2_b=g^2_c=(\frac{5}{3}g^2_Y)=\frac{16 \sqrt[4]{2} \pi  e^{\phi ^4}}{3 \sqrt{3}}$.}
	\label{tb:model6}
\end{table}

\begin{table}[!h]\scriptsize
	\begin{center}
		\begin{tabular}{|c||c|c||c|c|c|c|c|c|c|c|c|}
			\hline\rm{Model 21} & \multicolumn{11}{c|}{$U(4)\times U(2)_L\times U(2)_R\times USp(2)^3 $}\\
			\hline \hline			\rm{stack} & $N$ & $(n^1,l^1)\times(n^2,l^2)\times(n^3,l^3)$ & $n_{\Ysymm}$& $n_{\Yasymm}$ & $b$ & $b'$ & $c$ & $c'$ & 2 & 3 & 4\\
			\hline
			$a$ & 8 & $(1,1)\times (1,0)\times (1,-1)$ & 0 & 0  & 2 & 1 & 0 & -3 & 1 & 0 & -1\\
			$b$ & 4 & $(-1,0)\times (-2,1)\times (1,3)$ & -5 & 5  & - & - & 10 & 7 & 0 & -1 & 6\\
			$c$ & 4 & $(0,1)\times (-1,3)\times (-1,1)$ & 2 & -2  & - & - & - & - & 1 & 0 & 0\\
			\hline
			2 & 2 & $(1, 0)\times (0, -1)\times (0, 2)$& \multicolumn{9}{c|}{$x_A = \frac{1}{3}x_B = \frac{1}{18}x_C = \frac{1}{3}x_D$}\\
			3 & 2 & $(0, -1)\times (1, 0)\times (0, 2)$& \multicolumn{9}{c|}{$\beta^g_2=-3$, $\beta^g_3=-5$, $\beta^g_4=2$}\\
			4 & 2 & $(0, -1)\times (0, 1)\times (2, 0)$& \multicolumn{9}{c|}{$\chi_1=\frac{1}{3 \sqrt{2}}$, $\chi_2=\sqrt{2}$, $\chi_3=\frac{\sqrt{2}}{3}$}\\
			\hline
		\end{tabular}
	\end{center}
 	\caption{D6-brane configurations and intersection numbers of Model 21, and its MSSM gauge coupling relation is $g^2_a=\frac{54}{19}g^2_b=g^2_c=(\frac{5}{3}g^2_Y)=\frac{48}{19} \sqrt[4]{2} \pi  e^{\phi ^4}$.}
	\label{tb:model16}
\end{table}

\begin{table}[!h]\scriptsize
	\begin{center}
		\begin{tabular}{|c||c|c||c|c|c|c|c|c|c|c|c|c|}
			\hline\rm{Model 22} & \multicolumn{12}{c|}{$U(4)\times U(2)_L\times U(2)_R\times USp(2)^4 $}\\
			\hline \hline			\rm{stack} & $N$ & $(n^1,l^1)\times(n^2,l^2)\times(n^3,l^3)$ & $n_{\Ysymm}$& $n_{\Yasymm}$ & $b$ & $b'$ & $c$ & $c'$ & 1 & 2 & 3 & 4\\
			\hline
			$a$ & 8 & $(1,1)\times (1,0)\times (1,-1)$ & 0 & 0  & 2 & 1 & 0 & -3 & 0 & 1 & 0 & -1\\
			$b$ & 4 & $(-1,0)\times (-1,1)\times (1,3)$ & -2 & 2  & - & - & 4 & 4 & 0 & 0 & -1 & 3\\
			$c$ & 4 & $(0,1)\times (-1,3)\times (-1,1)$ & 2 & -2  & - & - & - & - & -3 & 1 & 0 & 0\\
			\hline
			1 & 2 & $(1, 0)\times (1, 0)\times (2, 0)$& \multicolumn{10}{c|}{$x_A = \frac{1}{3}x_B = \frac{1}{9}x_C = \frac{1}{3}x_D$}\\
			2 & 2 & $(1, 0)\times (0, -1)\times (0, 2)$& \multicolumn{10}{c|}{$\beta^g_1=-3$, $\beta^g_2=-3$, $\beta^g_3=-5$, $\beta^g_4=-1$}\\
			3 & 2 & $(0, -1)\times (1, 0)\times (0, 2)$& \multicolumn{10}{c|}{$\chi_1=\frac{1}{3}$, $\chi_2=1$, $\chi_3=\frac{2}{3}$}\\
			4 & 2 & $(0, -1)\times (0, 1)\times (2, 0)$& \multicolumn{10}{c|}{}\\
			\hline
		\end{tabular}
	\end{center}
 	\caption{D6-brane configurations and intersection numbers of Model 22, and its MSSM gauge coupling relation is $g^2_a=\frac{9}{5}g^2_b=g^2_c=(\frac{5}{3}g^2_Y)=\frac{12}{5} \sqrt{2} \pi  e^{\phi ^4}$.}
	\label{tb:model5}
\end{table}

\begin{table}[!h]\scriptsize
	\begin{center}
		\begin{tabular}{|c||c|c||c|c|c|c|c|c|c|}
			\hline\rm{Model 23} & \multicolumn{9}{c|}{$U(4)\times U(2)_L\times U(2)_R\times USp(4) $}\\
			\hline \hline			\rm{stack} & $N$ & $(n^1,l^1)\times(n^2,l^2)\times(n^3,l^3)$ & $n_{\Ysymm}$& $n_{\Yasymm}$ & $b$ & $b'$ & $c$ & $c'$ & 4\\
			\hline
			$a$ & 8 & $(-2,1)\times (-1,0)\times (1,1)$ & -1 & 1  & 3 & 0 & 0 & -3 & 2\\
			$b$ & 4 & $(-1,2)\times (-1,1)\times (-1,1)$ & 2 & 6  & - & - & 4 & 0 & 1\\
			$c$ & 4 & $(1,0)\times (1,-3)\times (1,1)$ & 2 & -2  & - & - & - & - & 1\\
			\hline
			4 & 4 & $(0, -1)\times (0, 1)\times (2, 0)$& \multicolumn{7}{c|}{$x_A = \frac{4}{3}x_B = 8x_C = \frac{8}{3}x_D$}\\
			& & & \multicolumn{7}{c|}{$\beta^g_4=0$}\\
			& & & \multicolumn{7}{c|}{$\chi_1=4$, $\chi_2=\frac{2}{3}$, $\chi_3=4$}\\
			\hline
		\end{tabular}
	\end{center}
 	\caption{D6-brane configurations and intersection numbers of Model 23, and its MSSM gauge coupling relation is $g^2_a=\frac{13}{6}g^2_b=\frac{1}{2}g^2_c=\frac{5}{8}(\frac{5}{3}g^2_Y)=\frac{16}{5} \sqrt{\frac{2}{3}} \pi  e^{\phi ^4}$.}
	\label{tb:model21}
\end{table}

\begin{table}[!h]\scriptsize
	\begin{center}
		\begin{tabular}{|c||c|c||c|c|c|c|c|c|c|}
			\hline{Model 24} & \multicolumn{9}{c|}{$U(4)\times U(2)_L\times U(2)_R\times USp(4) $}\\
			\hline \hline			\rm{stack} & $N$ & $(n^1,l^1)\times(n^2,l^2)\times(n^3,l^3)$ & $n_{\Ysymm}$& $n_{\Yasymm}$ & $b$ & $b'$ & $c$ & $c'$ & 3\\
			\hline
			$a$ & 8 & $(1,1)\times (-1,0)\times (-1,1)$ & 0 & 0  & 3 & 0 & 0 & -3 & 0\\
			$b$ & 4 & $(1,0)\times (2,-3)\times (1,1)$ & 1 & -1  & - & - & 8 & 0 & -3\\
			$c$ & 4 & $(-1,4)\times (0,1)\times (-1,1)$ & 3 & -3  & - & - & - & - & 1\\
			\hline
			3 & 4 & $(0, -1)\times (1, 0)\times (0, 2)$& \multicolumn{7}{c|}{$x_A = \frac{1}{6}x_B = \frac{1}{4}x_C = \frac{1}{6}x_D$}\\
			& & & \multicolumn{7}{c|}{$\beta^g_3=-2$}\\
			& & & \multicolumn{7}{c|}{$\chi_1=\frac{1}{2}$, $\chi_2=\frac{1}{3}$, $\chi_3=1$}\\
			\hline
		\end{tabular}
	\end{center}
 	\caption{D6-brane configurations and intersection numbers of Model 24, and its MSSM gauge coupling relation is $g^2_a=2g^2_b=\frac{2}{3}g^2_c=\frac{10}{13}(\frac{5}{3}g^2_Y)=\frac{16}{5} \sqrt{\frac{2}{3}} \pi  e^{\phi ^4}$.}
	\label{tb:model13}
\end{table}

\begin{table}[!h]\scriptsize
	\begin{center}
		\begin{tabular}{|c||c|c||c|c|c|c|c|c|c|c|c|}
			\hline\rm{Model 25} & \multicolumn{11}{c|}{$U(4)\times U(2)_L\times U(2)_R\times USp(2)^3 $}\\
			\hline \hline			\rm{stack} & $N$ & $(n^1,l^1)\times(n^2,l^2)\times(n^3,l^3)$ & $n_{\Ysymm}$& $n_{\Yasymm}$ & $b$ & $b'$ & $c$ & $c'$ & 2 & 3 & 4\\
			\hline
			$a$ & 8 & $(1,1)\times (1,0)\times (1,-1)$ & 0 & 0  & 3 & 0 & 0 & -3 & 1 & 0 & -1\\
			$b$ & 4 & $(2,-1)\times (1,-1)\times (1,1)$ & -2 & -6  & - & - & 7 & 0 & -1 & -2 & 2\\
			$c$ & 4 & $(-1,4)\times (0,1)\times (-1,1)$ & 3 & -3  & - & - & - & - & 0 & 1 & 0\\
			\hline
			2 & 2 & $(1, 0)\times (0, -1)\times (0, 2)$& \multicolumn{9}{c|}{$x_A = \frac{1}{9}x_B = \frac{1}{4}x_C = \frac{1}{9}x_D$}\\
			3 & 2 & $(0, -1)\times (1, 0)\times (0, 2)$& \multicolumn{9}{c|}{$\beta^g_2=-3$, $\beta^g_3=-3$, $\beta^g_4=-2$}\\
			4 & 2 & $(0, -1)\times (0, 1)\times (2, 0)$& \multicolumn{9}{c|}{$\chi_1=\frac{1}{2}$, $\chi_2=\frac{2}{9}$, $\chi_3=1$}\\
			\hline
		\end{tabular}
	\end{center}
 	\caption{D6-brane configurations and intersection numbers of Model 25, and its MSSM gauge coupling relation is $g^2_a=\frac{17}{9}g^2_b=\frac{4}{9}g^2_c=\frac{4}{7}(\frac{5}{3}g^2_Y)=\frac{32 \pi  e^{\phi ^4}}{15}$.}
	\label{tb:model19}
\end{table}

\begin{table}[!h]\scriptsize
	\begin{center}
		\begin{tabular}{|c||c|c||c|c|c|c|c|c|c|}
			\hline\rm{Model 26} & \multicolumn{9}{c|}{$U(4)\times U(2)_L\times U(2)_R\times USp(2) $}\\
			\hline \hline			\rm{stack} & $N$ & $(n^1,l^1)\times(n^2,l^2)\times(n^3,l^3)$ & $n_{\Ysymm}$& $n_{\Yasymm}$ & $b$ & $b'$ & $c$ & $c'$ & 3\\
			\hline
			$a$ & 8 & $(1,1)\times (1,0)\times (1,-1)$ & 0 & 0  & 3 & 0 & 0 & -3 & 0\\
			$b$ & 4 & $(2,-1)\times (1,-1)\times (1,1)$ & -2 & -6  & - & - & 8 & 0 & -2\\
			$c$ & 4 & $(-2,5)\times (0,1)\times (-1,1)$ & 3 & -3  & - & - & - & - & 2\\
			\hline
			3 & 2 & $(0, -1)\times (1, 0)\times (0, 2)$& \multicolumn{7}{c|}{$x_A = \frac{1}{6}x_B = \frac{2}{5}x_C = \frac{1}{6}x_D$}\\
			& & & \multicolumn{7}{c|}{$\beta^g_3=-2$}\\
			& & & \multicolumn{7}{c|}{$\chi_1=\sqrt{\frac{2}{5}}$, $\chi_2=\frac{\sqrt{\frac{5}{2}}}{6}$, $\chi_3=2 \sqrt{\frac{2}{5}}$}\\
			\hline
		\end{tabular}
	\end{center}
 	\caption{D6-brane configurations and intersection numbers of Model 26, and its MSSM gauge coupling relation is $g^2_a=\frac{11}{6}g^2_b=\frac{5}{6}g^2_c=\frac{25}{28}(\frac{5}{3}g^2_Y)=\frac{8 \sqrt[4]{2} 5^{3/4} \pi  e^{\phi ^4}}{7 \sqrt{3}}$.}
	\label{tb:model8}
\end{table}

\begin{table}[!h]\scriptsize
	\begin{center}
		\begin{tabular}{|c||c|c||c|c|c|c|c|c|c|c|c|}
			\hline\rm{Model 27} & \multicolumn{11}{c|}{$U(4)\times U(2)_L\times U(2)_R\times USp(2)^2\times USp(4) $}\\
			\hline \hline			\rm{stack} & $N$ & $(n^1,l^1)\times(n^2,l^2)\times(n^3,l^3)$ & $n_{\Ysymm}$& $n_{\Yasymm}$ & $b$ & $b'$ & $c$ & $c'$ & 1 & 2 & 4\\
			\hline
			$a$ & 8 & $(1,-1)\times (-1,1)\times (1,-1)$ & 0 & 4  & 0 & 3 & 0 & -3 & -1 & 1 & 1\\
			$b$ & 4 & $(0,1)\times (-2,1)\times (-1,1)$ & -1 & 1  & - & - & 0 & -2 & -1 & 2 & 0\\
			$c$ & 4 & $(-1,0)\times (4,1)\times (-1,1)$ & 3 & -3  & - & - & - & - & 0 & 0 & -4\\
			\hline
			1 & 4 & $(1, 0)\times (1, 0)\times (2, 0)$& \multicolumn{9}{c|}{$x_A = 2x_B = \frac{5}{2}x_C = 10x_D$}\\
			2 & 2 & $(1, 0)\times (0, -1)\times (0, 2)$& \multicolumn{9}{c|}{$\beta^g_1=-3$, $\beta^g_2=-2$, $\beta^g_4=0$}\\
			4 & 2 & $(0, -1)\times (0, 1)\times (2, 0)$& \multicolumn{9}{c|}{$\chi_1=\frac{5}{\sqrt{2}}$, $\chi_2=2 \sqrt{2}$, $\chi_3=\sqrt{2}$}\\
			\hline
		\end{tabular}
	\end{center}
 	\caption{D6-brane configurations and intersection numbers of Model 27, and its MSSM gauge coupling relation is $g^2_a=\frac{10}{9}g^2_b=\frac{4}{9}g^2_c=\frac{4}{7}(\frac{5}{3}g^2_Y)=\frac{16}{27} 2^{3/4} \sqrt{5} \pi  e^{\phi ^4}$.}
	\label{tb:model9}
\end{table}

\begin{table}[!h]\scriptsize
	\begin{center}
		\begin{tabular}{|c||c|c||c|c|c|c|c|c|c|c|}
			\hline\rm{Model 28} & \multicolumn{10}{c|}{$U(4)\times U(2)_L\times U(2)_R\times USp(2)\times USp(4) $}\\
			\hline \hline			\rm{stack} & $N$ & $(n^1,l^1)\times(n^2,l^2)\times(n^3,l^3)$ & $n_{\Ysymm}$& $n_{\Yasymm}$ & $b$ & $b'$ & $c$ & $c'$ & 1 & 3\\
			\hline
			$a$ & 8 & $(1,1)\times (-1,0)\times (-1,1)$ & 0 & 0  & 3 & 0 & 0 & -3 & 0 & 0\\
			$b$ & 4 & $(1,0)\times (1,-3)\times (1,1)$ & 2 & -2  & - & - & 4 & 0 & 0 & -3\\
			$c$ & 4 & $(-1,4)\times (0,1)\times (-1,1)$ & 3 & -3  & - & - & - & - & -4 & 1\\
			\hline
			1 & 2 & $(1, 0)\times (1, 0)\times (2, 0)$& \multicolumn{8}{c|}{$x_A = \frac{1}{12}x_B = \frac{1}{4}x_C = \frac{1}{12}x_D$}\\
			3 & 4 & $(0, -1)\times (1, 0)\times (0, 2)$& \multicolumn{8}{c|}{$\beta^g_1=-2$, $\beta^g_3=-2$}\\
			& & & \multicolumn{8}{c|}{$\chi_1=\frac{1}{2}$, $\chi_2=\frac{1}{6}$, $\chi_3=1$}\\
			\hline
		\end{tabular}
	\end{center}
 	\caption{D6-brane configurations and intersection numbers of Model 28, and its MSSM gauge coupling relation is $g^2_a=g^2_b=\frac{1}{3}g^2_c=\frac{5}{11}(\frac{5}{3}g^2_Y)=\frac{16 \pi  e^{\phi ^4}}{5 \sqrt{3}}$.}
	\label{tb:model10}
\end{table}

\begin{table}[!h]\scriptsize
	\begin{center}
		\begin{tabular}{|c||c|c||c|c|c|c|c|c|c|c|}
			\hline\rm{Model 29} & \multicolumn{10}{c|}{$U(4)\times U(2)_L\times U(2)_R\times USp(2)\times USp(4) $}\\
			\hline \hline			\rm{stack} & $N$ & $(n^1,l^1)\times(n^2,l^2)\times(n^3,l^3)$ & $n_{\Ysymm}$& $n_{\Yasymm}$ & $b$ & $b'$ & $c$ & $c'$ & 1 & 4\\
			\hline
			$a$ & 8 & $(1,-1)\times (-2,1)\times (1,-1)$ & 0 & 8  & 0 & 3 & 0 & -3 & -1 & 2\\
			$b$ & 4 & $(-2,1)\times (1,1)\times (-1,1)$ & -2 & -6  & - & - & 0 & -2 & -1 & -2\\
			$c$ & 4 & $(-4,-1)\times (1,0)\times (-1,1)$ & 3 & -3  & - & - & - & - & 0 & -4\\
			\hline
			1 & 4 & $(1, 0)\times (1, 0)\times (2, 0)$& \multicolumn{8}{c|}{$x_A = \frac{13}{2}x_B = \frac{13}{8}x_C = 26x_D$}\\
			4 & 2 & $(0, -1)\times (0, 1)\times (2, 0)$& \multicolumn{8}{c|}{$\beta^g_1=-3$, $\beta^g_4=4$}\\
			& & & \multicolumn{8}{c|}{$\chi_1=\sqrt{\frac{13}{2}}$, $\chi_2=2 \sqrt{26}$, $\chi_3=\frac{\sqrt{\frac{13}{2}}}{2}$}\\
			\hline
		\end{tabular}
	\end{center}
 	\caption{D6-brane configurations and intersection numbers of Model 29, and its MSSM gauge coupling relation is $g^2_a=\frac{7}{6}g^2_b=\frac{1}{6}g^2_c=\frac{1}{4}(\frac{5}{3}g^2_Y)=\frac{16}{135} 26^{3/4} \pi  e^{\phi ^4}$.}
	\label{tb:model18}
\end{table}

\begin{table}[!h]\scriptsize
	\begin{center}
		\begin{tabular}{|c||c|c||c|c|c|c|c|c|c|c|c|}
			\hline\rm{Model 30} & \multicolumn{11}{c|}{$U(4)\times U(2)_L\times U(2)_R\times USp(2)^3 $}\\
			\hline \hline			\rm{stack} & $N$ & $(n^1,l^1)\times(n^2,l^2)\times(n^3,l^3)$ & $n_{\Ysymm}$& $n_{\Yasymm}$ & $b$ & $b'$ & $c$ & $c'$ & 1 & 2 & 4\\
			\hline
			$a$ & 8 & $(1,1)\times (1,0)\times (1,-1)$ & 0 & 0  & 2 & 1 & 0 & -3 & 0 & 1 & -1\\
			$b$ & 4 & $(-1,0)\times (-1,1)\times (1,3)$ & -2 & 2  & - & - & 2 & 5 & 0 & 0 & 3\\
			$c$ & 4 & $(0,1)\times (-2,3)\times (-1,1)$ & 1 & -1  & - & - & - & - & -3 & 2 & 0\\
			\hline
			1 & 2 & $(1, 0)\times (1, 0)\times (2, 0)$& \multicolumn{9}{c|}{$x_A = \frac{2}{3}x_B = \frac{2}{9}x_C = \frac{2}{3}x_D$}\\
			2 & 2 & $(1, 0)\times (0, -1)\times (0, 2)$& \multicolumn{9}{c|}{$\beta^g_1=-3$, $\beta^g_2=-2$, $\beta^g_4=-1$}\\
			4 & 2 & $(0, -1)\times (0, 1)\times (2, 0)$& \multicolumn{9}{c|}{$\chi_1=\frac{\sqrt{2}}{3}$, $\chi_2=\sqrt{2}$, $\chi_3=\frac{2 \sqrt{2}}{3}$}\\
			\hline
		\end{tabular}
	\end{center}
 	\caption{D6-brane configurations and intersection numbers of Model 30, and its MSSM gauge coupling relation is $g^2_a=\frac{27}{11}g^2_b=2g^2_c=\frac{10}{7}(\frac{5}{3}g^2_Y)=\frac{48}{11} \sqrt[4]{2} \pi  e^{\phi ^4}$.}
	\label{tb:model22}
\end{table}

\begin{table}[!h]\scriptsize
	\begin{center}
		\begin{tabular}{|c||c|c||c|c|c|c|c|c|c|c|}
			\hline\rm{Model 31} & \multicolumn{10}{c|}{$U(4)\times U(2)_L\times U(2)_R\times USp(2)\times USp(4) $}\\
			\hline \hline			\rm{stack} & $N$ & $(n^1,l^1)\times(n^2,l^2)\times(n^3,l^3)$ & $n_{\Ysymm}$& $n_{\Yasymm}$ & $b$ & $b'$ & $c$ & $c'$ & 1 & 3\\
			\hline
			$a$ & 8 & $(1,1)\times (-1,0)\times (-1,1)$ & 0 & 0  & 2 & 1 & 0 & -3 & 0 & 0\\
			$b$ & 4 & $(1,0)\times (1,-1)\times (1,3)$ & -2 & 2  & - & - & 8 & 4 & 0 & -1\\
			$c$ & 4 & $(-1,4)\times (0,1)\times (-1,1)$ & 3 & -3  & - & - & - & - & -4 & 1\\
			\hline
			1 & 2 & $(1, 0)\times (1, 0)\times (2, 0)$& \multicolumn{8}{c|}{$x_A = \frac{3}{4}x_B = \frac{1}{4}x_C = \frac{3}{4}x_D$}\\
			3 & 4 & $(0, -1)\times (1, 0)\times (0, 2)$& \multicolumn{8}{c|}{$\beta^g_1=-2$, $\beta^g_3=-4$}\\
			& & & \multicolumn{8}{c|}{$\chi_1=\frac{1}{2}$, $\chi_2=\frac{3}{2}$, $\chi_3=1$}\\
			\hline
		\end{tabular}
	\end{center}
 \caption{D6-brane configurations and intersection numbers of Model 31, and its MSSM gauge coupling relation is $g^2_a=\frac{13}{5}g^2_b=3g^2_c=\frac{5}{3}(\frac{5}{3}g^2_Y)=\frac{16}{5} \sqrt{3} \pi  e^{\phi ^4}$.}
	\label{tb:model25}
\end{table}

\begin{table}[!h]
\scriptsize
	\begin{center}
		\begin{tabular}{|c||c|c||c|c|c|c|c|c|c|}
			\hline\rm{Model 32} & \multicolumn{9}{c|}{$U(4)\times U(2)_L\times U(2)_R\times USp(4) $}\\
			\hline \hline			\rm{stack} & $N$ & $(n^1,l^1)\times(n^2,l^2)\times(n^3,l^3)$ & $n_{\Ysymm}$& $n_{\Yasymm}$ & $b$ & $b'$ & $c$ & $c'$ & 4\\
			\hline
			$a$ & 8 & $(2,1)\times (1,0)\times (1,-1)$ & 1 & -1  & 2 & 1 & 0 & -3 & -2\\
			$b$ & 4 & $(1,0)\times (1,-1)\times (1,3)$ & -2 & 2  & - & - & 0 & 4 & 3\\
			$c$ & 4 & $(-1,2)\times (-1,1)\times (-1,1)$ & 2 & 6  & - & - & - & - & 1\\
			\hline
			4 & 4 & $(0, -1)\times (0, 1)\times (2, 0)$& \multicolumn{7}{c|}{$x_A = 2x_B = \frac{4}{3}x_C = 4x_D$}\\
			& & & \multicolumn{7}{c|}{$\beta^g_4=2$}\\
			& & & \multicolumn{7}{c|}{$\chi_1=2 \sqrt{\frac{2}{3}}$, $\chi_2=\sqrt{6}$, $\chi_3=2 \sqrt{\frac{2}{3}}$}\\
			\hline
		\end{tabular}
	\end{center}
 \caption{D6-brane configurations and intersection numbers of Model 32, and its MSSM gauge coupling relation is $g^2_a=\frac{21}{10}g^2_b=\frac{7}{2}g^2_c=\frac{7}{4}(\frac{5}{3}g^2_Y)=\frac{8}{5} 6^{3/4} \pi  e^{\phi ^4}$.}
	\label{tb:model27}
\end{table}

\begin{table}[!h]\scriptsize
	\begin{center}
		\begin{tabular}{|c||c|c||c|c|c|c|c|c|c|}
			\hline\rm{Model 33} & \multicolumn{9}{c|}{$U(4)\times U(2)_L\times U(2)_R\times USp(4) $}\\
			\hline \hline			\rm{stack} & $N$ & $(n^1,l^1)\times(n^2,l^2)\times(n^3,l^3)$ & $n_{\Ysymm}$& $n_{\Yasymm}$ & $b$ & $b'$ & $c$ & $c'$ & 3\\
			\hline
			$a$ & 8 & $(1,1)\times (-1,0)\times (-1,1)$ & 0 & 0  & 2 & 1 & 0 & -3 & 0\\
			$b$ & 4 & $(1,0)\times (2,-1)\times (1,3)$ & -5 & 5  & - & - & 16 & 8 & -1\\
			$c$ & 4 & $(-1,4)\times (0,1)\times (-1,1)$ & 3 & -3  & - & - & - & - & 1\\
			\hline
			3 & 4 & $(0, -1)\times (1, 0)\times (0, 2)$& \multicolumn{7}{c|}{$x_A = \frac{3}{2}x_B = \frac{1}{4}x_C = \frac{3}{2}x_D$}\\
			& & & \multicolumn{7}{c|}{$\beta^g_3=-4$}\\
			& & & \multicolumn{7}{c|}{$\chi_1=\frac{1}{2}$, $\chi_2=3$, $\chi_3=1$}\\
			\hline
		\end{tabular}
	\end{center}
 \caption{D6-brane configurations and intersection numbers of Model 33, and its MSSM gauge coupling relation is $g^2_a=\frac{26}{5}g^2_b=6g^2_c=2(\frac{5}{3}g^2_Y)=\frac{16}{5} \sqrt{6} \pi  e^{\phi ^4}$.}
	\label{tb:model32}
\end{table}

\FloatBarrier

\section{The Evolution for The Gauge Couplings}
\label{apdx-RGE}

\begin{figure}[H]\centering
    \subfigure[]{\includegraphics[width=0.45\linewidth]{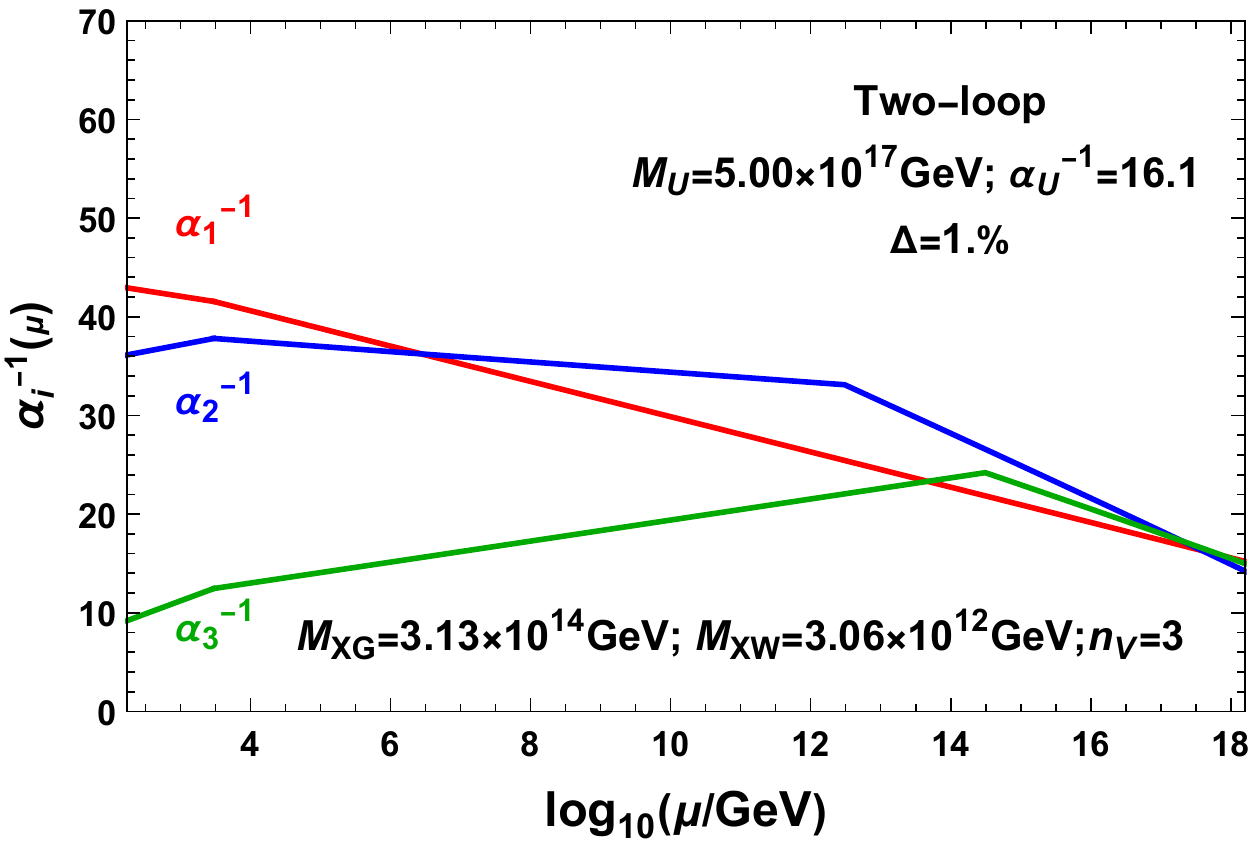}}\qquad
    \subfigure[]{\includegraphics[width=0.45\linewidth]{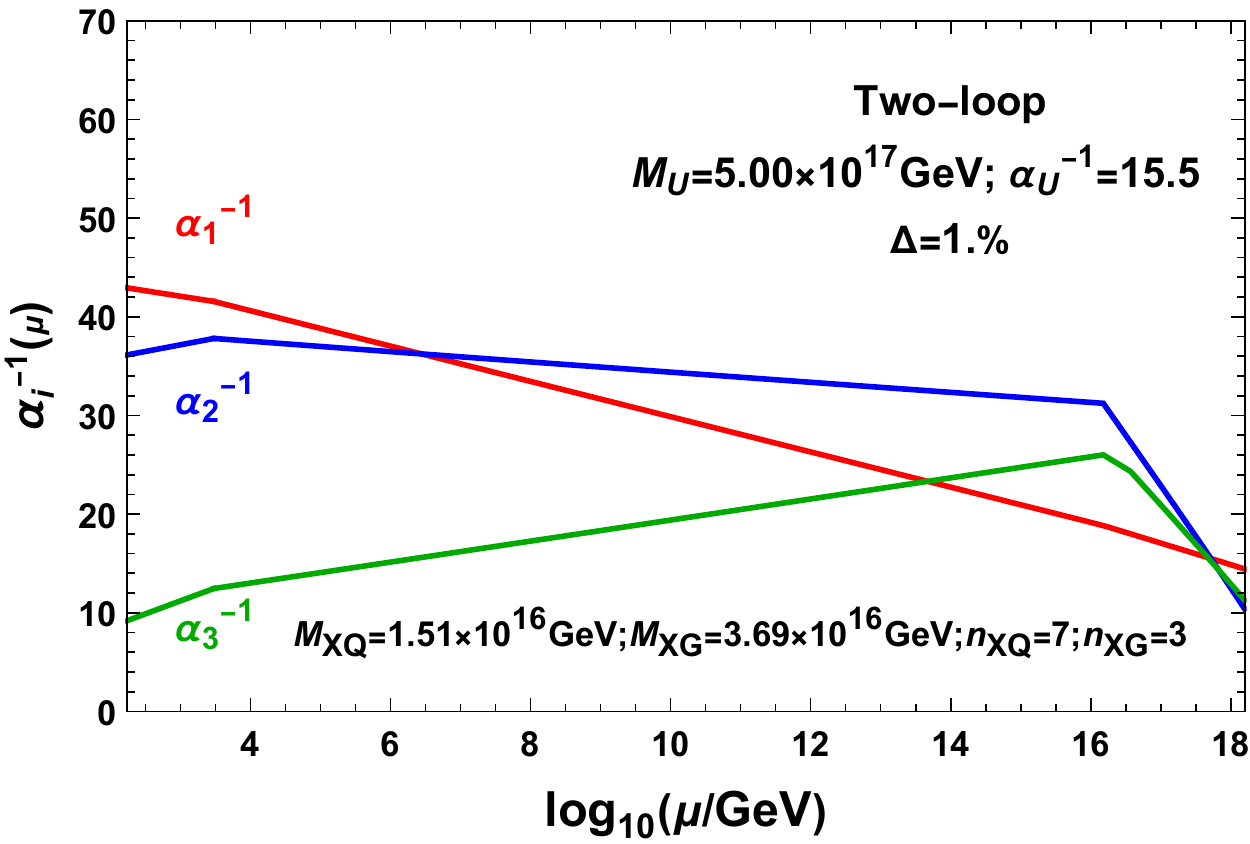}}
    \caption{Two-loop evolution of gauge couplings for the \textbf{Model 6} with vector-like particles. In the model, $k_Y=1\times\frac{11}{8}$ and $k_2=\frac{5}{6}$. The string-scale gauge coupling relations can be achieved by adding $3(XG+XW)$ (a) and $7(XQ+\overline{XQ})+3XG$ (b).} \label{fig:model17}
\end{figure}

\begin{figure}[H]\centering
    \subfigure[]{\includegraphics[width=0.45\linewidth]{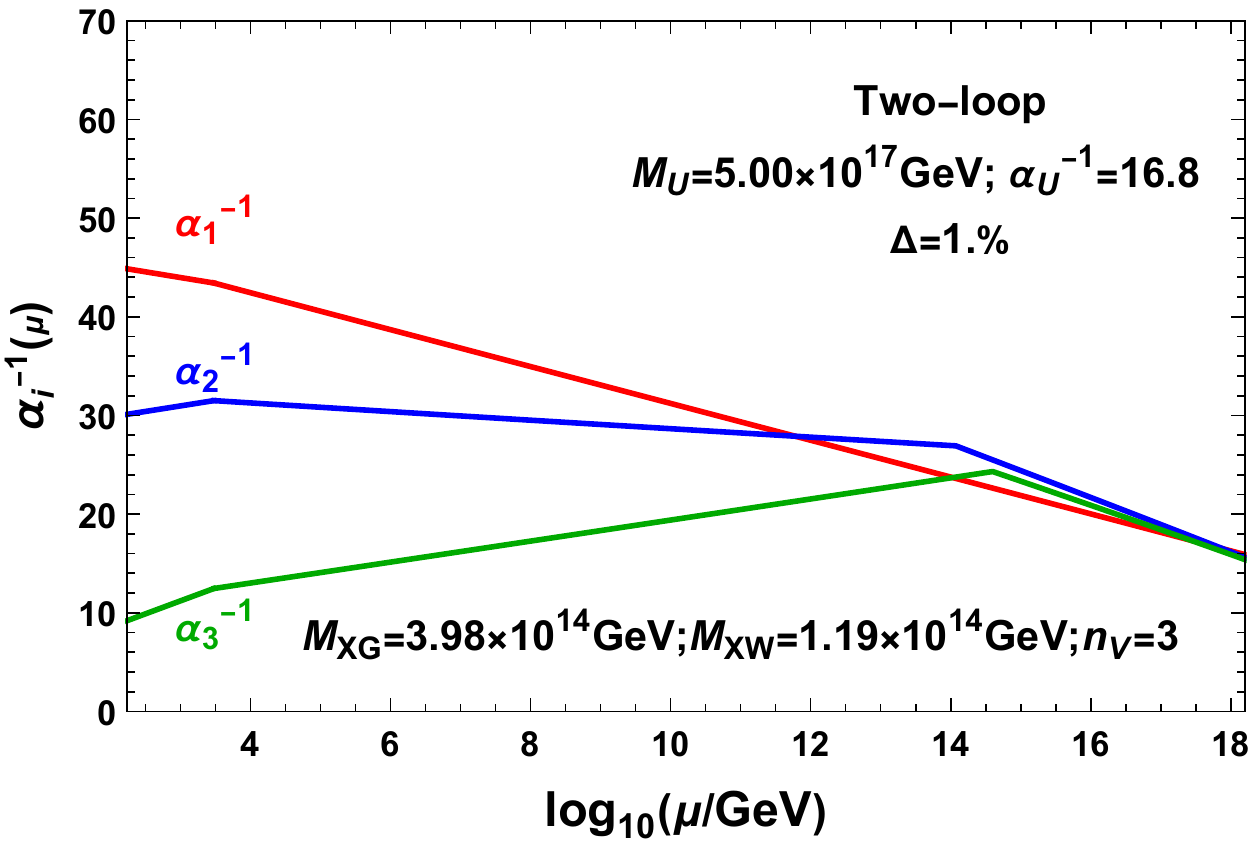}}\qquad
    \subfigure[]{\includegraphics[width=0.45\linewidth]{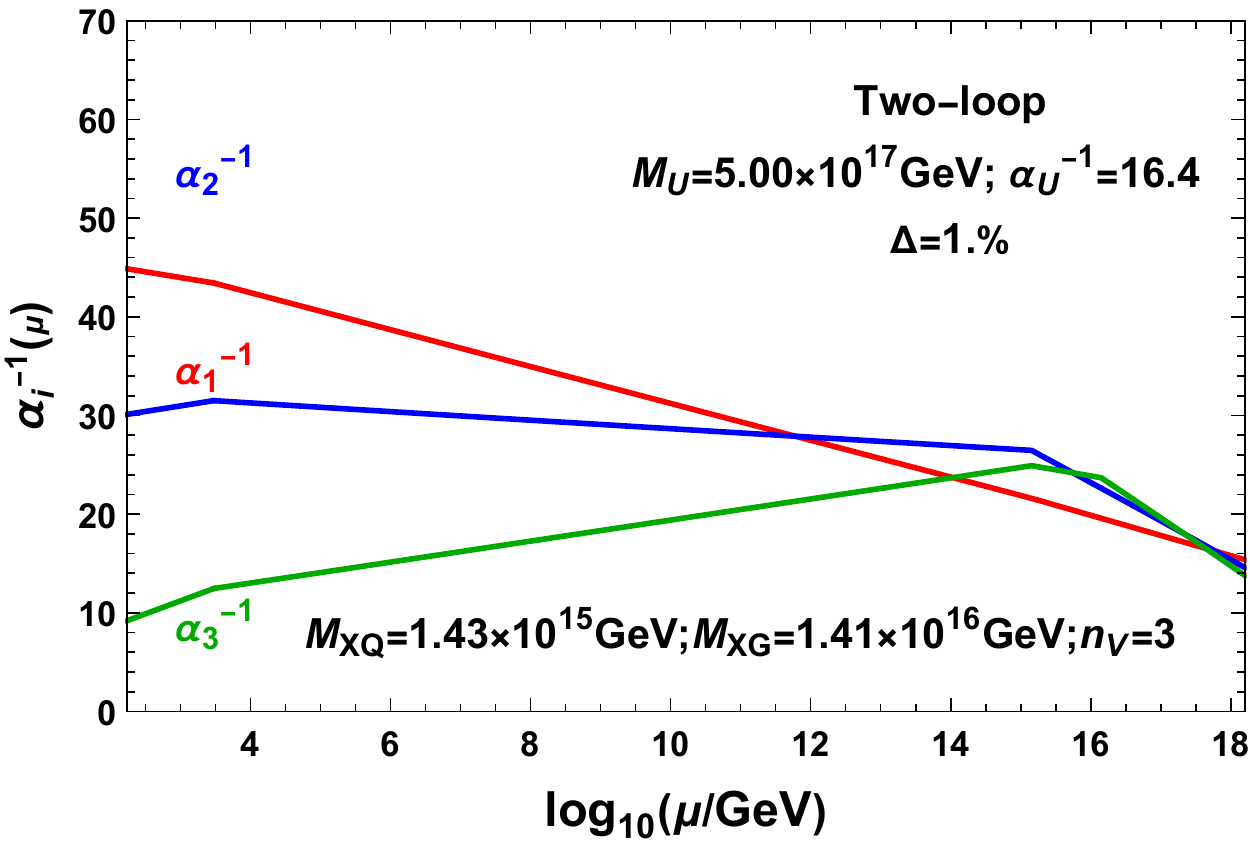}}
    \caption{Two-loop evolution of gauge couplings for the \textbf{Model 7} with vector-like particles. In the model, $k_Y=\frac{25}{19}\times\frac{5}{3}$ and $k_2=1$. The string-scale gauge coupling relations can be achieved by adding $3(XG+XW)$ (a) and $3(XQ+\overline{XQ})+3XG$(b).}\label{fig:model7}
\end{figure}

\begin{figure}[H]\centering
    \subfigure[]{\includegraphics[width=0.45\linewidth]{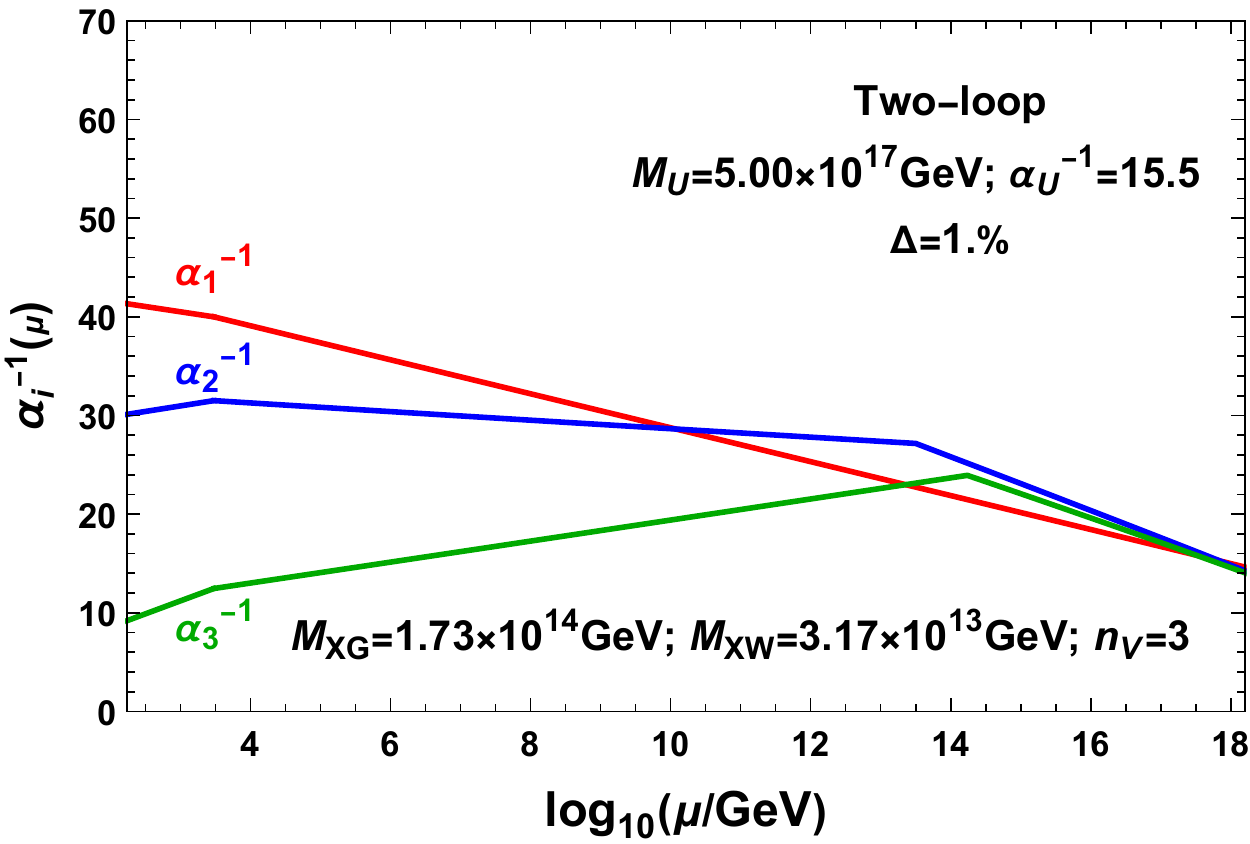}}\qquad
    \subfigure[]{\includegraphics[width=0.45\linewidth]{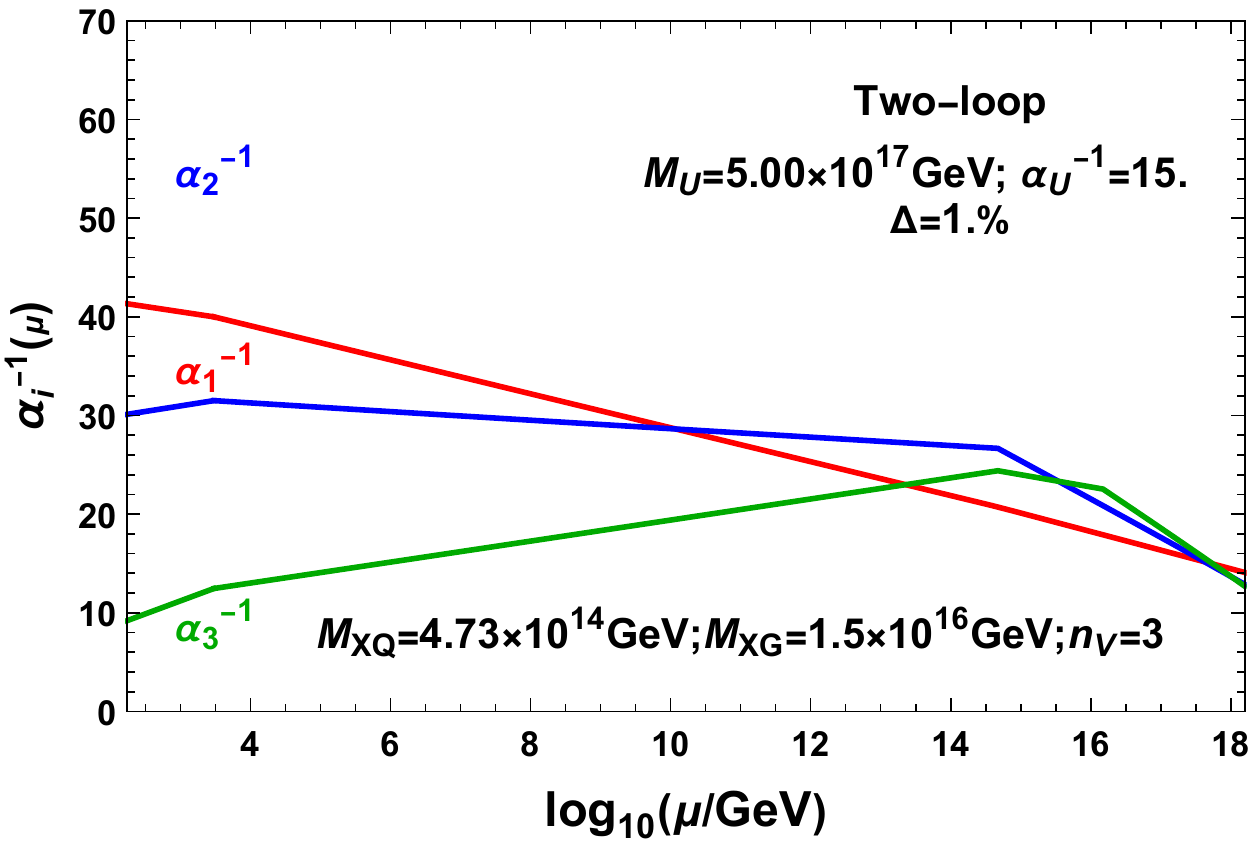}}
    \caption{Two-loop evolution of gauge couplings for the \textbf{Model 8} with vector-like particles. In the model,  $k_Y=\frac{10}{7}\times\frac{5}{3}$ and $k_2=1$. The string-scale gauge coupling relations can be achieved by adding $3(XG+XW)$ (a) and $3(XQ+\overline{XQ})+3XG$(b).} 	\label{fig:model14}
\end{figure}

\begin{figure}[H]\centering
	\subfigure[]{\includegraphics[width=0.45\linewidth]{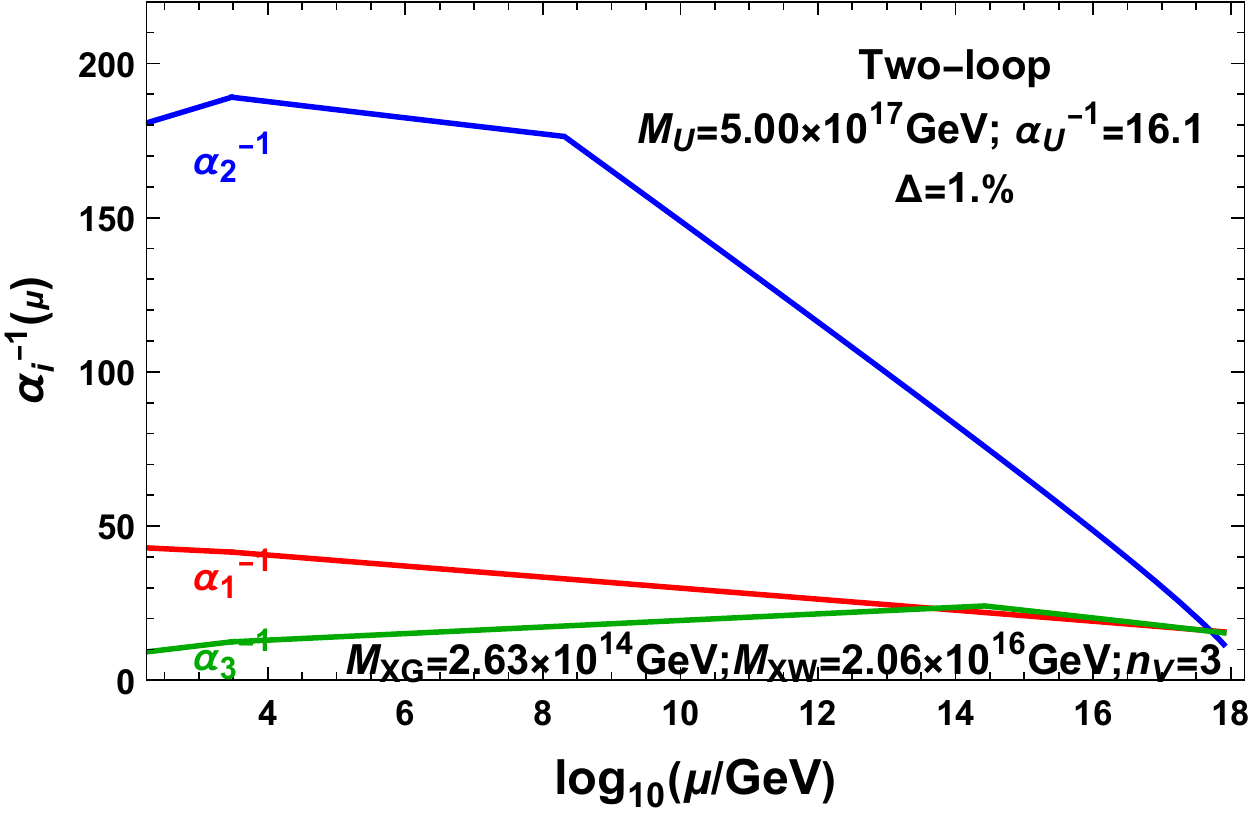}}\qquad
    \subfigure[]{\includegraphics[width=0.45\linewidth]{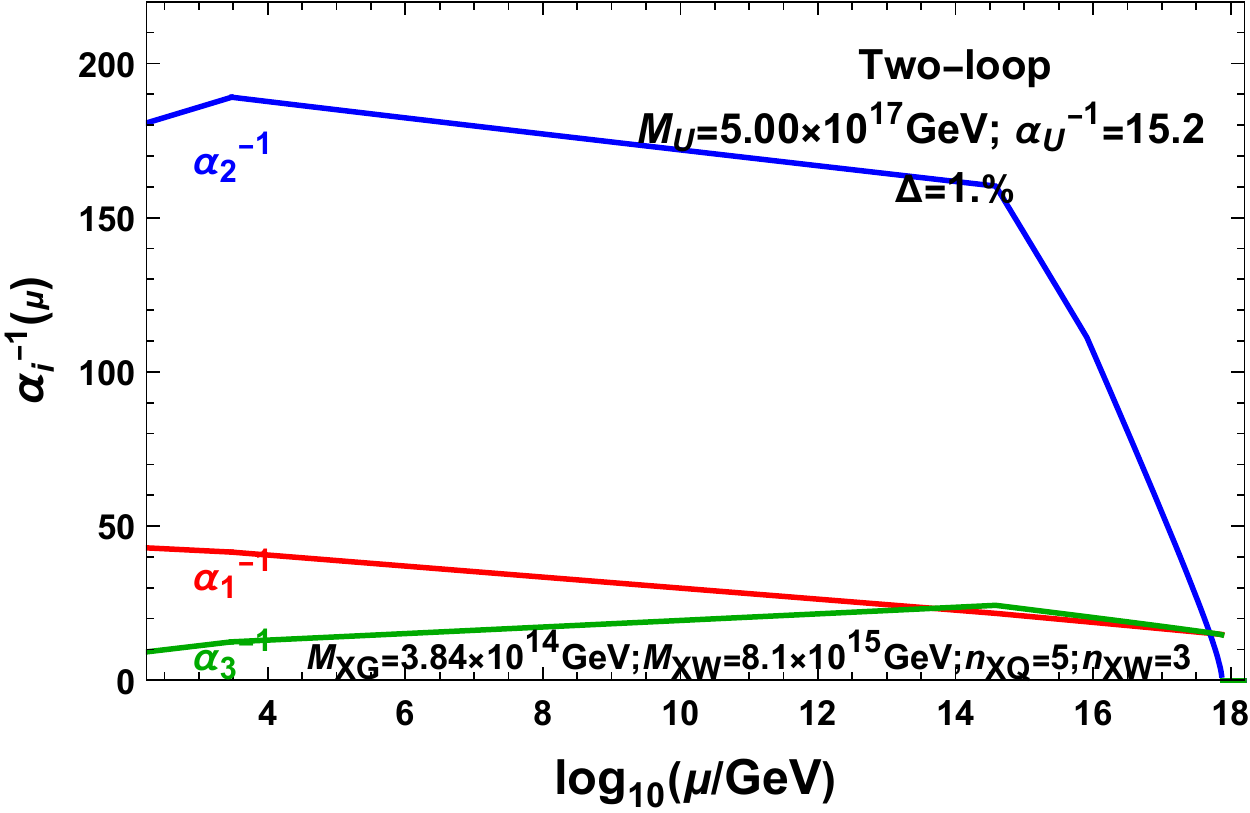}}
	\caption{Two-loop evolution of gauge couplings for the \textbf{Model 9} with vector-like particles. In the model, $k_Y=\frac{11}{8}\times\frac{5}{3}$ and $k_2=\frac{1}{6}$. The string-scale gauge coupling relation  can be achieved by adding $3(XG+XW)$ (a) and $5(XQ+\overline{XQ})+3XW$(b).}
	\label{fig:model33}
\end{figure}

\begin{figure}[H]\centering
	\subfigure[]{\includegraphics[width=0.45\linewidth]{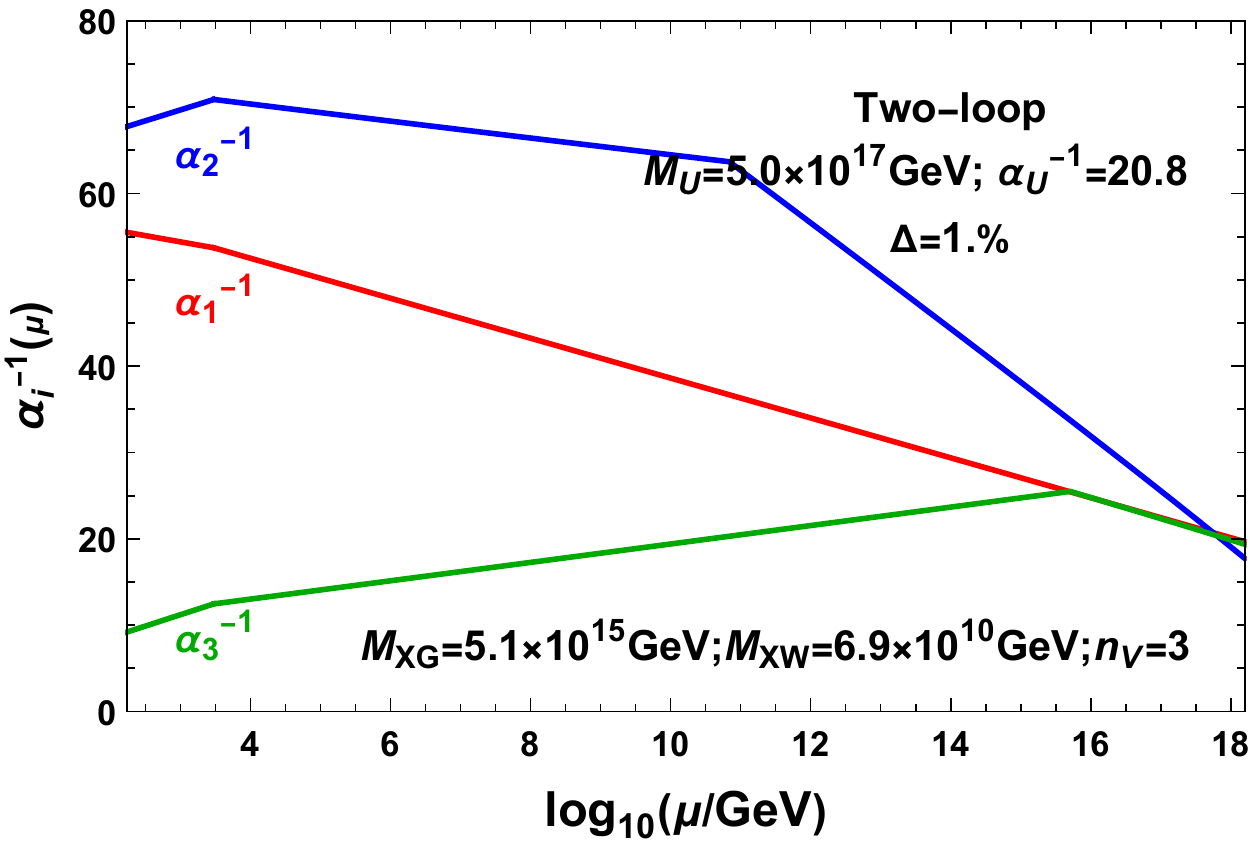}}\qquad
    \subfigure[]{\includegraphics[width=0.45\linewidth]{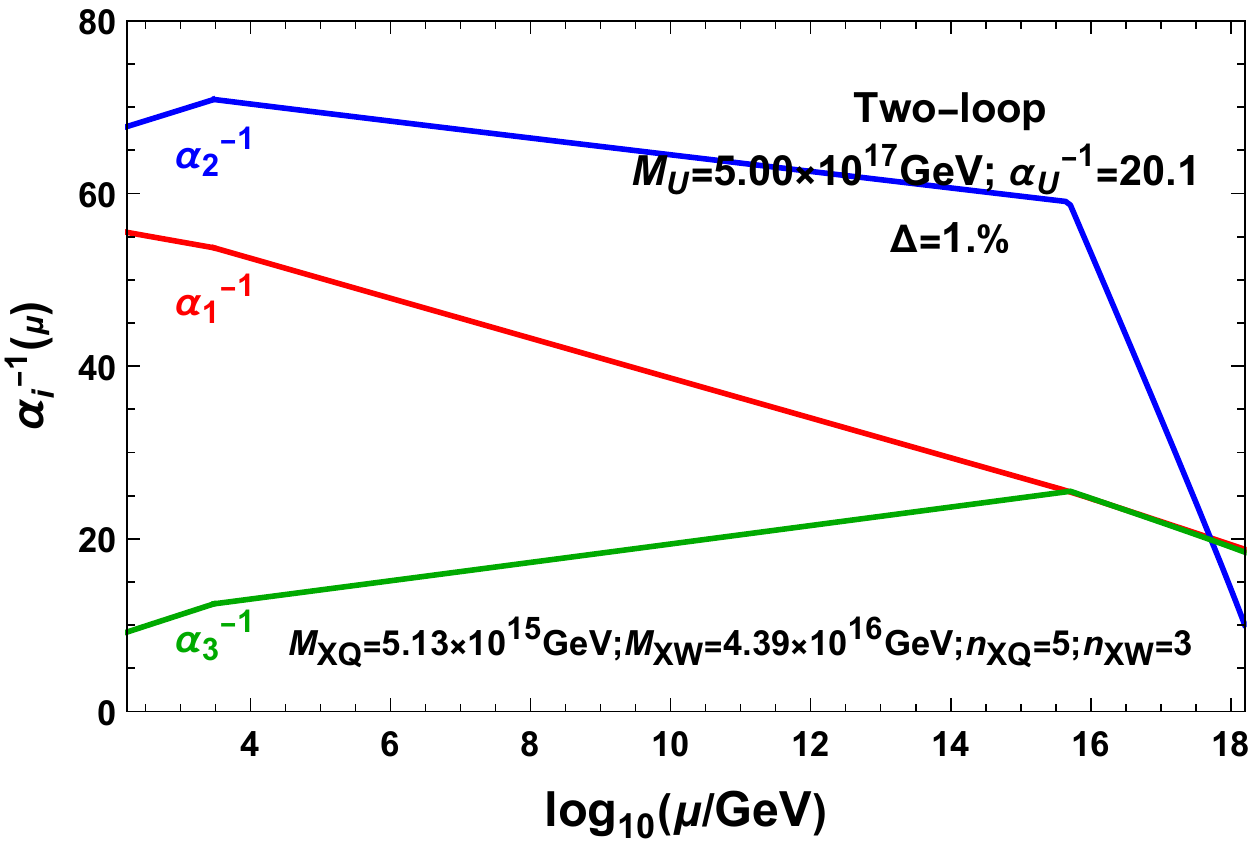}}
	\caption{Two-loop evolution of gauge couplings for the \textbf{Model 10} with vector-like particles. In the model,  $k_Y=\frac{50}{47}\times\frac{5}{3}$ and $k_2=\frac{4}{9}$. The string-scale gauge coupling relation  can be achieved by adding $3(XG+XW)$ (a) and $5(XQ+\overline{XQ})+3XW$ (b).}
	\label{fig:model12}
\end{figure}

\begin{figure}[H]\centering
	\subfigure[]{\includegraphics[width=0.45\linewidth]{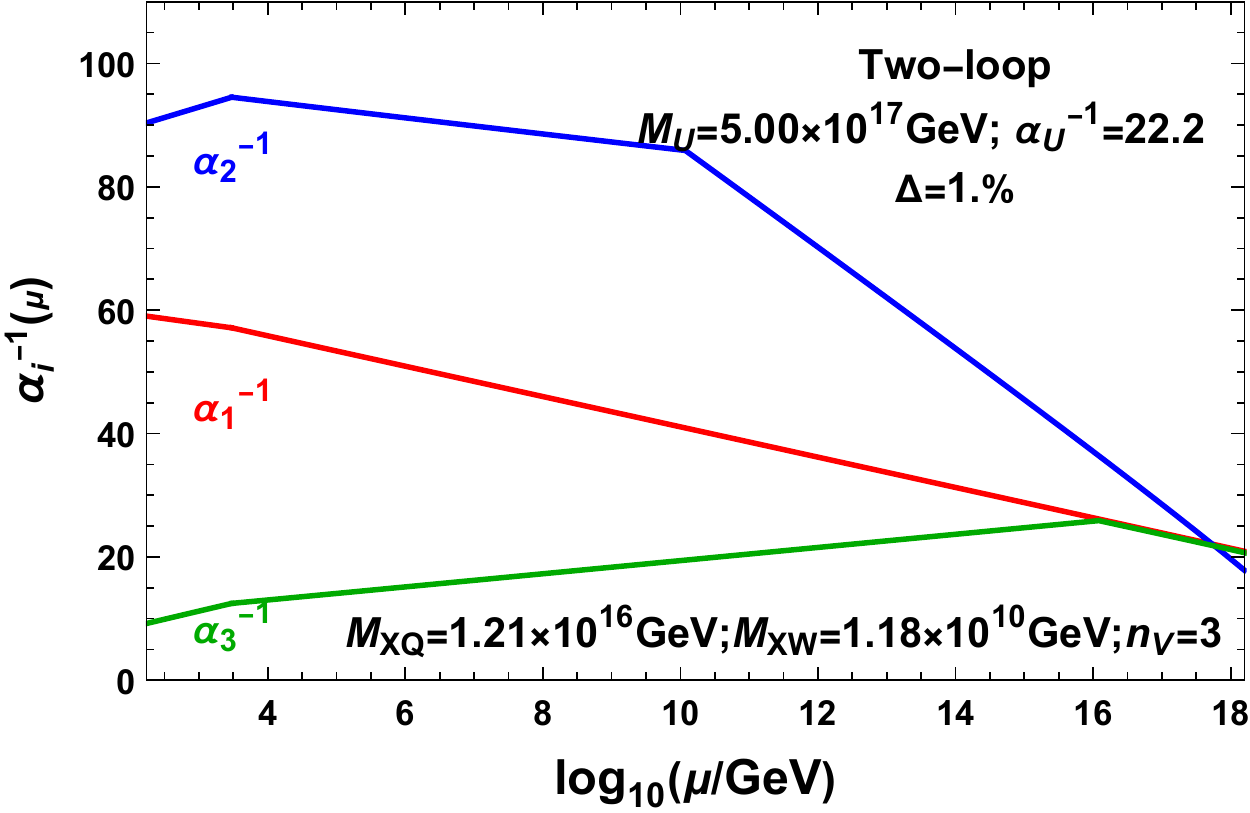}}\qquad
	\subfigure[]{\includegraphics[width=0.45\linewidth]{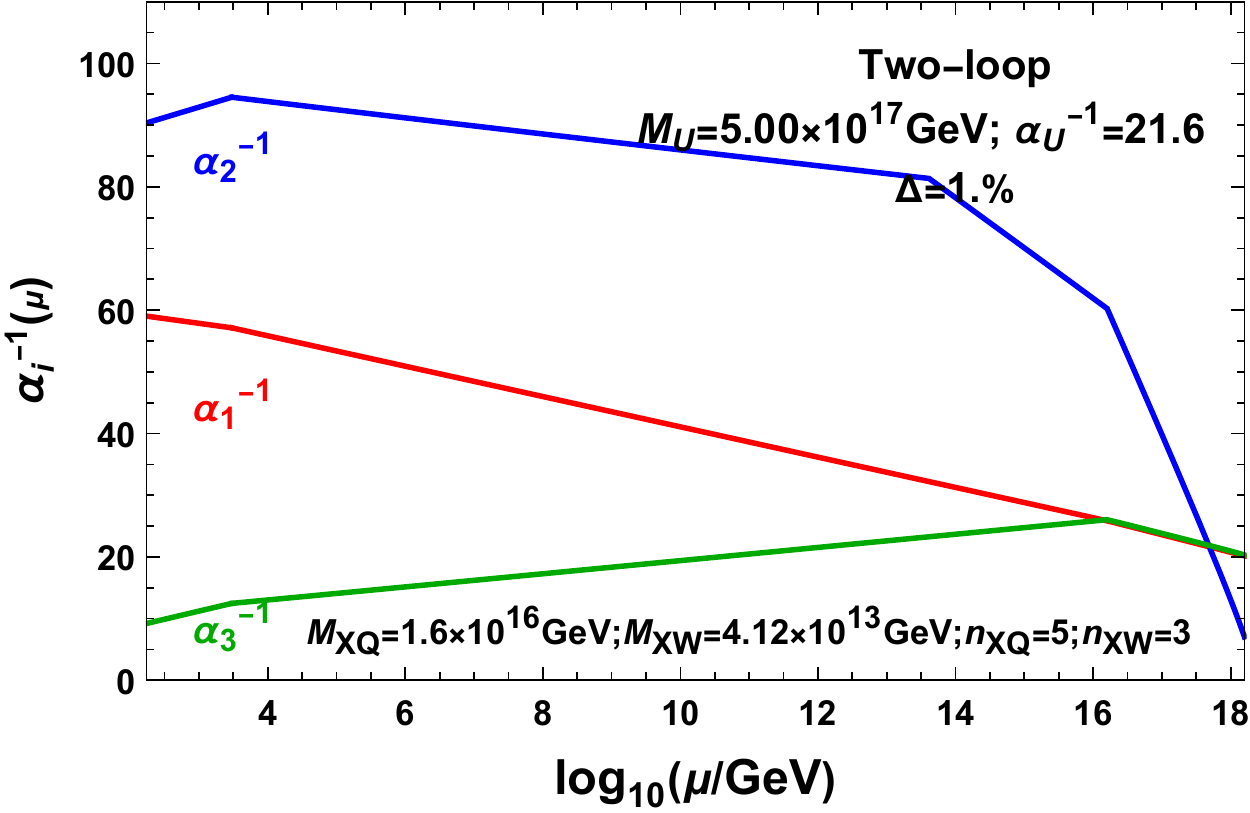}}
	\caption{Two-loop evolution of gauge couplings for the \textbf{Model 11} without and with vector-like particles. In the model, $k_Y=1\times\frac{5}{3}$ and $k_2=\frac{1}{3}$. The string-scale gauge coupling relation  can be achieved by adding $3(XG+XW)$ (a) and $5(XQ+\overline{XQ})+3XW$ (b). }
	\label{fig:model20}
\end{figure}

\begin{figure}[H]\centering
	\subfigure[]{\includegraphics[width=0.45\linewidth]{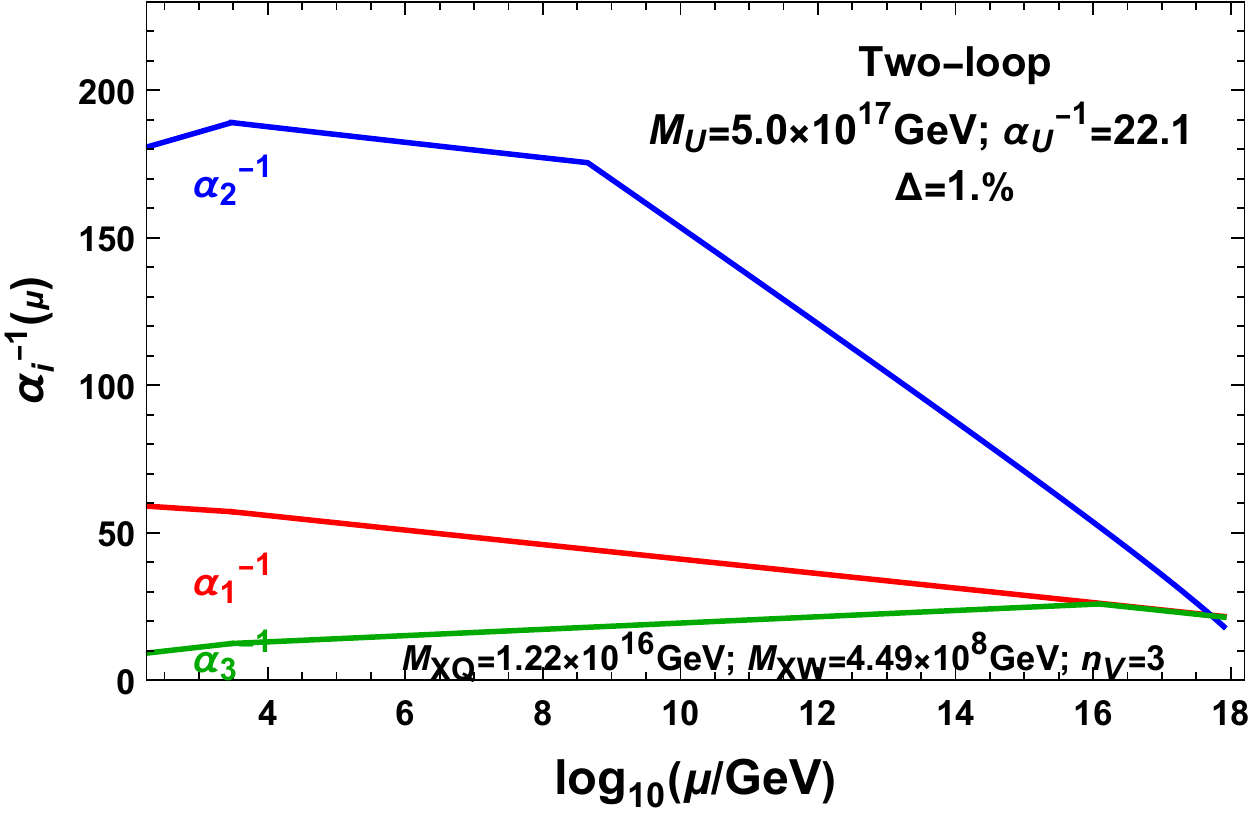}}\qquad
	\subfigure[]{\includegraphics[width=0.45\linewidth]{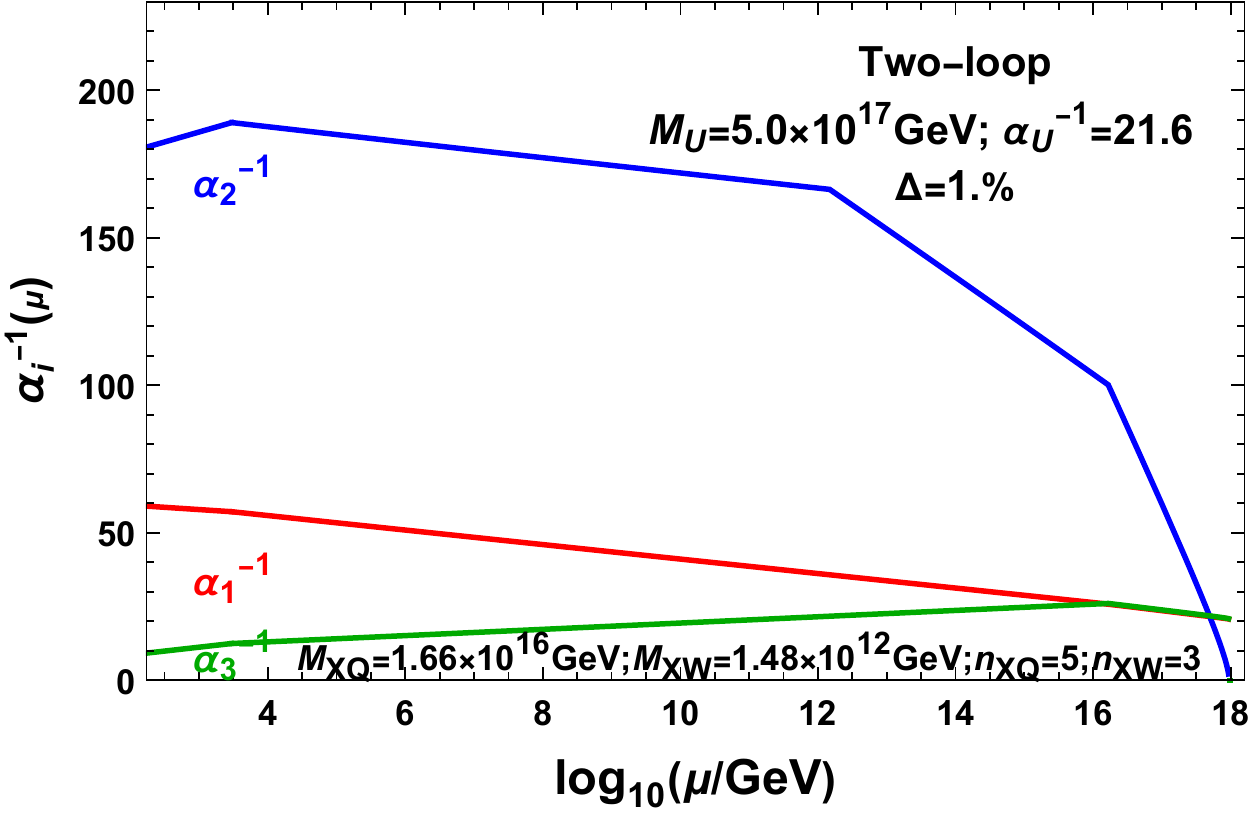}}
	\caption{Two-loop evolution of gauge couplings for the \textbf{Model 12} with vector-like particles. In the model,  $k_Y=\frac{35}{32}\times\frac{5}{3}$ and $k_2=\frac{1}{6}$. The string-scale gauge coupling relation  can be achieved by adding $3(XG+XW)$ (a) and $5(XQ+\overline{XQ})+3XW$ (b).}
	\label{fig:model31}
\end{figure}

\begin{figure}[H]\centering
	\subfigure[]{\includegraphics[width=0.45\linewidth]{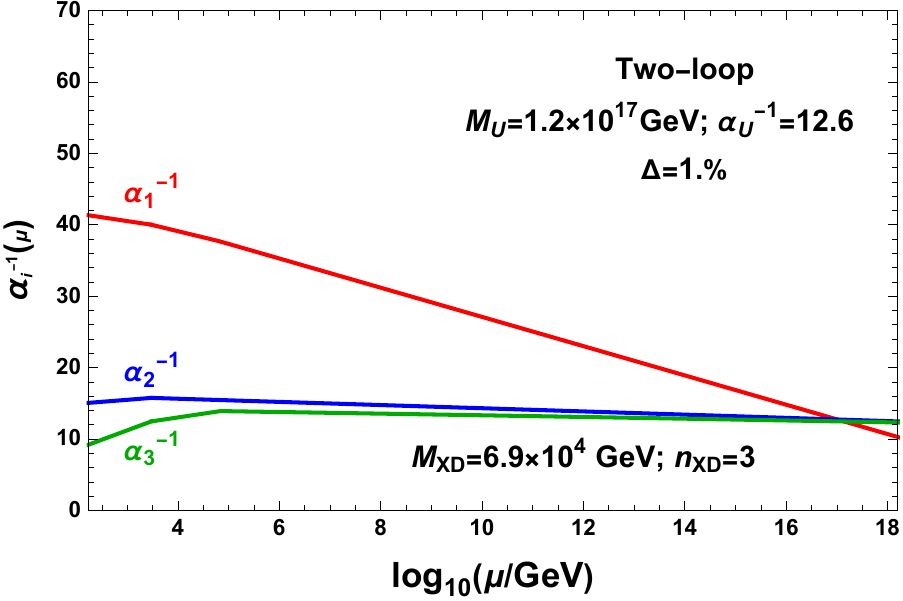}}\qquad
	\subfigure[]{\includegraphics[width=0.45\linewidth]{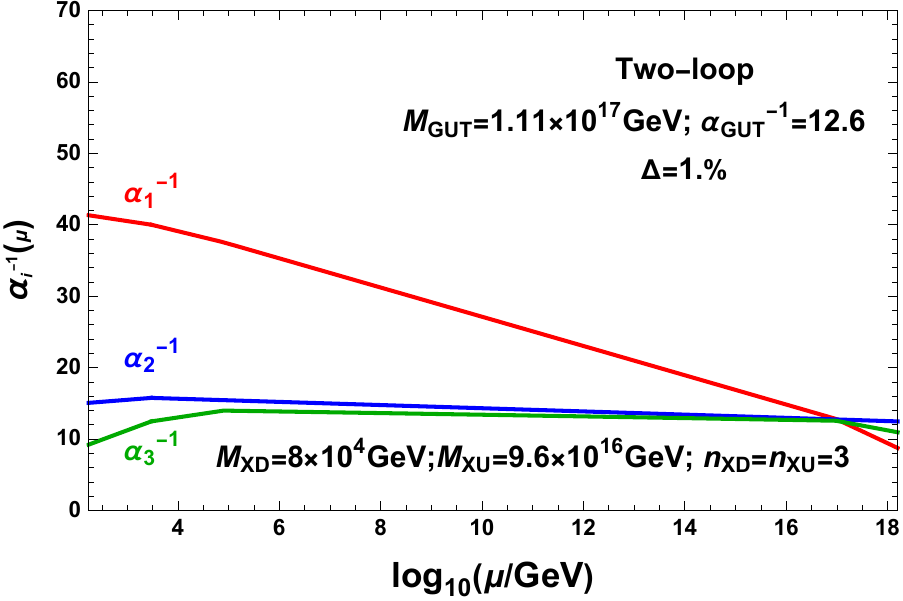}}
	\caption{Two-loop evolution of gauge couplings for the \textbf{Model 16} with vector-like particles. In this model, $k_Y=\frac{10}{7}\times\frac{5}{3}$ and $k_2=2$. The string-scale gauge coupling relation  can be achieved by adding $3(XD+\overline{XD})$ (a) as well as $3(XU+\overline{XU})$ (b).} 
	\label{fig:model15}
\end{figure}

\begin{figure}[H]\centering
    \subfigure[]{\includegraphics[width=0.45\linewidth]{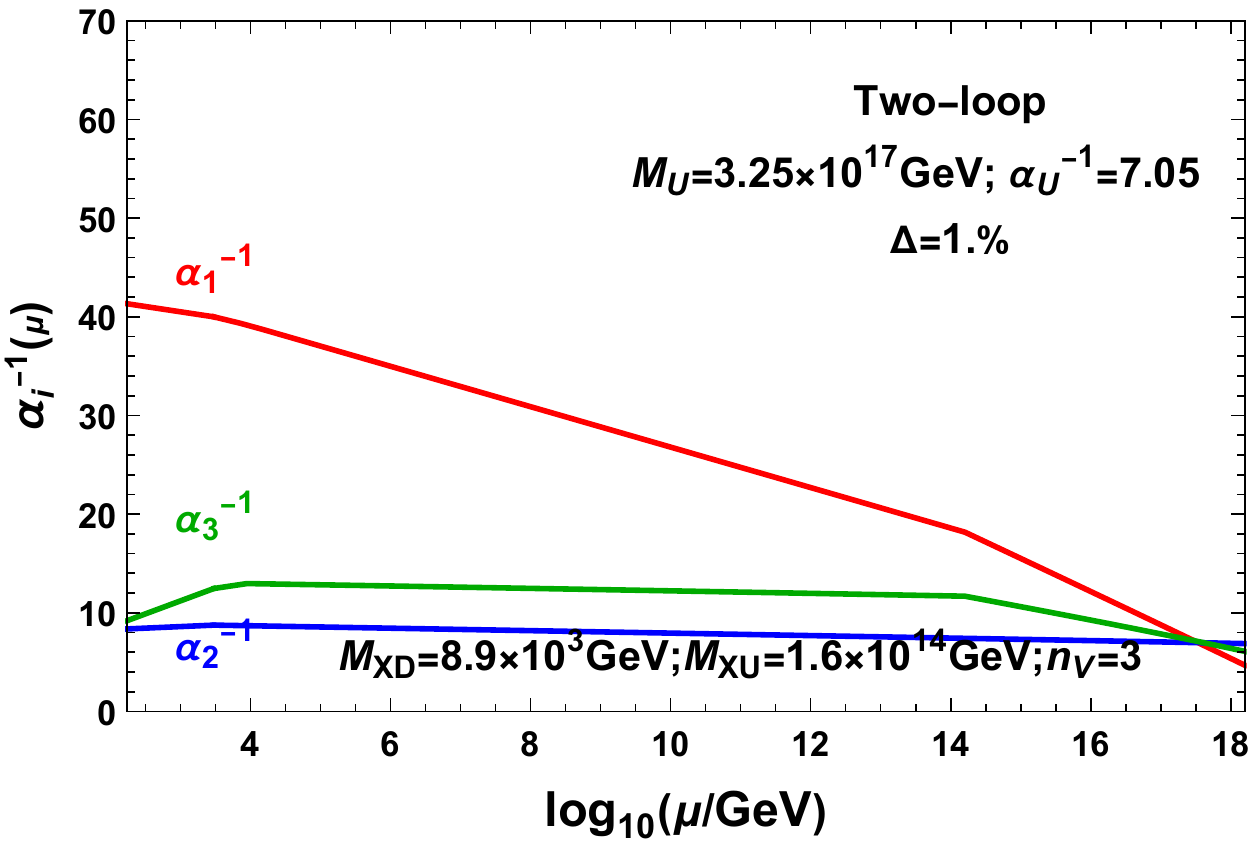}}\qquad
    \subfigure[]{\includegraphics[width=0.45\linewidth]{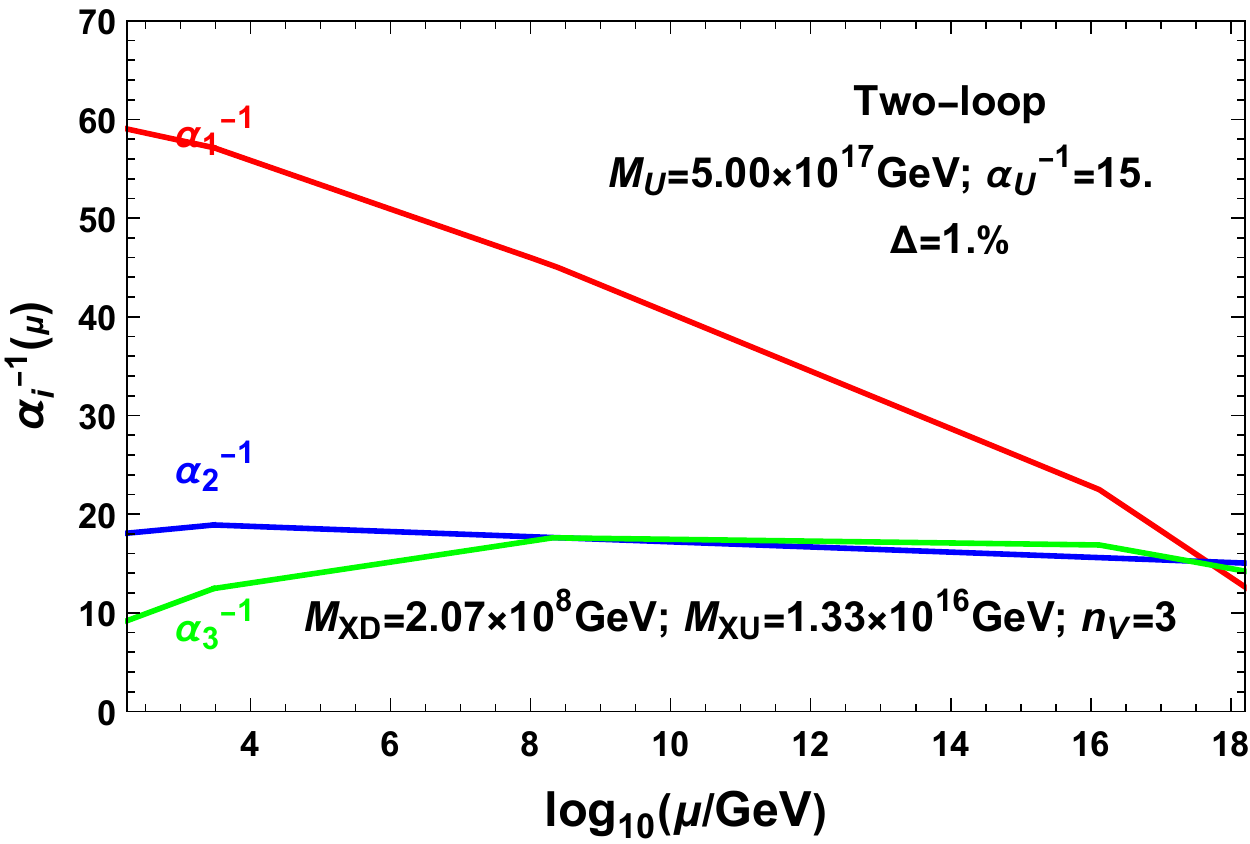}}
    \subfigure[]{\includegraphics[width=0.45\linewidth]{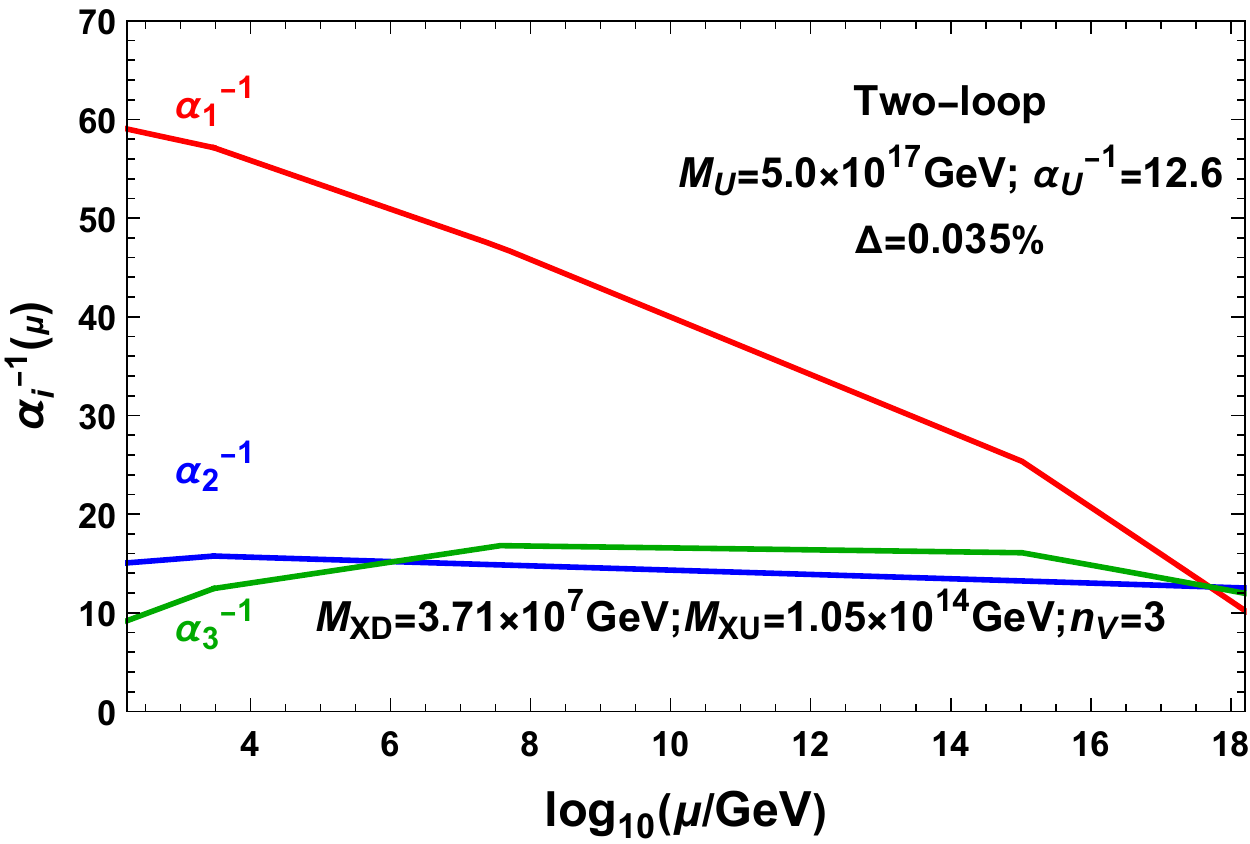}}\qquad
    \subfigure[]{\includegraphics[width=0.45\linewidth]{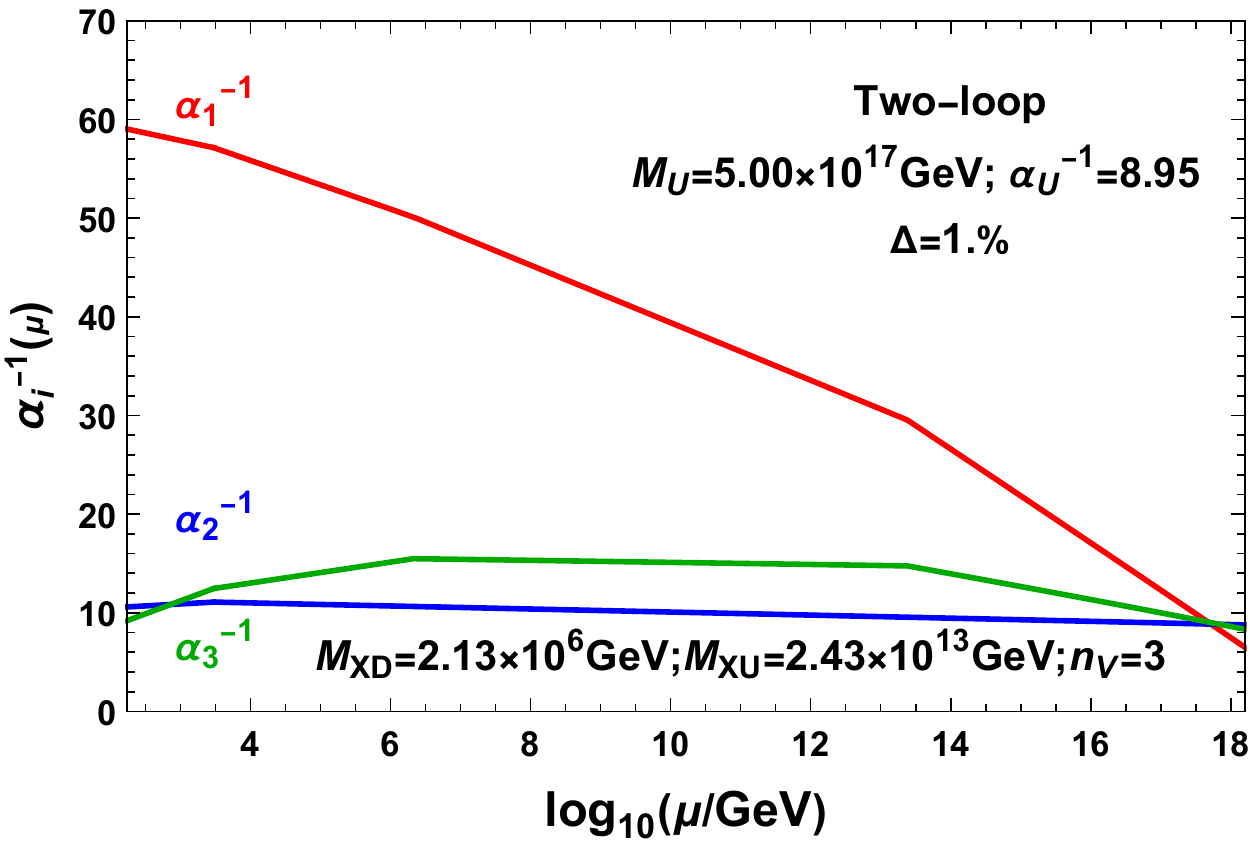}}
    \subfigure[]{\includegraphics[width=0.45\linewidth]{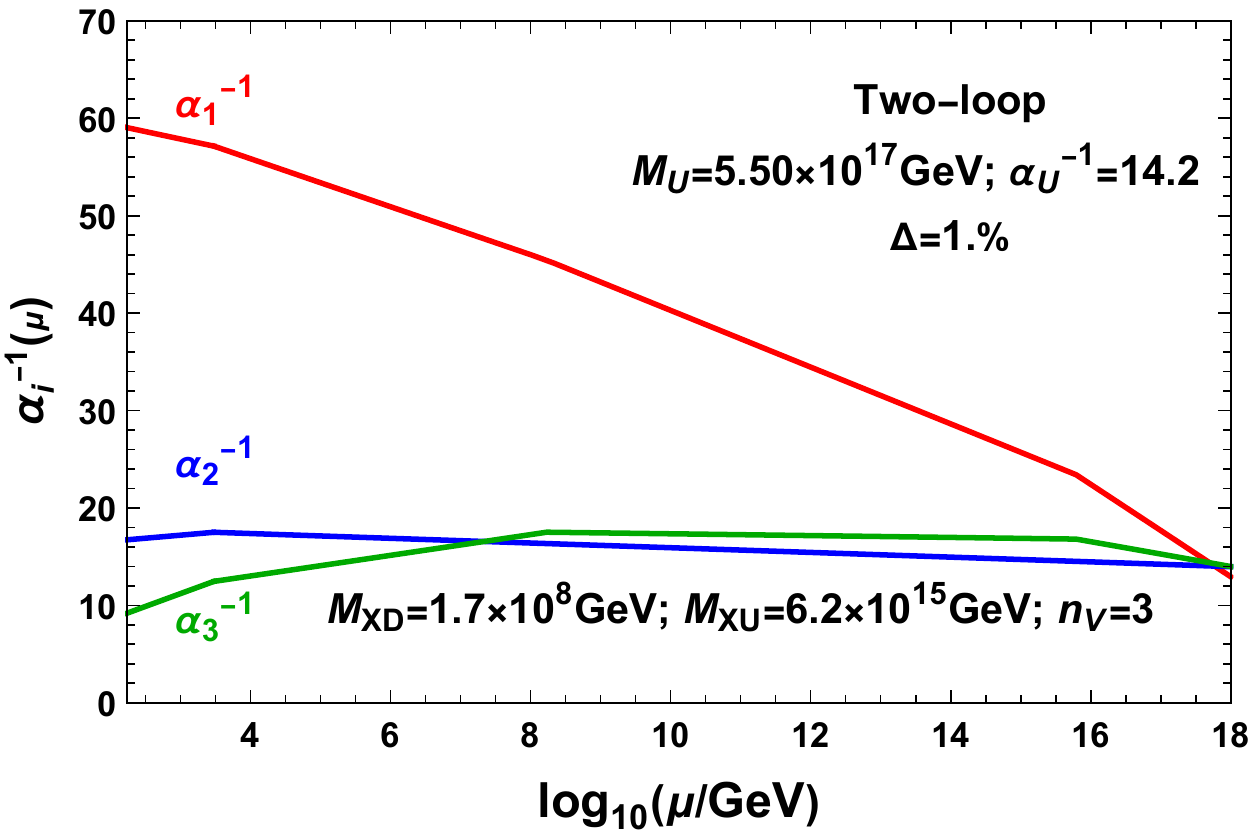}}\qquad
    \subfigure[]{\includegraphics[width=0.45\linewidth]{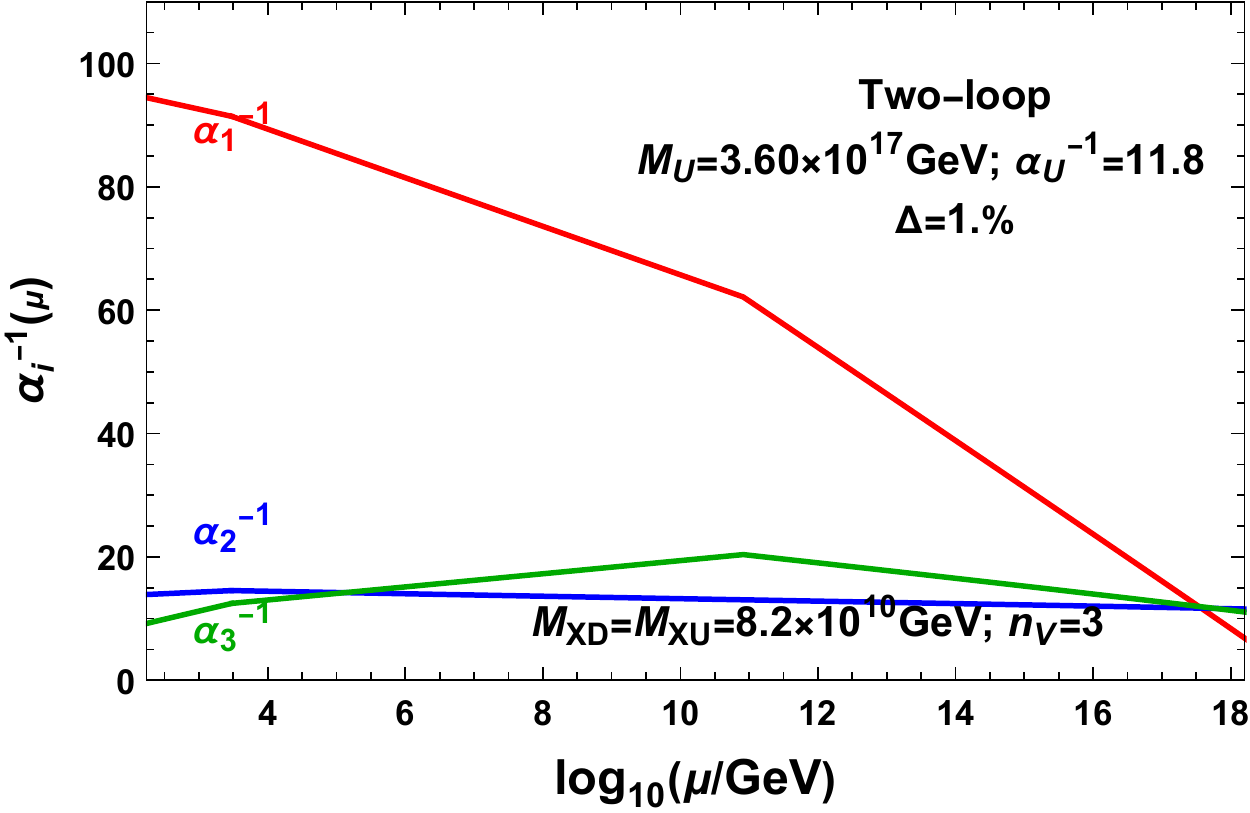}}
    \caption{Two-loop evolution of gauge couplings for the \textbf{Model 18} (a), \textbf{Model 19} (b), \textbf{Model 20} (c), \textbf{Model 21} (d), \textbf{Model 22} (e) and \textbf{Model 23} (f) with vector-like particles. The string-scale gauge coupling relation  can be achieved by adding $3(XD+\overline{XD})+3(XU+\overline{XU})$.}	\label{fig:model4-6-16-5-21}
\end{figure}

\begin{figure}[H]\centering
	\subfigure[]{\includegraphics[width=0.45\linewidth]{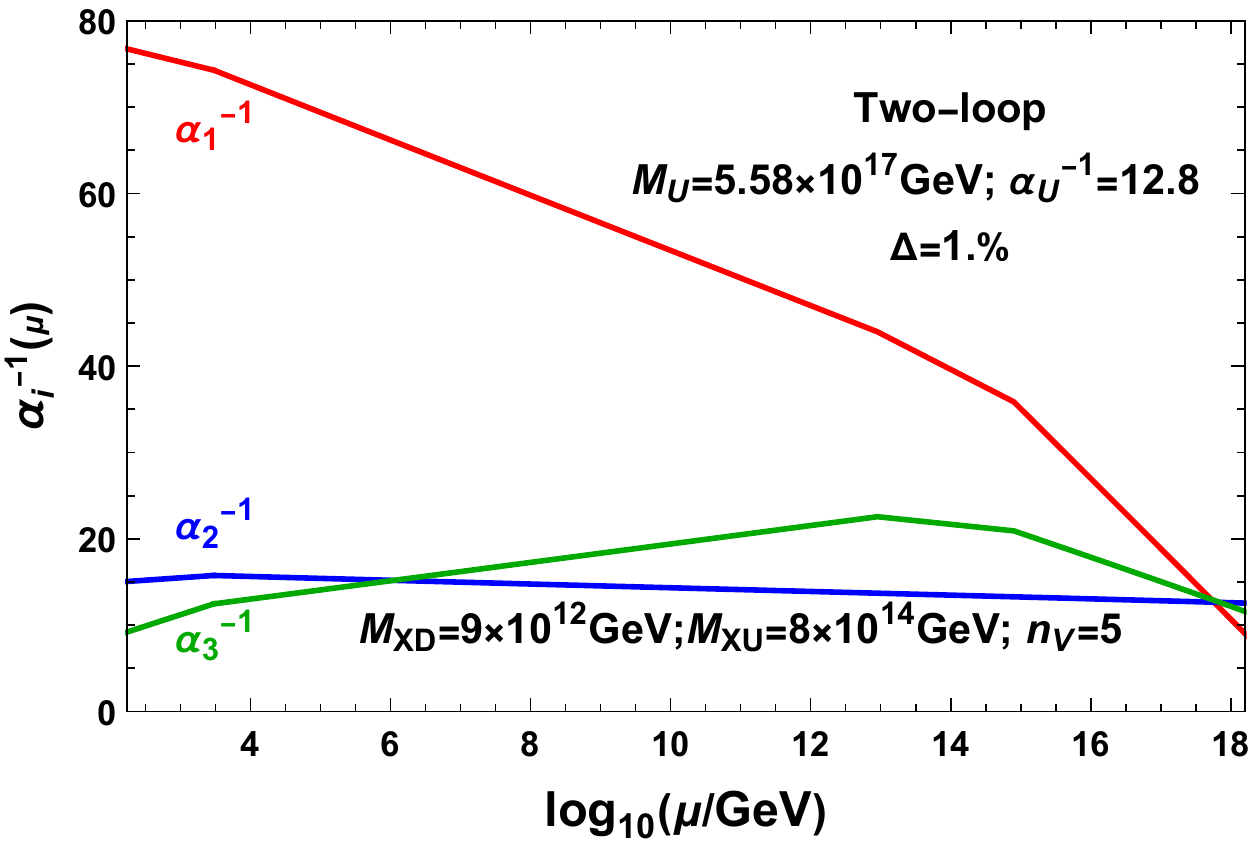}}\qquad   
    \subfigure[]{\includegraphics[width=0.45\linewidth]{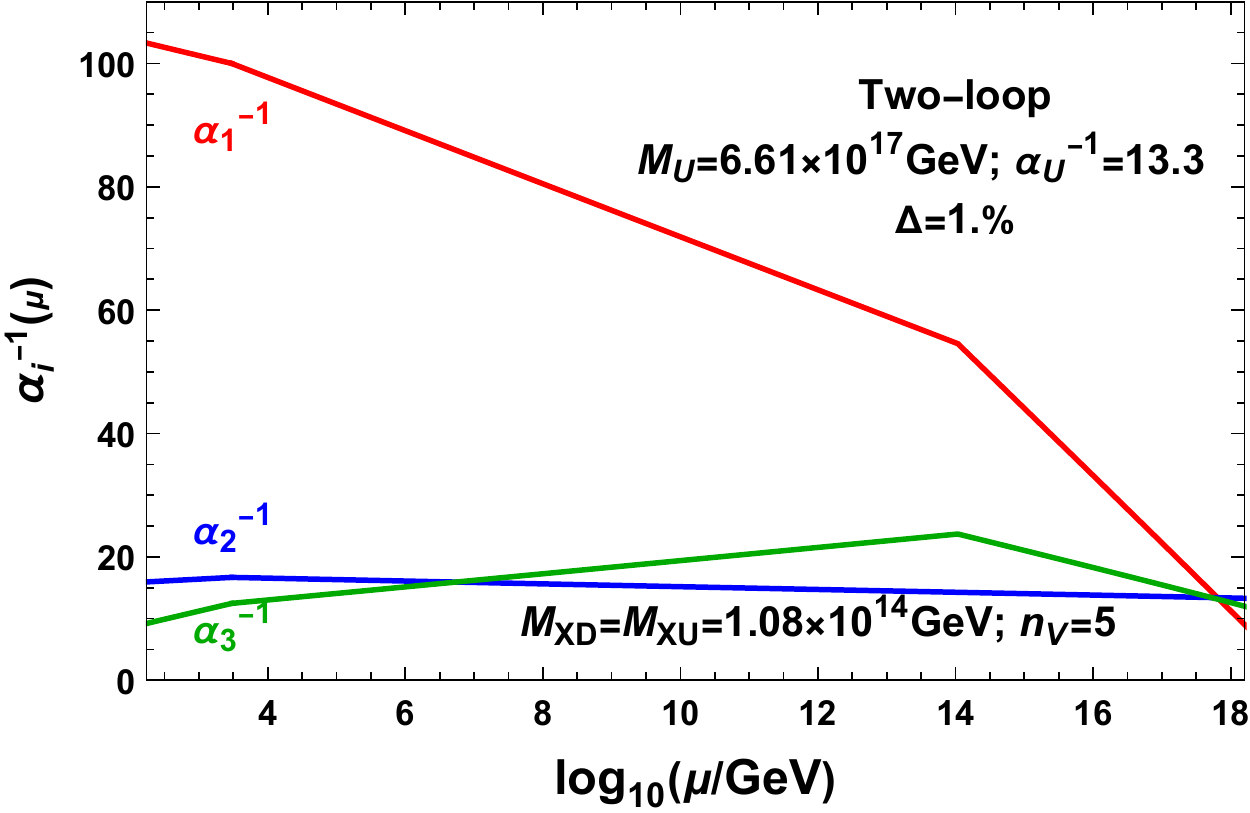}}
	\caption{Two-loop evolution of gauge couplings for the \textbf{Model 24} (a) and \textbf{Model 25} (b) with vector-like particles. In this model,  $k_Y=\frac{10}{13}\times\frac{5}{3}$ and $k_2=2$. The string-scale gauge coupling relation  can be achieved by adding $5(XD+\overline{XD})+5(XU+\overline{XU})$. }
	\label{fig:model13-19}
\end{figure}

\begin{figure}[H]\centering
    \subfigure{\includegraphics[width=0.45\linewidth]{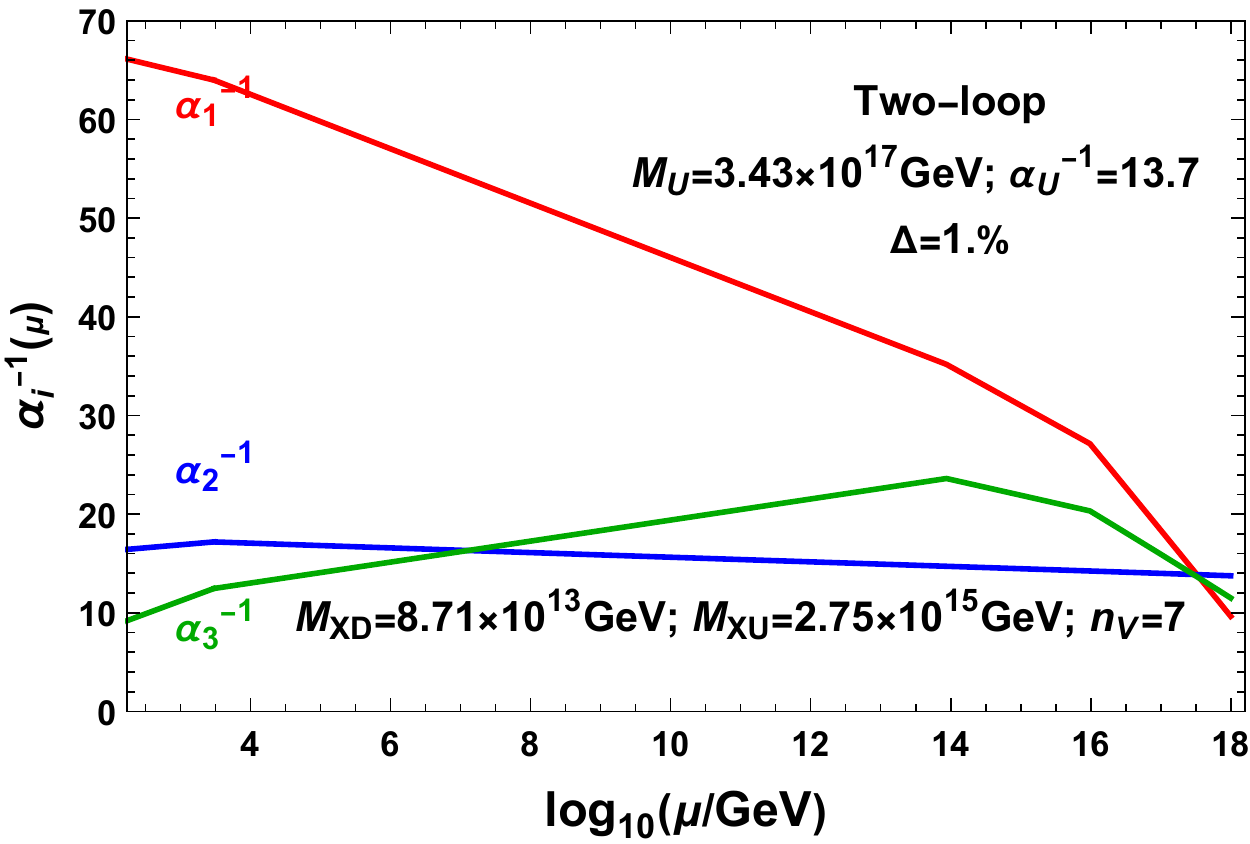}}  
	\caption{Two-loop evolution of gauge couplings for the \textbf{Model 26} with vector-like particles. In this model, $k_Y=\frac{25}{28}\times\frac{5}{3}$ and $k_2=\frac{11}{6}$. The string-scale gauge coupling relation  can be achieved by adding $7(XD+\overline{XD})+7(XU+\overline{XU})$.} 
	\label{fig:model8}
\end{figure}

\begin{figure}[H]\centering
	\subfigure[]{\includegraphics[width=0.45\linewidth]{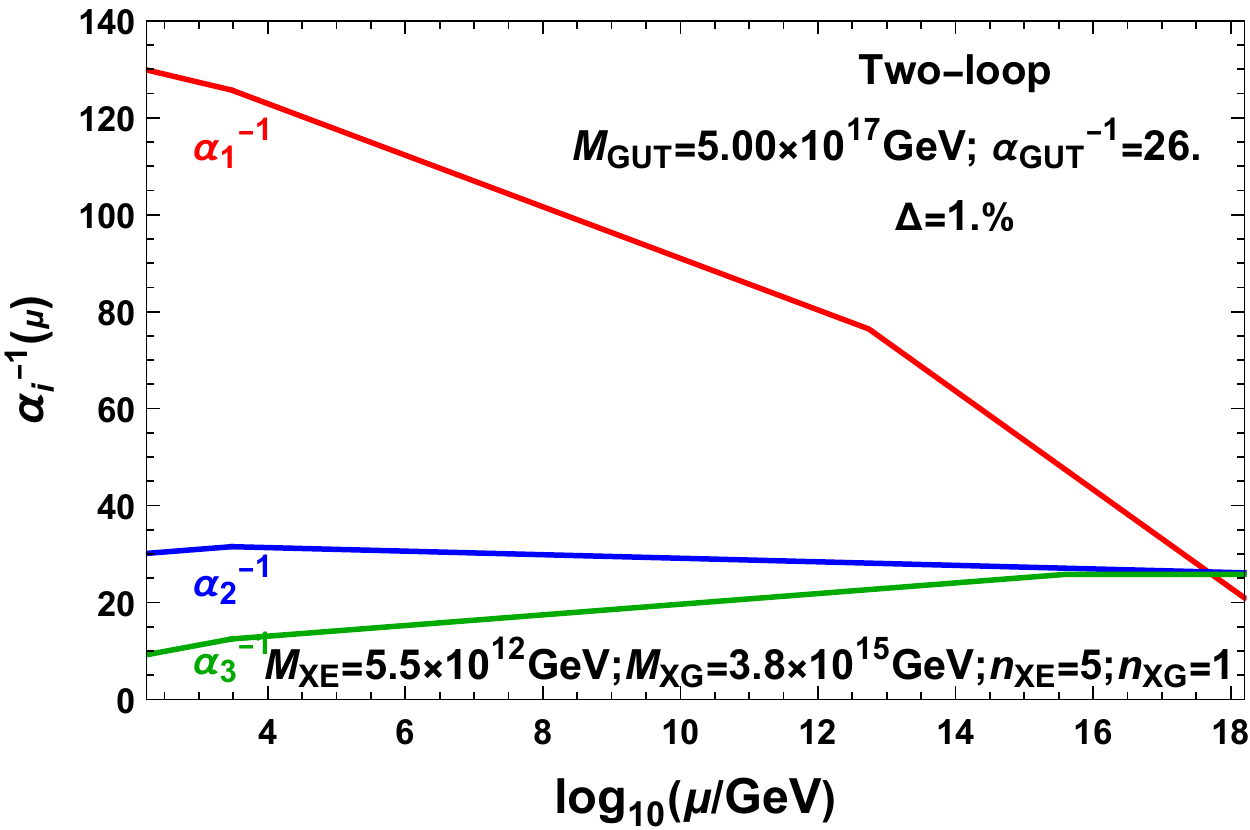}}\qquad
	\subfigure[]{\includegraphics[width=0.45\linewidth]{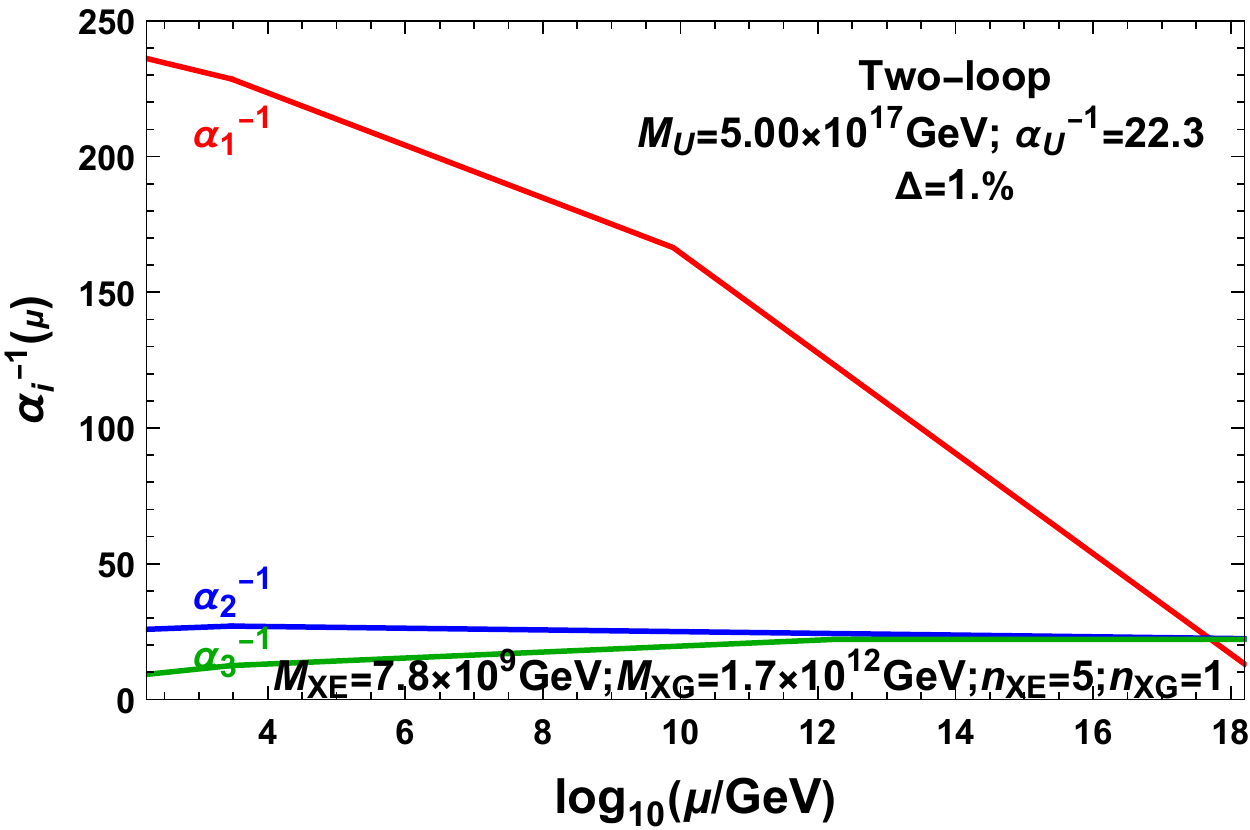}}
	\caption{Two-loop evolution of gauge couplings for the \textbf{Model 28} (a) and \textbf{Model29} (b) with vector-like particles. The string-scale gauge coupling relation  can be achieved by adding $5(XE+\overline{XE})+XG$. }
	\label{fig:model10-18}
\end{figure}

\begin{figure}[H]\centering
	\subfigure[]{\includegraphics[width=0.45\linewidth]{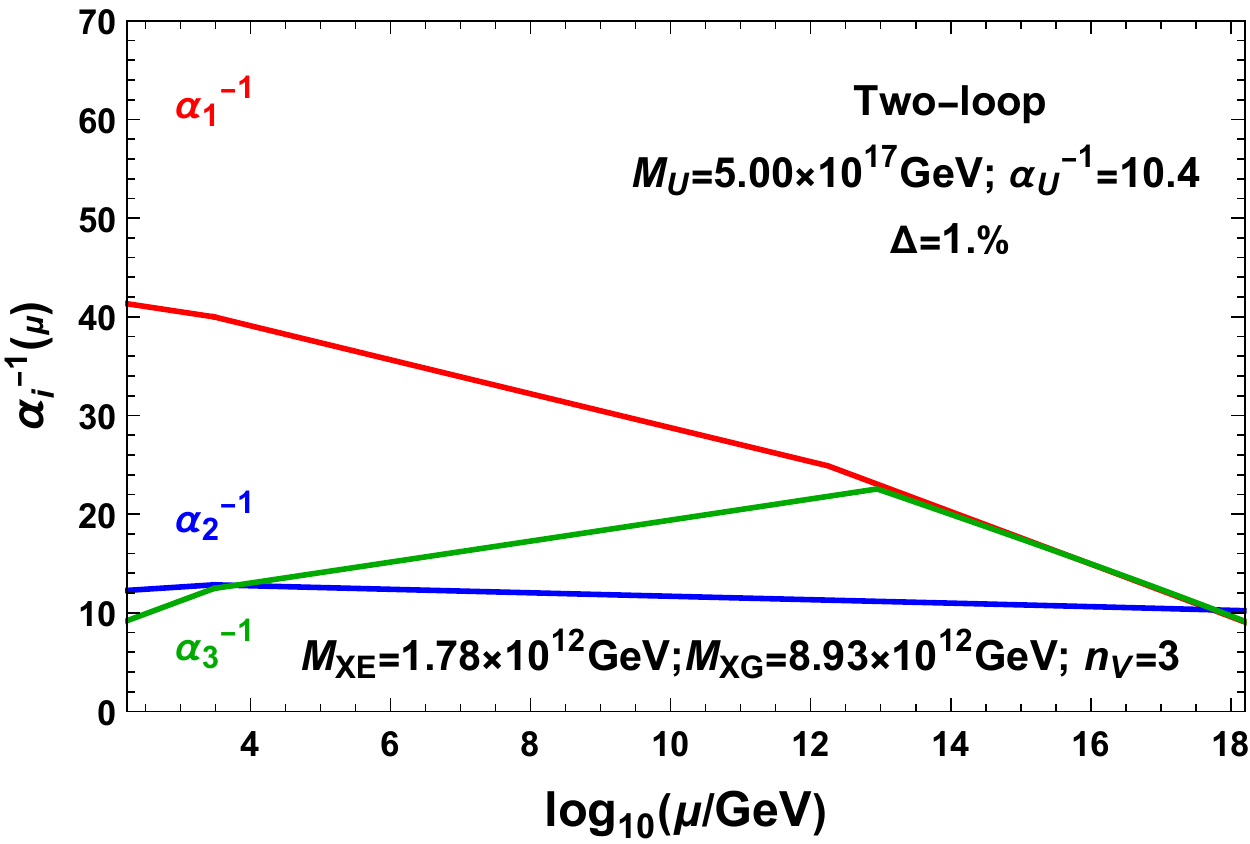}}\qquad
	\subfigure[]{\includegraphics[width=0.45\linewidth]{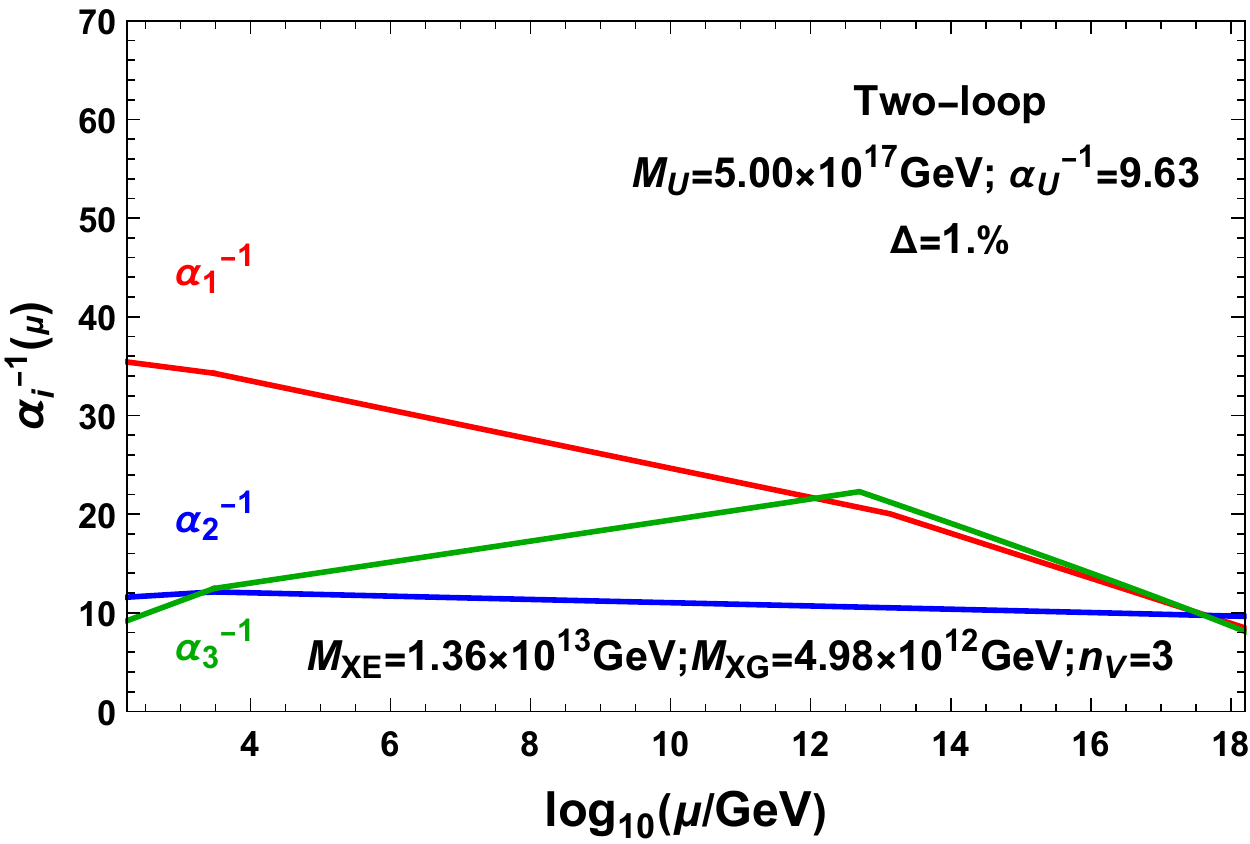}}
	\subfigure[]{\includegraphics[width=0.45\linewidth]{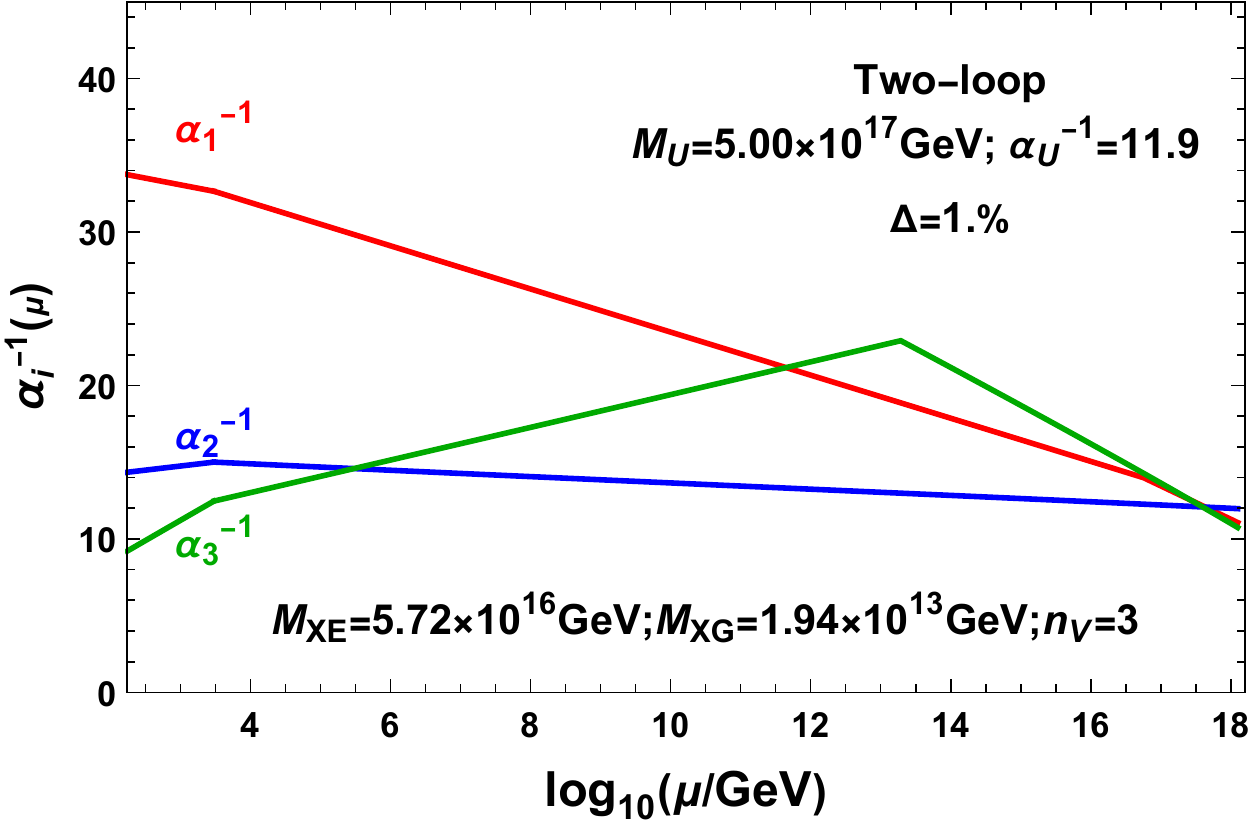}}\qquad
    \subfigure[]{\includegraphics[width=0.45\linewidth]{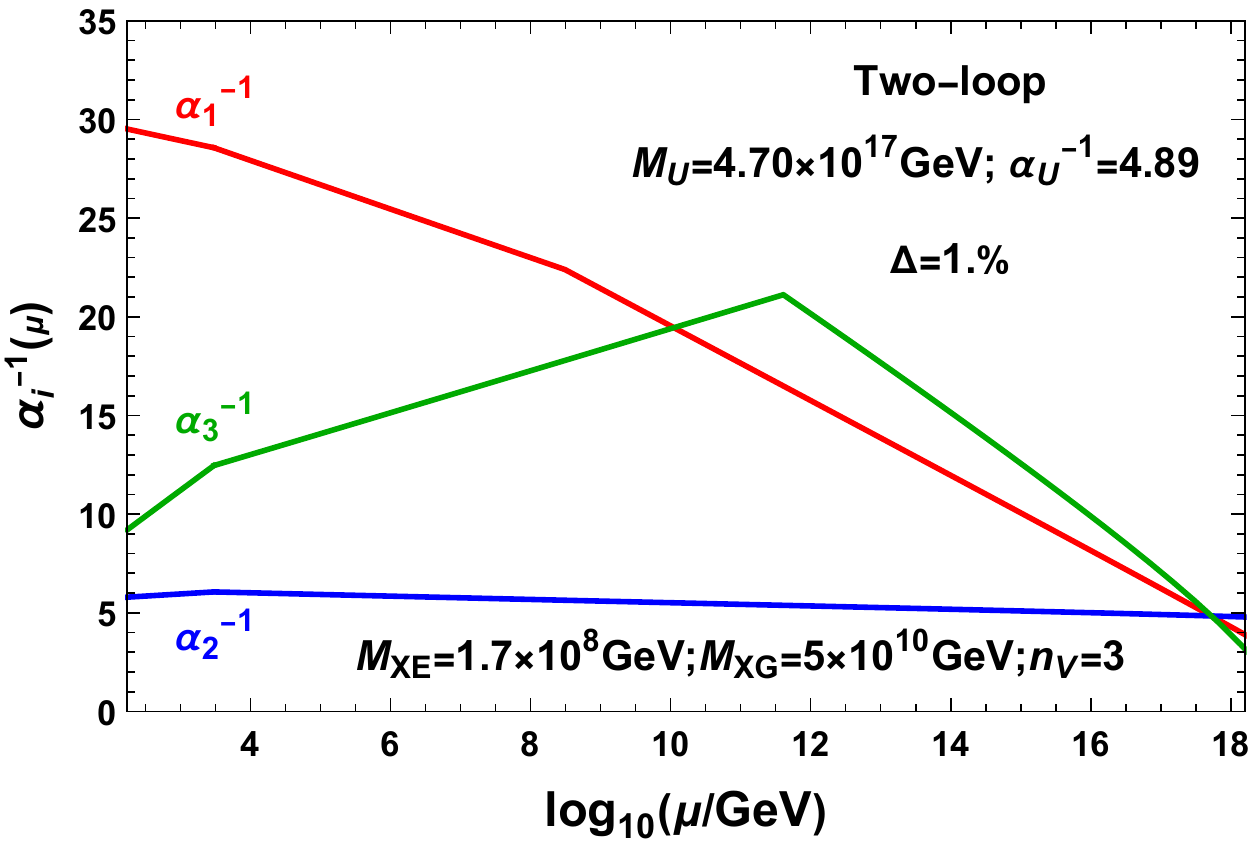}}
	\caption{Two-loop evolution of gauge couplings for the \textbf{Model 30} (a), \textbf{Model 31} (b), \textbf{Model 32} (c) and \textbf{Model 33} (d) with vector-like particles. The string-scale gauge coupling relation  can be achieved by adding $3(XE+\overline{XE})+3XG$.} 
	\label{fig:model22-25-27-32}
\end{figure}

\newpage
\bibliography{reference}

\end{document}